\documentclass[universe,review,accept,pdftex,moreauthors]{Definitions/mdpi} 

\firstpage{1} 
\pubvolume{1}
\issuenum{1}
\articlenumber{0}
\pubyear{2023}
\copyrightyear{2023}
\externaleditor{Academic Editor: Firstname Lastname}
\datereceived{ } 
\daterevised{ } 
\dateaccepted{ } 
\datepublished{ } 
\hreflink{https://doi.org/} 

\newcommand{\ttbar} {t\bar{t}}

\def  \st {{\it single top}}

\def\met  {$\not\!\!E_T$}

\def\ttbar{$t\bar{t}$}

\Title{Single-Top Quark Physics at the LHC: From Precision Measurements to Rare Processes and Top Quark Properties}

\TitleCitation{Single-Top Quark Physics at the LHC: From Precision Measurements to Rare Processes and Top Quark Properties}

\Author{J\'er\'emy Andrea $^{1}$\orcidA{}, Nicolas Chanon $^{2,}$*\orcidB{}}

\AuthorNames{J\'er\'emy Andrea, Nicolas Chanon}

\AuthorCitation{Andrea, J.; Chanon, N.}

\address{%
$^{1}$ \quad Universit\'e de Strasbourg, CNRS, IPHC UMR7178, 67000 Strasbourg, France; jeremy.andrea@cern.ch\\
$^{2}$ \quad Univ. Lyon, Univ. Claude Bernard Lyon 1, CNRS/IN2P3, IP2I Lyon, F-69622 Villeurbanne, France}

\corres{Correspondence: nicolas.pierre.chanon@cern.ch}

\abstract{Since the initial measurements of single-top quark production at the Tevatron in 2009, tremendous progress has been made at the LHC. While LHC Run 1 marked the beginning of a precision era for the single-top quark measurements in some of the main production mechanisms, LHC Run 2 witnessed the emergence and exploration of new processes associating top quark production with a neutral boson. In this paper, we review the measurements of the three main production mechanisms ($t$-channel, $s$-channel, and $tW$ production), and of the associated production with a photon, a $Z$ boson, or a Higgs boson. Differential cross-sections are measured for several of these processes and compared with theoretical predictions. The top quark properties that can be measured in single-top quark processes are scrutinized, such as $Wtb$ couplings and top quark couplings with neutral bosons, and the polarizations of both the $W$ boson and top quark. The effective field theory framework is emerging as a standard for interpreting property measurements. Perspectives for LHC Run 3 and the HL-LHC are discussed in the conclusions.}


\begin{document}

\section{\label{sec:singletopProd}Introduction}

After the discovery of the top quark~\cite{CDF:discovery,D0:discovery} in 1995 at the Fermilab Tevatron, the~CERN LHC era opened up many opportunities to investigate top quark processes.
Both at the LHC and the Tevatron, the~processes with the largest cross-sections for producing top quarks in proton--proton or proton--antiproton collisions are the \ttbar\ production modes. 
In addition to the \ttbar\ production, which arises from quantum chromodynamics (QCD) interactions, top quarks can be singly produced through electroweak interactions. This leads to the so-called single-top quark channels. 
The single-top quark production features many interesting properties owing to the V--A structure of the electroweak interaction. 
It shows specific sensitivities to parton density functions (PDFs),~the $V_{tb}$ matrix element of the CKM matrix, $Wtb$ coupling beyond the standard model (SM), and~top quark polarization, to name a few examples. 
Measuring inclusive cross-sections and differential cross-sections for single-top quark processes serves as an interesting test of perturbative QCD (pQCD). 
The associated production of a single-top quark with a boson offers insights into the coupling between the top quark and bosons, complementing the associated production of a boson with a \ttbar~pair.

\begin{figure}[H]
  \resizebox{11cm}{!}{\includegraphics{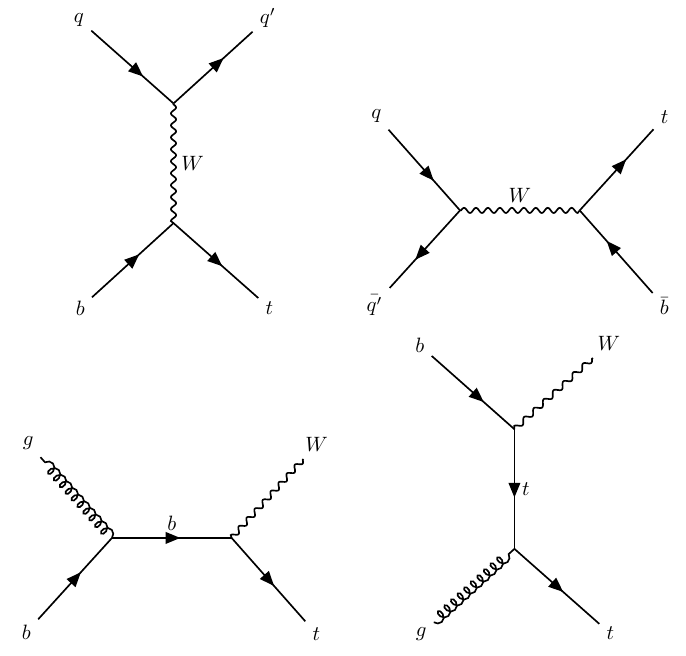}}
  \caption{Examples of Feynman diagrams for single-top production at LO: $t$-channel (\textbf{top left}), $s$-channel (\textbf{top right}), and $tW$ production (\textbf{bottom}).} 
  \label{fig:stopdiag}
\end{figure}

Three main production modes for single-top quark processes can be distinguished: production via the exchange of a virtual $W$ boson in the $t$- and $s$-channels, and~the associated production with a $W$ boson (tW production). 
The corresponding diagrams in the leading order (LO) in pQCD are presented in Figure~\ref{fig:stopdiag}.

The first observation of single-top electroweak production ($t$- and $s$-channels combined) made at the Tevatron ~\cite{D0:tchandiscovery,CDF:tchandiscovery} in 2009, followed by the observation of the $t$-channel~\cite{D0:2011aco}. 
The CDF and D0 collaborations performed simultaneous measurements of the $s$- and $t$-channel processes~\cite{D0:2013tnv,CDF:2014vfc}. 
The p-$\bar{\mathrm{p}}$ collisions at the Tevatron provided a unique setting for measuring the $s$-channel since the initial state of this process predominantly involves a light quark and a light antiquark, taken from the valence partons in the proton and antiproton. To date, the $s$-channel has been observed solely at the Tevatron~\cite{CDF:2014uma}; it remains to be observed at the LHC, although~there have been reports suggesting evidence of this process at both 8 TeV~\cite{ATLAS:2015jmq} and 13 TeV~\cite{ATLAS:2022wfk}.
At the LHC, the~largest cross-sections at \mbox{$\sqrt{s}=$ 13~TeV} (the center-of-mass energy of Run 2) are obtained for the $t-$channel ($214.2^{+4.1}_{-2.6}$ pb at NNLO with MCFM~\cite{Campbell:2020fhf}), followed by the $tW$ production ($79.3^{+2.9}_{-2.8}$ pb at NLO+NNLL~\cite{Kidonakis:2021vob}), and~the $s-$channel ($10.3^{+0.4}_{-0.4}$ pb at NLO with Hathor v2.1~\cite{Aliev:2010zk,Kant:2014oha}). 

\vspace{-6pt}

At the LHC, many new processes involving single-top quarks were measured in p--p collisions. The~$t$-channel production has been frequently measured and has been thoroughly investigated by evaluating differential cross-sections. Most of the top quark properties probed with single-top production channels are measured using the $t$-channel production since it yields the largest cross-section at the LHC among all the production mechanisms. The~LHC is able to observe the associated $tW$ production, for~which the differential cross-sections are even measured. This channel is of particular interest because at next-to-LO (NLO) in pQCD, it features interference with the \ttbar\ process. Understanding the nuances of this interference is still a focal point in the field.
As noted in an earlier review (Ref.~\cite{Giammanco:2015bxk}), Run 1 marked the start of a precision epoch in single-top quark measurements for~those main production mechanisms. This is ongoing, with~remarkable scrutiny focused on the $t$-channel and $tW$ production.

In addition, single-top quarks can be produced in tandem with neutral bosons. Those processes yield relatively low cross-sections; however, the additional boson in the lepton channel offers invaluable experimental leverage for measuring couplings or searching for new physics. 
This class of rare processes covers the production of a single-top quark with a photon ($t\gamma$),~a $Z$ boson ($tZ$), or a Higgs boson ($tH$). For~each of these processes, the~single-top quark can be produced via the $t$-channel, $tW$ production, or $s$-channel, with~the boson emitted from a quark line or a $W$ boson exchange. The~$t\gamma$ process was observed only at the end of Run 2. There was anticipation for the discovery of the $tZ$ process at the onset of Run 2; nowadays, it is being measured differentially and is utilized for property measurements. 
The analysis methodology employed in $tZ$ measurements is close to that of the searches for the $tH$ final states. 
With great similarity in the $tH$ and $t\bar{t}H$ final states and their common sensitivity to the top quark Yukawa coupling, the~$tH$ processes constitute a special case and are searched for simultaneously with the $t\bar{t}H$ production. 
Because of the destructive interference between processes where the Higgs boson emerges from a $W$ boson or from a top-quark line, the~cross-section for the $tH$ production is so small that evidence for such processes remains elusive.
However, it is already considered in several analyses because of its unique sensitivity to the sign of the top quark Yukawa coupling, which could lead to a large enhancement of its cross-section. 
When the review in Ref.~\cite{Giammanco:2017xyn} was published at the outset of Run 2, it signified the dawn of an era where the processes associating the production of a top quark with a neutral boson began to be measured. Run 2 saw the in-depth exploration of these processes, with~a particular emphasis on the $tZ$ production.

\begin{figure}[H]
  \resizebox{12cm}{!}{\includegraphics{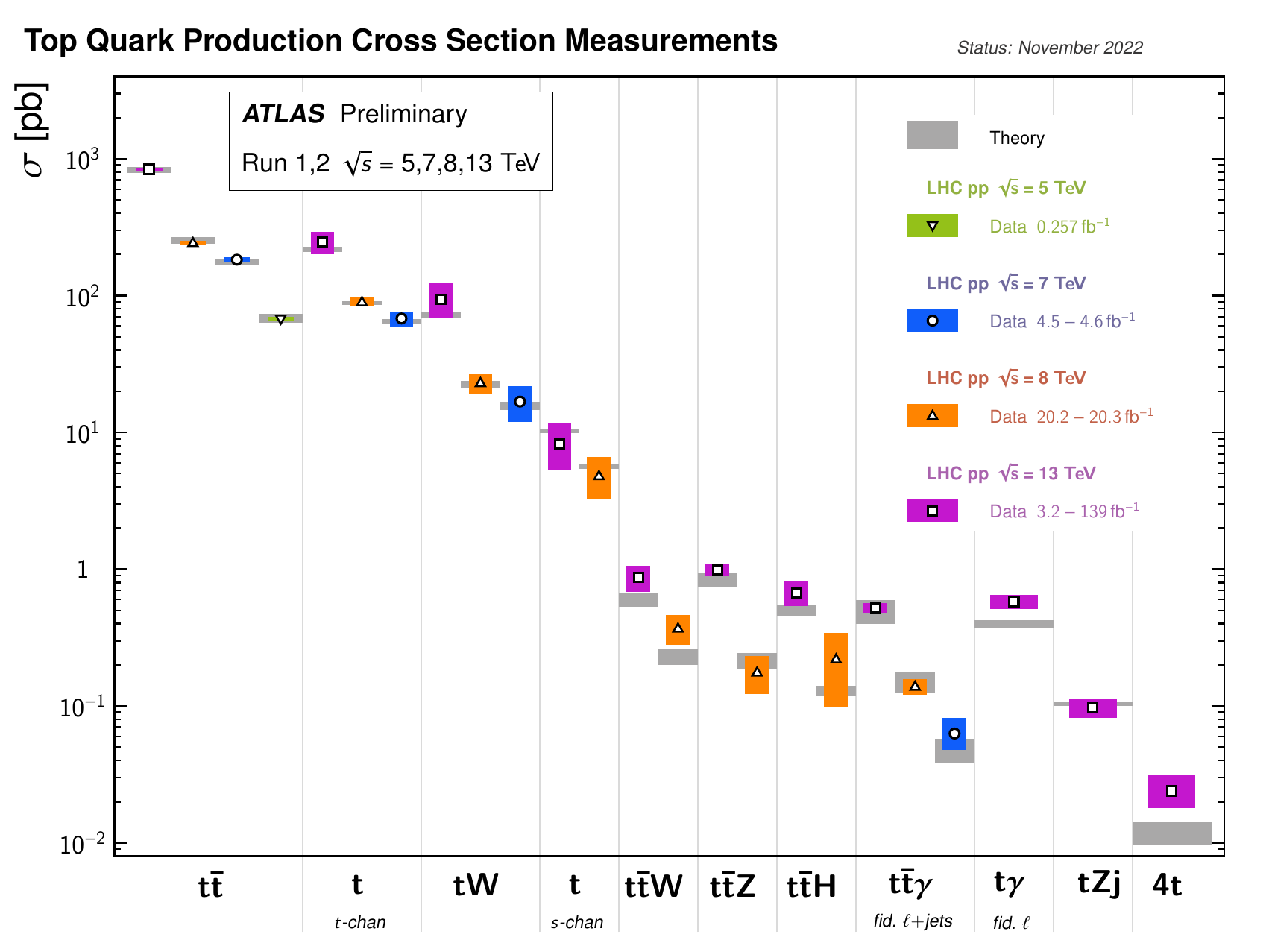}}
  \caption{Summary of cross-sections for top quark processes measured by ATLAS~\cite{ATL-PHYS-PUB-2023-014} and compared with SM~predictions.} 
  \label{fig:ATLAS_TopCrossSection}
\end{figure}

The cross-sections for all SM top quark processes measured by ATLAS are compared with theoretical predictions in Figure~\ref{fig:ATLAS_TopCrossSection}. The~cross-sections for single-top production (\mbox{$t$ + X}) are generally less than those for top pair production ($t\bar{t}$ + X). This top pair production acts as a large background in single-top quark~searches. 

\begin{figure}[H]
  \resizebox{11.5cm}{!}{\includegraphics{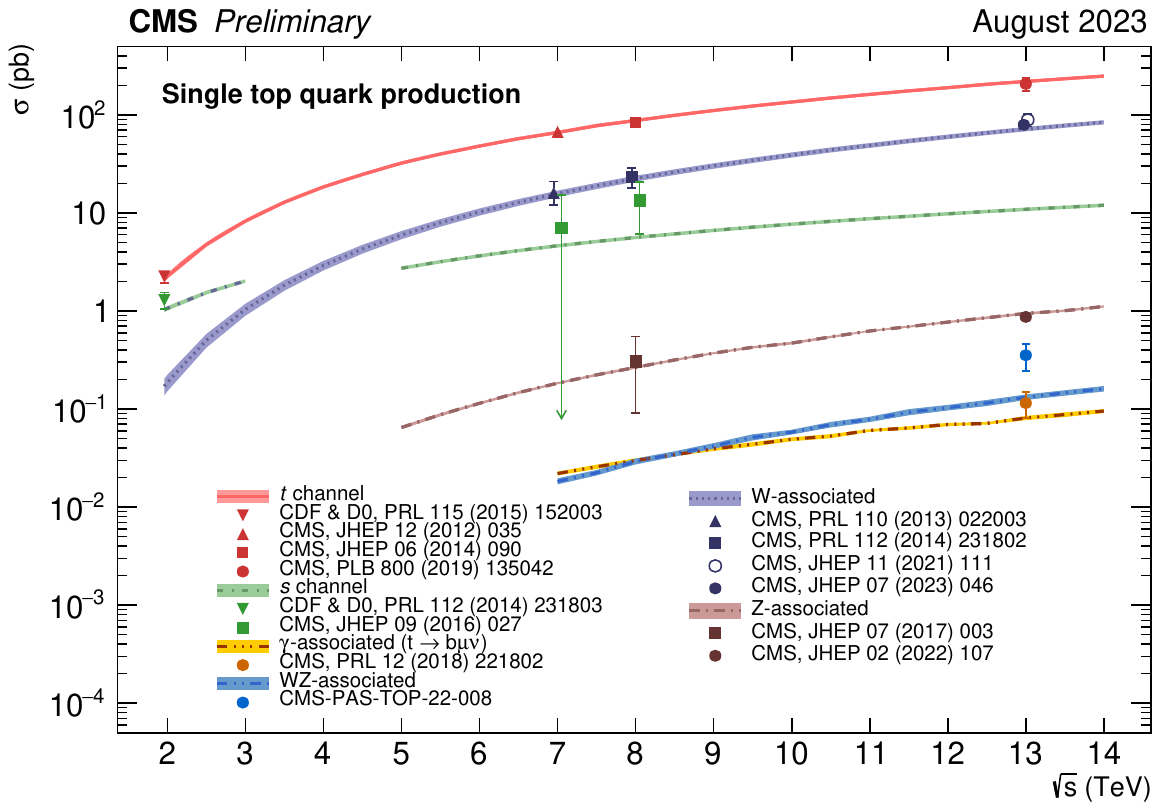}}
  \caption{Summary of measured cross-sections for single-top quark production at CMS~\cite{CMSsummaryFigures}. Theoretical calculations for the $t$-channel, $s$-channel, and~$W$-associated production have been provided by N. Kidonakis to the CMS collaboration.} 
  \label{fig:CMS_SingleTopCrossSection}
\end{figure}

A summary of the cross-section for single-top quark processes as measured at CMS is compared with theoretical predictions and presented as~a function of the center-of-mass energy, as seen in~Figure~\ref{fig:CMS_SingleTopCrossSection}. It can be observed that the cross-section for the $s$-channel process does not grow as fast as that of the $t$-channel process as a function of the energy, which makes the search for the $s$-channel more difficult with recent LHC runs. The~production of a single-top quark associated with a photon or $Z$ boson results in cross-sections that are lower than those observed in the $t$-channel, $s$-channel, or $tW$ production.

\vspace{-4pt}
\unskip

We will now focus on the top quark property measurements. It is notable that the cross-sections for single-top quark production are directly proportional to the square of $|V_{tb}|$. Therefore, it is possible to determine $|V_{tb}|$ from the measurements of single-top quark cross-sections. If~one assumes that $|V_{td}|, |V_{ts}| \ll |V_{tb}|$, the~$|V_{tb}|$ matrix element can be extracted~from the following:
\begin{linenomath}
\begin{equation}
|V_{tb}| = \sqrt{ \sigma_{st}/\sigma_{st} ({theo, |V_{tb}|=1})}, \label{equ:vtb}
\end{equation}
\end{linenomath}
where $\sigma_{st}$ is the measured cross-section and $\sigma_{st}^{theo, V_{tb}=1}$ is the expected cross-section for \mbox{$|V_{tb}|=1$}. Equation (\ref{equ:vtb}) also assumes that no new physics effect modifies the V--A structure of the $tWb$ interaction~vertex.

In single-top quark processes, the~$Wtb$ vertex appears in the top quark production and its decay, while in \ttbar\ production, it appears twice in the top quark decay. Therefore, the~Lorentz structure of the $Wtb$ coupling can be investigated in detail using decay information. 
The single-top quark production is also sensitive to the CP property of the $Wtb$ vertex (it is much more difficult to measure in \ttbar~production, where the CP symmetry is probed preferentially in the top quark--gluon coupling). The~$W$ boson polarization and top quark polarization can also be probed. 
For all of these properties, the~$t$-channel process is usually employed as a probe because of its large cross-section. Within~the $t$-channel, as well as in the associated production with a boson, modern tools, such as the SM effective field theory (EFT), are increasingly used to parametrize deviations from the SM in an almost model-independent way. This systematic approach of searching for signs of new physics is a novelty of Run~2.

The single-top quark production is indeed a sensitive probe in physics beyond the SM. 
The $t$-channel signature can occur via the exchange of a supersymmetric particle~\cite{Tait:2000sh}, resulting from the decay of a new heavy resonance, like a color-octet scalar~\cite{Drueke:2014pla} or a new resonance in technicolor models~\cite{Burdman:1999sr}. 
The $s$-channel shares the same final state as the possible decay of a $W'$ boson, which is predicted in many models beyond the SM, such as supersymmetric models with $R$-parity violation~\cite{Oakes:1997zg}, or~within the paradigm of universal extra dimensions~\cite{Burdman:2006gy}. 
The $tW$ final state, along with the $tZ$ and $tH$ states, are typical products of vector-like quark decays~\cite{Aguilar-Saavedra:2009xmz}. 
Excited top quarks, predicted in Randall--Sundrum models, can decay in the $t\gamma$ final state~\cite{Hassanain:2009at}.
The $tZ$, $t\gamma$, and $tH$ processes can also be modified by flavor-changing neutral currents (FCNCs) in the top quark production or decay, as predicted in several extensions of the SM, like the two-Higgs doublet model~\cite{Atwood:1996vj} (2HDM), supersymmetry~\cite{Cao:2007dk}, or in warped extra dimensions~\cite{Agashe:2006wa}. 
In the top quark property domain, for example, CP violation is predicted in the top-Higgs boson coupling within the complex 2HDM~\cite{Fontes:2017zfn}, and~in $Wtb$ coupling within supersymmetric models~\cite{Bi:1999is}. 
The~so-called mono-top quark signatures, designating the associated production of a single-top quark with a dark matter candidate, are areas that are being focused on in the quest for new physics (for a review, see~\cite{Behr:2023nch}).

The outline of this review is as follows. In~Section~\ref{SingleTopCrossSections}, after~a brief note on the generation of each single-top quark process, the~measurements of the three main single-top quark production mechanisms are presented: $t$-channel, $tW$ production, and $s$-channel. The \mbox{Section~\ref{SingleTopBoson}} will discuss the measurements of single-top quark production in association with a neutral boson (a photon, a~$Z$ boson, or a Higgs boson). Top quark property measurements with single-top quark production will be reviewed in Section~\ref{SingleTopProperties}, with~a focus on $V_{tb}$, the~$W$ boson, top quark polarization, the~structure of the $Wtb$ vertex, and the interpretation in terms of the SM EFT. The conclusions of this review will be presented in Section~\ref{Conclusions}.

\section{\label{SingleTopCrossSections}Precise and Differential Measurements of Single-Top Quark~Processes}
\unskip

\subsection{\label{t-channel}The $t$-Channel Process: The Production~Mode With The Largest Statistics}
\unskip

\subsubsection{Features of the $t$-Channel~Process}

\begin{figure}[H]
\includegraphics[width=0.7\textwidth]{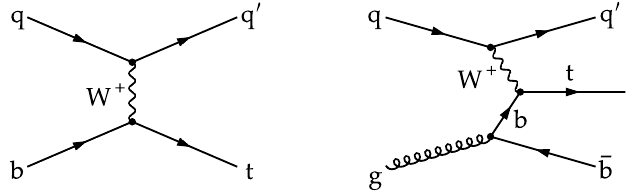}
\caption{\label{T-channel:Diagrams}Feynman diagrams for single-top $t$-channel production at the LO in pQCD~\cite{Sirunyan_2020_TOP_17_023}, in~the 5-flavor scheme (\textbf{left}), and 4-flavor scheme (\textbf{right}).}
\end{figure}

The so-called $t$-channel production mode features the largest cross-section among all single-top quark production modes. Top quarks produced in the $t$-channel are accompanied by a high $p_T$ light quark that is predominantly produced in the forward region of the detector ($|\eta| > 2.5$), and~of a low $p_T$ b-quark that often fails the minimum jet $p_T$ requirements in the analysis; as a result, it often remains experimentally invisible. Feynman diagrams are presented in Figure~\ref{T-channel:Diagrams}.

\vspace{-3pt}

The initial $b$ quark is accounted for in the theoretical calculation according to the 5-flavor scheme (5FS) or the 4-flavor scheme (4FS), as~shown in the Feynman diagrams in Figure~\ref{T-channel:Diagrams}. Each one considers a $b$ quark PDF within the proton (5FS); one can consider that the proton is made of light-flavored quarks in the sea (4FS), in~which case, the b-quark arises from virtual gluons. In~the 5FS, the~uncertainty associated with the PDF can be relatively large because~b-quark PDFs are not necessarily well known. On~the other hand,~4FS calculations usually suffer from higher sensitivity to QCD renormalization and factorization scales. The~decision as to whether to employ the 4FS or the 5FS is particularly important for the $t$-channel signature, where the additional jet (the so-called recoiling or spectator jet) is relatively forward, and its pseudorapidity ($\eta$) distribution is sensitive to the PDF.
It has been observed that the $\eta$ distribution of the recoiling jet ($\eta(j')$) in data is actually better described using the 4FS, while inclusive cross-sections are more accurately described with the~5FS.

With single-top quark production in the $t-$channel (as well in the $s-$channel), one of the incoming light quarks can be a valence quark of the proton, depending on whether a top or an antitop quark is being produced. This leads to a larger cross-section for top quark production ($134.2^{+2.6}_{-1.7}$ pb at $\sqrt{s}=$13 TeV, calculated at NNLO with MCFM~\cite{Campbell:2020fhf}) than for antitop quark production ($80.0^{+1.8}_{-1.4}$ pb).

\subsubsection{Experimental Techniques for the $t$-Channel~Measurement}

The $t-$channel process was the first single-top production mode observed at the LHC~\cite{bib:CMSFirstSingleTop,ATLASFirstSingleTop}, thanks to its large cross-section and its manageable signal-over-background ratio. For~this channel, clear discriminating observables exist between the signal and background, such as the $\eta(j')$ distribution.
Most of the $t$-channel analyses share a lot of common features in the event selection, background estimation, separation of the signal from the background, and~signal extraction.
The following paragraphs provide general descriptions of the analysis methods applied in $t$-channel measurements, and~are valid, to a large extent, to the other single-top quark measurements discussed in this~paper.

The top quark decays at almost 100\% to a $W$ boson and a b-quark. Top quark decays are said to be leptonic ($t\rightarrow b W \rightarrow b l \nu$) or hadronic ($t\rightarrow b W \rightarrow b q q'$). The~hadronic decay of top quarks produced in the single-top $t$-channel leads to a signature with several jets, and suffers from an overwhelming QCD multijet background. For~this reason, only the leptonic decay of the top quark is usually studied. The~experimental signature for the analysis presented here targets the leptonic decay products from the $W$ bosons: a charged lepton (electron or muon potentially arising from tau lepton decay), and~the presence of a significant missing transverse energy \met\ originating from a neutrino. Leptons are accompanied by a (mainly) forward light-quark jet and a b-quark jet arising from the top quark~decay.

The data sample considered usually selects events with a trigger requiring at least one lepton with a large $p_T$ isolated from hadronic activity. The~usage of b-quark identification (``b-tagging'') at the trigger level was investigated in earlier analyses~\cite{tchanBtagTrigger}, but~was found to add a significant complexity for a limited gain, especially with increasing luminosity. 
To summarize a typical event selection, the~presence of only one high $p_T$-charged lepton (electron or muon with $p_T>20$ GeV) is required, with~a significant missing transverse energy (\met$>$ 30 GeV) and the~presence of at least two high $p_T$ jets ($p_T>$ 30 GeV), with one of them from a b-quark and the other failing this requirement, while possibly being detected in the forward region ($|\eta(j')|<4.7$).

The backgrounds can be classified as arising from two main sources: events containing a charged lepton produced from a boson decay (referred to as prompt lepton), and~events with hadronic objects misidentified as prompt leptons. Given that prompt leptons are typically distanced from significant hadronic activity while non-prompt leptons are surrounded by hadrons, a potent method to reject the prevalent QCD multijet background is to require the charged lepton to be isolated. An isolation variable is devised by accumulating the hadronic energy around charged leptons, and this needs to be small. As~the modeling of a non-prompt background is hardly well-simulated, non-prompt backgrounds are usually estimated directly from the data, possibly leading to large systematic uncertainties. 
This estimate is performed, for~instance, by~inverting the lepton isolation requirement, thus enriching the events in QCD multijet processes. The~shape of a distribution of interest is then used as a data-driven estimate of the non-prompt lepton~background. 

The major prompt lepton background events are chiefly from \ttbar\ production with semi-leptonic decays, where jets are not well reconstructed or do not pass the b-tagging requirements. The~\ttbar\ process has been extensively studied; precise measurements have been confronted with theoretical predictions. This process is well described by the state-of-the-art Monte Carlo (MC) generators, such that single-top measurements rely on simulations to describe \ttbar\ kinematics, while the normalization is usually estimated or constrained from~data.

The associated production of a single $W$ boson with additional jets, referred to as ``$W$+jets'' 
 in the following, constitutes the second main source of background events. The $W$+jets processes were measured at the LHC, and~the event kinematics show a good agreement between data and MC predictions. However, the~kinematics of the $W$+jets process varies slightly depending on the flavor of the additional jets. For~this reason, several analyses actually split the $W$+jets simulation into $W$+b, c, or light jets, measuring the normalization of each contribution~separately.

Finally, other subdominant processes after the selection contribute to the background events, such as the Drell--Yan production, when one of the two leptons is not reconstructed or does not pass the lepton selection. These processes are usually estimated from~simulations.

\begin{figure}[H]
  \resizebox{6cm}{!}{\includegraphics{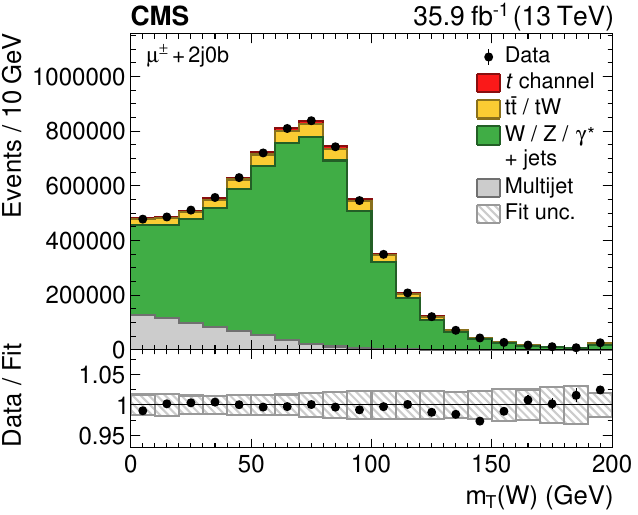}}
  \resizebox{6cm}{!}{\includegraphics{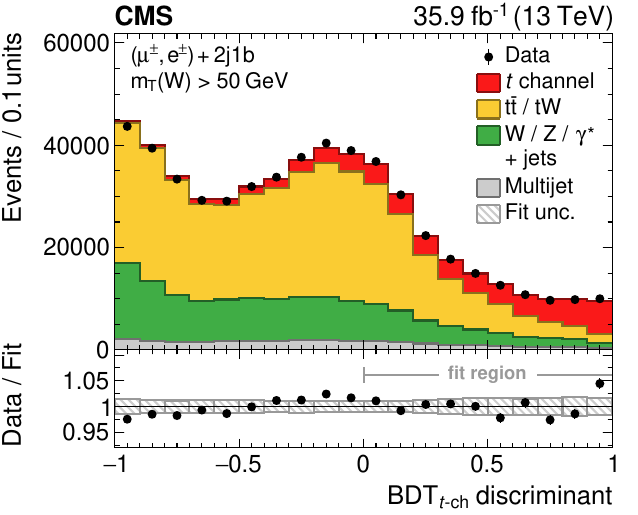}}
  \caption{Distribution of the transverse mass of the $W$ boson in the muon channel of the $2j0t$ region~(\textbf{left}), the BDT discriminant in the $2j1t$ category (\textbf{right})~\cite{Sirunyan_2020_TOP_17_023}.} 
  \label{fig:t-channel_mTW_BDT_CMS}
\end{figure}

The $t$-channel analyses capitalized on the rise of the LHC profile likelihood \linebreak \mbox{method~\cite{plr2,CMS-NOTE-2011-005}} to simultaneously estimate the background contributions and constrain the systematic uncertainties from the data. 
Background normalization is adjusted within the fit, possibly using control regions enriched in background events, usually defined by jet and b-tagged jet multiplicities. 
For instance, the~\ttbar\ background can be controlled by fitting events with at least three jets and two b-tagged jets ($3j,2t$).
The $W$+jet background can be controlled with events containing two jets and no b-tagged jet ($2j0t$), using the distribution in the transverse mass of the $W$ boson ($m_T(W)$), showing a broad resonance for $W$ bosons, as~shown in Figure~\ref{fig:t-channel_mTW_BDT_CMS}.
The signal events are mainly expected in the region defined by asking for two jets, one of which is a b-tagged jet ($2j1t$). 
The signal is extracted from a combined fit in ($3j,2t$), ($2j0t$), and the ($2j1t$) regions.
Discriminating observables in each of these regions are fitted together with common nuisance parameters representing the systematic~uncertainties. 

Several distributions can be used to discriminate the signals from backgrounds. In~the early versions of analyses, the~most obvious observables included the pseudorapidity of the recoiling jet or the reconstructed top-quark mass. 
In the most precise measurements, the~discriminating variables in the ($3j,2t$) and ($2j1t$) regions are constructed from multivariate analyses, such as boosted decision trees (BDTs) or neural networks (NNs), using various kinematic observables as input. An~example is shown in Figure~\ref{fig:t-channel_mTW_BDT_CMS}.
In the latest published cross-section measurement at 13 TeV~\cite{Sirunyan_2020_TOP_17_023}, the~BDTs are trained using input variables related to the absolute value of the pseudorapidity of the untagged jet, $|\eta(j')|$, the~reconstructed top quark mass, the~transverse $W$ boson mass, $m_T(W)$, the~distance in the $\eta-\phi$ space between the b-tagged and the untagged jet, $\Delta R(b, j')$, the~absolute difference in the pseudorapidity between the b-tagged jet used to reconstruct the top quark and the selected lepton, $|\Delta \eta(b, l)|$. 

\begin{figure}[H]
  \resizebox{6cm}{!}{\includegraphics{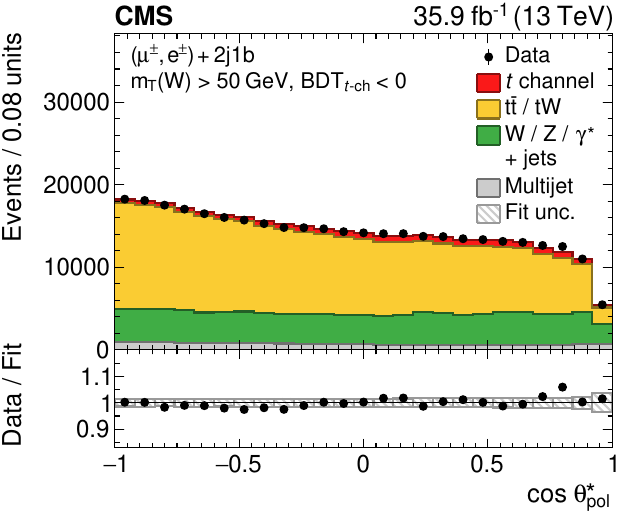}}
  \resizebox{6cm}{!}{\includegraphics{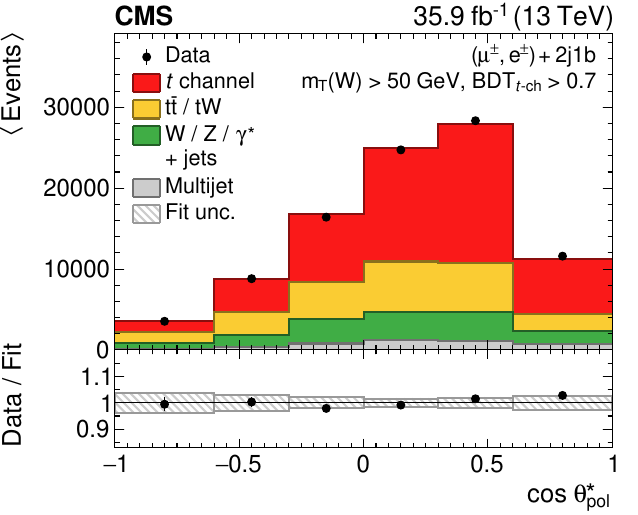}}
  \caption{Distributions in the cosine of the top quark polarization angle with a background-dominated selection (\textbf{left}) and a signal-enriched selection (\textbf{right}) for events in the $2j1t$ region~\cite{Sirunyan_2020_TOP_17_023}.} 
  \label{fig:tcahdiff_cossthetastar}
\end{figure}

Thanks to the large amount of integrated luminosity collected at the LHC, the~uncertainties related to the $t$-channel measurements are no longer statistically dominated. 
Remarkably, one can even select a relatively pure sample of $t$-channel events by applying stringent requirements on the BDT discriminants, as~illustrated in 
Figure~\ref{fig:tcahdiff_cossthetastar} (taken from~\cite{Sirunyan_2020_TOP_17_023}), showing the distribution in the cosine of the top quark polarization angle $\cos{\theta^*}$ in a background-enriched region (with the requirement of $BDT_{t-ch}<0$) and in a signal-enriched region ($BDT_{t-ch}>0.7$). The~sample can be vastly enriched in signal events while still providing a large event~yield.

The main sources of systematic uncertainties impacting $t$-channel measurements can be summarized as~follows:
\begin{itemize}
	\item Integrated luminosity: Typically a few percent (depending on the dataset).
	\item Signal and background modeling from SM theoretical predictions: Uncertainties in the modeling of signal acceptance and in the modeling of distributions used as discriminant observables are usually major sources of systematic uncertainty in top quark physics. This includes renormalization and factorization scale variations (accounting for missing higher-order contributions in pQCD), parton shower and hadronization, PDFs, the~choice of a matching scheme between fixed-order predictions and the parton shower, the~choice of flavor scheme (4FS or 5FS), and~MC statistical uncertainty. These uncertainties are treated by generating various MC samples, or~including various event weights in the MC samples, with~generation parameters varied up and down. The~same uncertainties are also included for most backgrounds, which are estimated from simulations. 
	\item Data-driven background estimate: Due to the inadequate representation in simulations of jets misidentified as leptons, the~non-prompt lepton background is directly estimated from the data. 
	Usually, these estimations are complicated and rather imprecise. It is rare to lower the relative systematic uncertainty below 30\% before~any constraints from the fit.
	\item Simulation-to-data corrections: Several corrections (so-called scale factors) to the reconstructed objects are applied to the simulation to improve its agreement with the data. These corrections are derived from dedicated analyses estimating the associated systematic uncertainties. The~corrections are typically related to trigger and lepton selections, jet energy scale and resolution, and~b-tagging. 
\end{itemize}

In the most recent analyses, the~statistical uncertainty provides a small contribution to the total uncertainty (less than 5\%). 
The relative sizes of the systematic uncertainties depend on the analysis strategy; for instance, the~choice of the discriminating observable matters. 
The use of the $|\eta_{j'}|$ distribution naturally leads to large uncertainties on the jet energy scale and resolution (up to about 5\%), since controlling such corrections in the forward part of the detector is difficult. 
Using a multivariate discriminant can significantly reduce the jet energy scale and resolution uncertainties to a few percent, most likely due to the higher signal purity and increased constraining power. 
Another large source of systematic uncertainty arises from the signal modeling, which can be lowered by performing a fiducial measurement, as~described in Section~\ref{sec:fiducial}. Fiducial measurements are performed within a generator-level acceptance to avoid the extrapolation from the visible phase space to the full process phase space, thus reducing the modeling~uncertainties.

\subsubsection{Summary of the Latest Measurements of the $t$-Channel-Inclusive Cross-Section}

\begin{figure}[H]
\begin{adjustwidth}{-\extralength}{0cm}
\centering
\includegraphics[width=14.5cm]{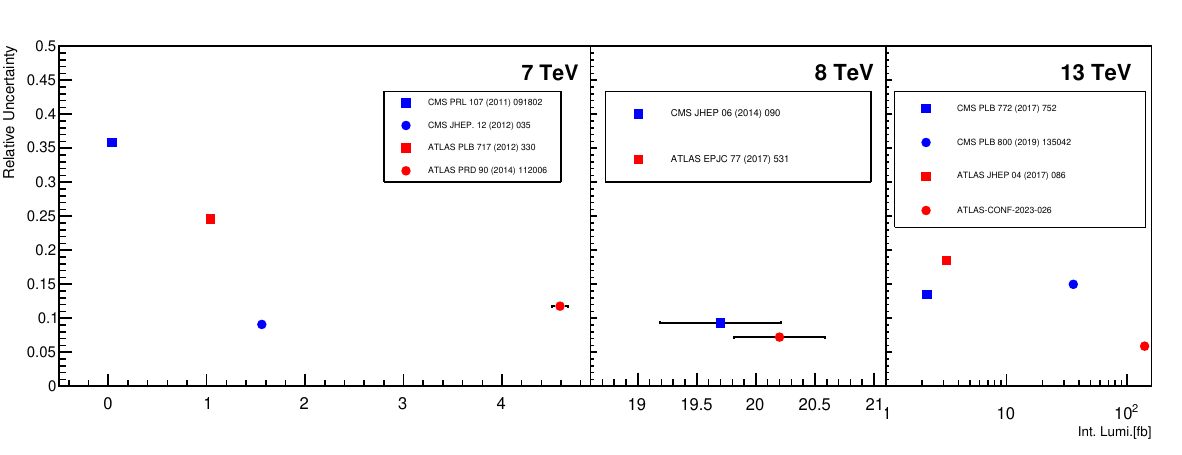}
\end{adjustwidth}
\caption{\label{fig:tchan_RelativUncertEvolution}Evolution of the relative total uncertainty in the inclusive $t$-channel cross-section measurements, plotted as a function of the integrated luminosity at $\sqrt{s}=$7, 8, and 13~TeV.}
\end{figure}

A summary of the latest measurements of the cross-section for $t$-channel production at $\sqrt{s}=$7, 8, and 13 TeV from the ATLAS and CMS collaborations can be seen in Table~\ref{tab:singletop_t-channel_measurements}, where the combinations made by the LHC$top$WG are also shown when available. 
Figure~\ref{fig:tchan_RelativUncertEvolution} shows the relative total uncertainty of the $t$-channel cross-section $\Delta \sigma_{t-chan}/\sigma_{t-chan}$ as a function of the integrated luminosity accumulated at different center-of-mass energies. 
The most precise 13 TeV measurement was recently released, as~a conference note, by~the ATLAS collaboration~\cite{ATLAS-CONF-2023-026}. The~$t$-channel cross-section has also recently been measured at 5.02 TeV by the ATLAS collaboration~\cite{ATLAS-CONF-2023-033}.

In Table~\ref{tab:singletop_t-channel_measurements}, it quickly becomes evident that the statistical uncertainty soon turns into a secondary source of uncertainty, with precision measurements primarily dominated by systematic uncertainties. 
Comparing the most precise result measured at 7 TeV~\cite{JHEP05.2019.088} with the published measurement at 13 TeV~\cite{Sirunyan_2020_2}, it becomes clear that while several systematic uncertainties related to detector effects and background estimates have decreased, there is an increase in all the theory uncertainties related to the signal modeling.
While experimental systematic uncertainties can be reduced further, a~significant improvement in the total precision of the inclusive $t$-channel cross-section requires effort in the signal modeling involving the theory community. The~largest uncertainties that are common to both ATLAS and CMS are related to the parton shower used in the simulation samples of the $t$-channel and \ttbar~processes. Sources of large uncertainties can include the choice of the parton shower algorithm, the~matching scheme used to interface the NLO fixed-order matrix element with the parton shower, models of hadronization, or final state radiation. Prescriptions should be refined and agreed within ATLAS and CMS (an ongoing effort), and~work is needed to decrease the uncertainty based on physics arguments. For~instance, improved algorithms, such as antenna-based parton showers, could be tested~\cite{Fischer:2016vfv}. The~developments of parton showers at NLL~\cite{Dasgupta:2020fwr} or even higher accuracy~\cite{FerrarioRavasio:2023kyg} could bring about large improvements in the future. Eventually, exploring in situ constraints of these uncertainties from ancillary measurements in data~\cite{Corcella:2017rpt} is another path to consider.

\begin{table}[H] 
\caption{\label{tab:singletop_t-channel_measurements}
Summary of the most recent and precise $t$-channel cross-sections from the ATLAS and CMS collaborations, and~their combinations for 7 and 8 TeV.}
\tablesize{\footnotesize}
\begin{tabularx}{\textwidth}{CCC}
\toprule
 	& \textbf{Cross-Section (pb)}	& \boldmath{$\Delta \sigma_{t-chan}/\sigma_{t-chan}$}\\
\midrule
7 TeV  								  &        &  \\ 

 ATLAS~\cite{PhysRevD.90.112006}    &   $ 68 \pm 2 \pm 8 \pm 1$     & 0.122 \\  
 
 CMS~\cite{tchanBtagTrigger}  &   $67.2\pm 3.7 \pm 4.6 \pm 1.5$    & 0.091 \\ 

 Combination~\cite{JHEP05.2019.088}    &  $ 67.5 \pm 2.4 \pm 5.5 \pm 1.1 $   & 0.090\\ 

\midrule
8 TeV     &   								& \\  

 ATLAS~\cite{Aaboud_2017}    &  $89.6 \pm 1.2 ^{+6.8}_{-5.9} \pm 1.7$  & 0.076   \\  

 CMS.~\cite{Khachatryan_2014}  &  $83.6 \pm 2.3 \pm 7.1 \pm 2.2$    & 0.093    \\ 

 Combination~\cite{JHEP05.2019.088}     &  $87.7 \pm 1.1 \pm 5.5 \pm 1.5$   & 0.066 \\ 

\midrule
13 TeV     &   \\  

 ATLAS~\cite{Aaboud_2017_2}      &  $247 \pm 6 \pm 45 \pm5$     & 0.185   \\  

 ATLAS~\cite{ATLAS-CONF-2023-026}     &  $221 \pm 13$     & 0.059  \\  

 CMS~\cite{Sirunyan_2020_2}      &  $207 \pm 2 \pm 30 \pm 5$    & 0.147  \\ 
\bottomrule
\end{tabularx}
\end{table}
\unskip
\vspace{-12pt}

\unskip

\subsubsection{Measurements of Fiducial and Differential Cross-Sections\label{sec:fiducial} }

With the latest investigations into the $t$-channel process, leveraging the extensive statistics of LHC Run 2, differential cross-section distributions ~\cite{Sirunyan_2020_TOP_17_023} were measured. 
Differential cross-sections provide precious information on the theory modeling, and~can also be used to constrain the parameters of the EFT. Differential measurements are critical for a deeper understanding of the $t$-channel process and for identifying any deviations from SM~predictions.

The distributions measured at the reconstructed level are unfolded to theoretically well-defined observables,~correcting for detector and acceptance effects. 
The basic principle consists of determining corrections from simulations to infer the ``true'' top quark properties, by~accounting for the signal acceptance induced by the selection for~the detector resolution and the efficiencies. The~unfolded distributions can be compared in a robust way with theoretical predictions. 
Two fundamental unfolding levels are usually defined in top quark~physics:
\begin{itemize}
	\item \bf Parton level: 
 \rm Corresponds to the generated on-shell top quarks after~QCD radiative corrections.
	\item \bf Particle level: \rm Corresponds to (pseudo-)top quarks reconstructed from simulated particles after QED and QCD radiation, particle decay, and hadronization, with~a dedicated algorithm. 
\end{itemize}

\begin{figure}[H]
  \resizebox{6cm}{!}{\includegraphics{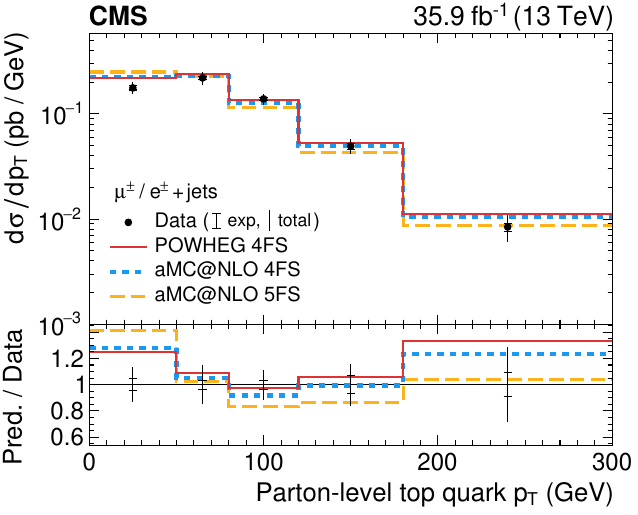}}
  \hspace{-6pt} \resizebox{6cm}{!}{\includegraphics{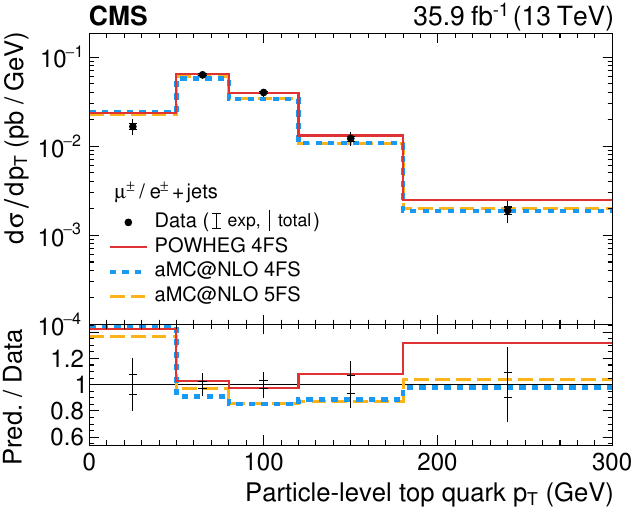}}
 \resizebox{6cm}{!}{\includegraphics{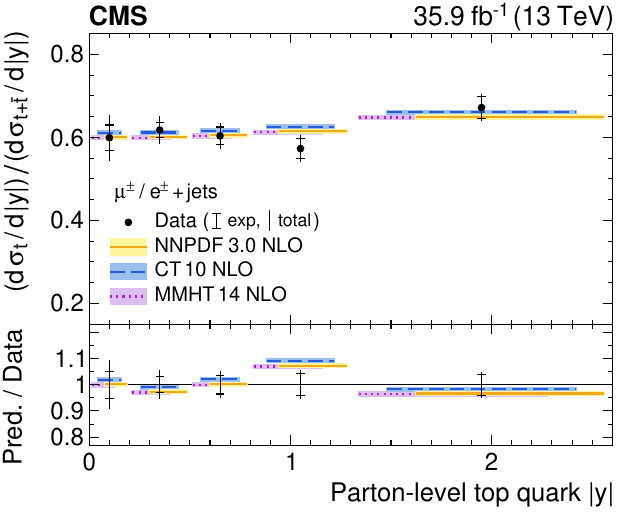}}
  \resizebox{6cm}{!}{\includegraphics{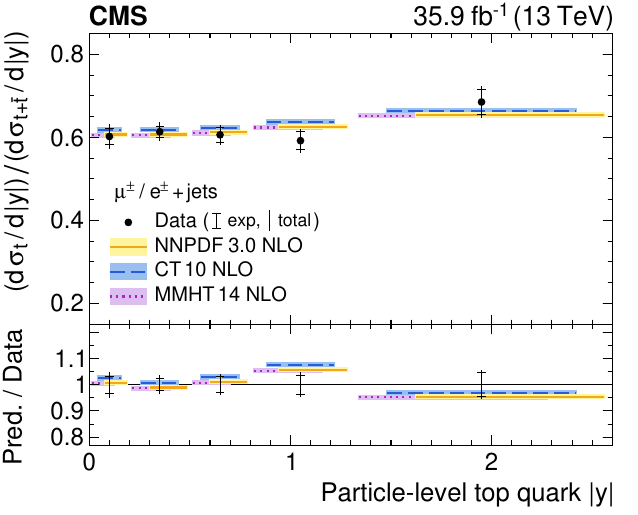}}
  \resizebox{6cm}{!}{\includegraphics{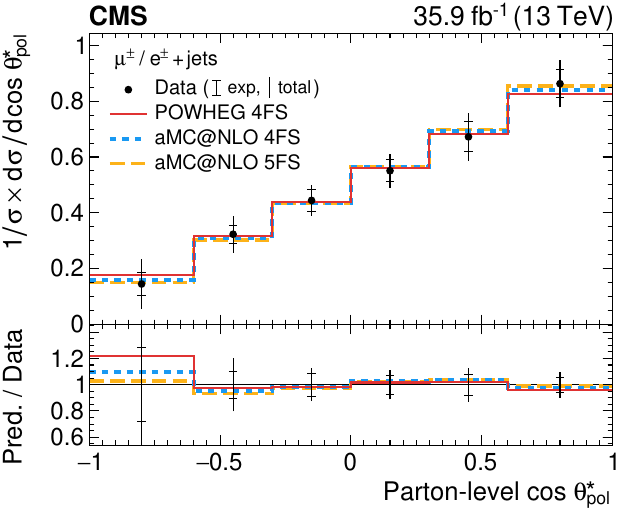}}
 \hspace{48pt}  \resizebox{6cm}{!}{\includegraphics{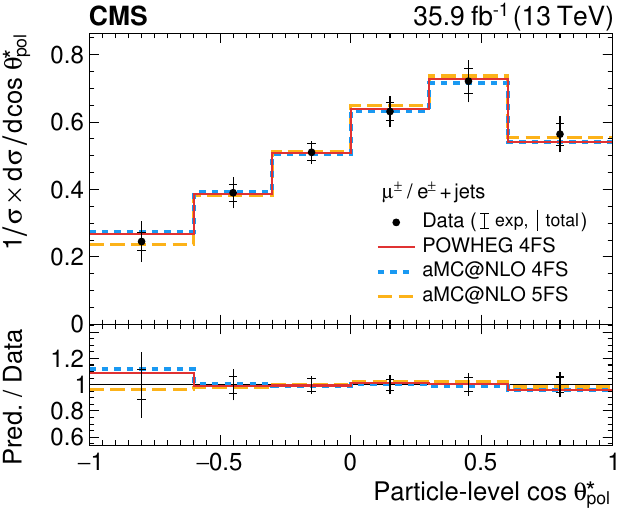}}
  \caption{Unfolded differential cross-section measurements: top quark $p_T$ (\textbf{upper row}), charge ratio as a function of the rapidity $y$ (middle row), and $\cos{\theta^*}$ (\textbf{bottom row}) at the parton level (left) and particle level (\textbf{right})~\cite{Sirunyan_2020_TOP_17_023}.} 
  \label{fig:tchan_unfolded}
\end{figure}

With the definitions adopted in~\cite{Sirunyan_2020_TOP_17_023}, measurements unfolded to parton and particle levels are confronted with NLO theoretical predictions for various observables, like the top-quark $p_T$, rapidity $y$, $\cos{\theta^*}$, or $W$ boson $p_T$. Beyond~the differential cross-sections, the~charge ratios of the cross-section $\sigma_t$ to $\sigma_{t+\bar{t}} $ are also measured. This observable is sensitive to the PDFs. 
Figure~\ref{fig:tchan_unfolded} presents examples of differential cross-sections and cross-section ratios. 
The measurements show good agreement between data and NLO predictions, validating our understanding of the electroweak interactions in the production of single-top quarks for most of the observables that were scrutinized. 
However, the precision reached (even in the differential cross-sections normalized to the total cross-section, thus canceling the impacts of several uncertainties) is not yet completely sufficient to unambiguously determine which generator agrees best with the data. 
As noted by the authors, a~few trends can still be highlighted. The predictions with the 4FS well describe the $W$ boson $p_T$ while the 5FS does not; neither  the 4FS nor 5FS predictions are able to nicely reproduce the entire distribution of the top quark $p_T$ in the data. 
This latter trend can be confirmed with deeper studies in the future since a possible mis-modeling could be of great importance for measurements of top quark properties employing the $t$-channel, and~in measurements or searches in which the SM $t$-channel is an important background (for instance, the $s$-channel searches). 
It should be noted that including $t$-channel differential distributions in PDF fits allows for reducing the gluon and light quark PDF uncertainties~\cite{Nocera:2019wyk}; for this purpose, it is essential to release experimental correlation matrices, which were not made public in the latest~measurements.

The so-called fiducial cross-section is defined at the particle level, and~is less sensitive than the inclusive cross-section to the systematic uncertainties arising from signal modeling. 
In inclusive cross-section measurements, the~number of signal events is measured in the visible phase space at the reconstructed level, defined by the detector acceptance and selection efficiencies. The~observed number of events is then extrapolated to the full phase space based on simulations. This extrapolation induces a large systematic uncertainty related to the modeling of signal events in the simulation. 
In contrast, the~measurement of the fiducial cross-section is performed in the visible phase space, and~extrapolated to the fiducial phase space volume that is defined as close as possible to the phase space of the selected dataset. 
The fiducial single-top $t$-channel cross-section was measured~\cite{Aaboud_2017} at $\sqrt{s}=$ 8 TeV and led to a reduction of about 2\% in the size of the systematic uncertainties related to the QCD scale and the NLO matching. This resulted in a significant improvement in the precision, and provided a robust method for comparing data with theoretical calculations.

\subsection{The $tW$ Process, and~Its Interplay with the $t\bar{t}$ Process}
\unskip

\subsubsection{Introduction to the $tW$ process}

\begin{figure}[H]
\includegraphics[width=0.25\textwidth]{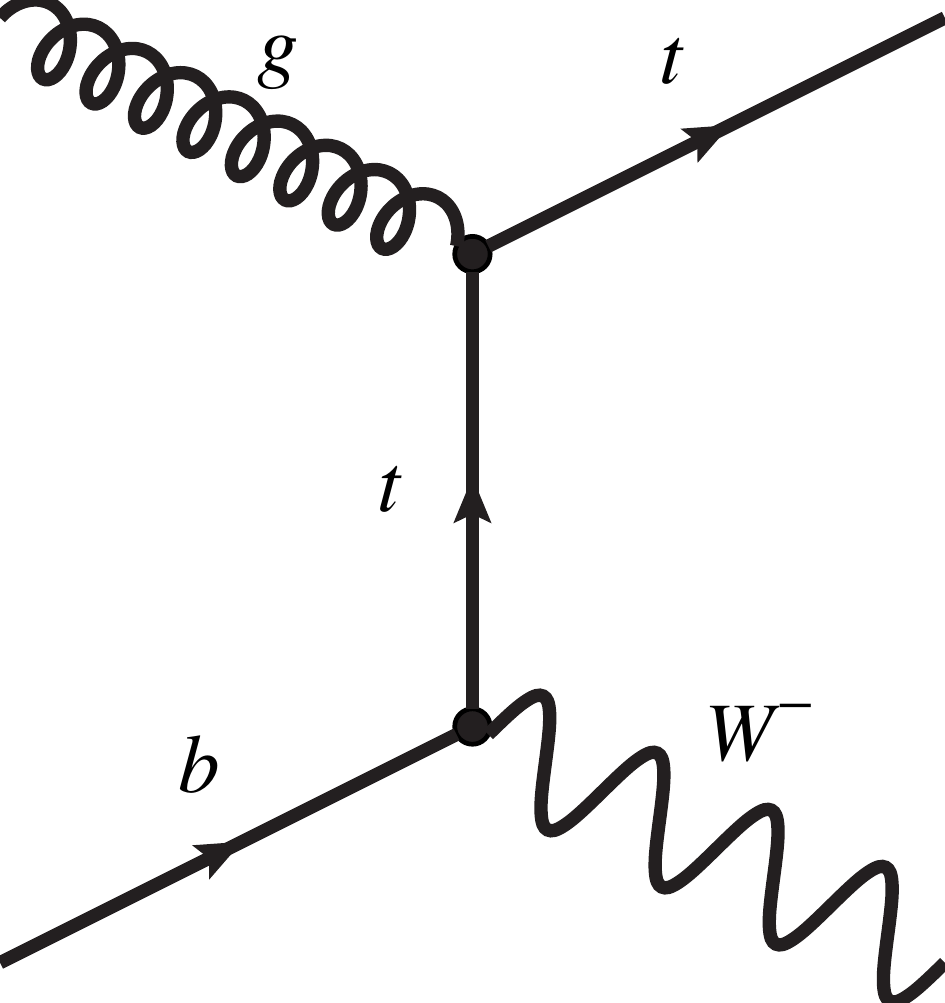}
\caption{\label{tWchannnel:diagram} Example of the Feynman diagram for $tW$ production at the LO in pQCD within~the 5FS~\cite{ATLAS:2020cwj}.}
\end{figure}

The $tW$ process features a top quark produced in association with a $W$ boson, either initiated by a gluon and a b-quark (in the 5FS, see Figure~\ref{tWchannnel:diagram}), or~with the b-quark produced by gluon splitting (in the 4FS). 
Because the PDFs for bottom and anti-bottom quarks in the proton in the 5FS are assumed to be the same, the~predicted cross-section for $tW^{-}$ and $\bar{t}W^{+}$ is identical at LO (and almost identical at a higher order)~\cite{Kidonakis:2021vob}.

\begin{figure}[H]
  \resizebox{12cm}{!}{\includegraphics{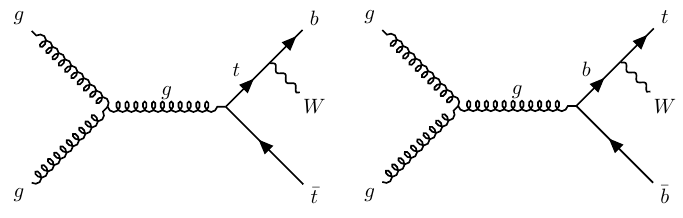}}
  \caption{Examples of Feynman diagrams for the production of $tWb$ with (\textbf{left}) and without (\textbf{right}) an on-shell~top.} 
  \label{fig:tWbdiag}
\end{figure}

There is some degree of overlap between the $tW$ process and the \ttbar\ process, since the $tW$ production at NLO in pQCD features resonant diagrams, which  interfere with LO diagrams of \ttbar\ production.
The NLO corrections to the production of $tW$ include $tWb$ processes, where the $Wb$ system can also arise from the decay of an on-shell top quark. 
Examples of LO Feynman diagrams for $tWb$ processes are shown in Figure~\ref{fig:tWbdiag}. 
Since the cross-section for \ttbar\ production is much higher than that of $tW$ production, these corrections are very large. As~a result, there is ambiguity in the way the $tW$ + 1 jet processes are defined. A~similar situation is occurring in the FCNC processes~\cite{universe8110609}.

\vspace{-3pt}

The definition of the $tW$ process, therefore, relies on the treatment of this interference and presents challenges at a theoretical level, depending on the choice of suppressing the interference to define independent simulation samples for the $tW$ process at NLO, or~including it in the simulation in a consistent way between $tW$ and the \ttbar\ process.
Two methods exist to suppress the interference~\cite{tWIsolating}. In~the diagram removal (DR) method, the~resonant \ttbar\ diagrams are excluded at the level of the matrix element calculation. In~the diagram subtraction (DS) approach, the~\ttbar\ resonant contributions are removed from the cross-section calculation by means of counter terms. Thus, a~comparison of the DR and DS prediction provides an estimation of the importance of the interference terms and  treatment, which is small for the usual kinematic selection applied~\cite{WtbAguilar}.

\subsubsection{Measurements of the $tW$ Process}

The $tW$ process has not been measured at the Tevatron, as its cross-section is small at the Tevatron center-of-mass energy in $p-\bar{p}$ collisions. 
The measurement of the $tW$ process is more challenging than that of the $t$-channel since a very large background from \ttbar\ events mimics the signal with almost the same experimental signature.
The ATLAS and CMS collaborations presented evidence for this process in the dilepton channel at \mbox{7 TeV~\cite{ATLAS:2012bqt,CMS:2012pxd}}, while the inclusive cross-section was measured at 8 TeV~\cite{ATLAS:2015igu,CMS:2014fut} and \mbox{13~TeV~\cite{ATLAS:2016ofl,CMS:2018amb}}. The~measurements at 13 TeV performed with larger collected data samples allowed measuring the differential cross-sections~\cite{ATLAS:2017quy,CMS:2022ytw} for the first time. 
The~$tW$ process was measured in the lepton+jets channel, more difficult owing to larger backgrounds, by ATLAS using 8 TeV collisions~\cite{ATLAS:2020cwj} and at CMS using 13 TeV collisions~\cite{CMS:2021vqm}.

\begin{figure}[H]
\includegraphics[width=0.4\textwidth]{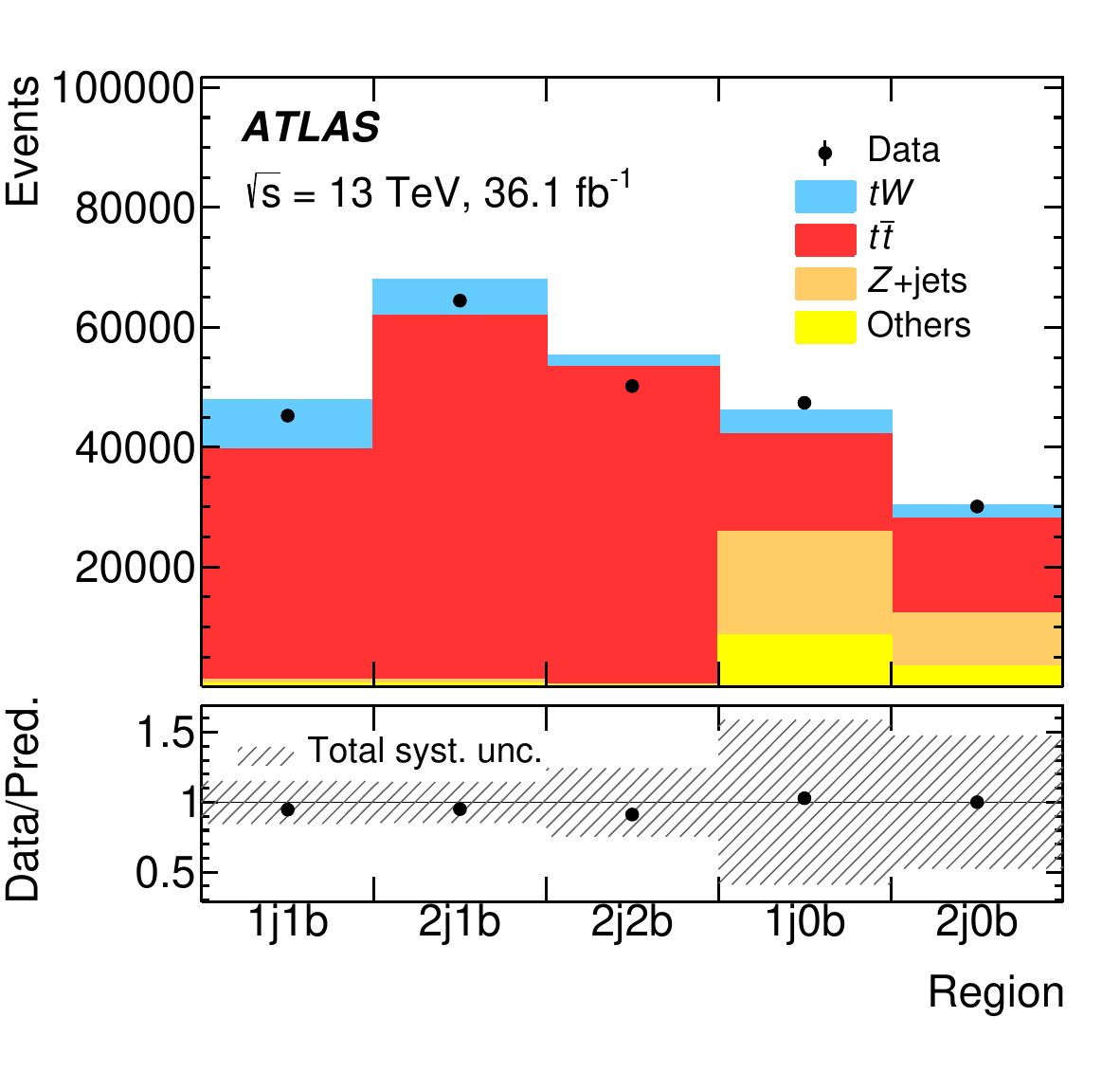}
\includegraphics[width=0.4\textwidth]{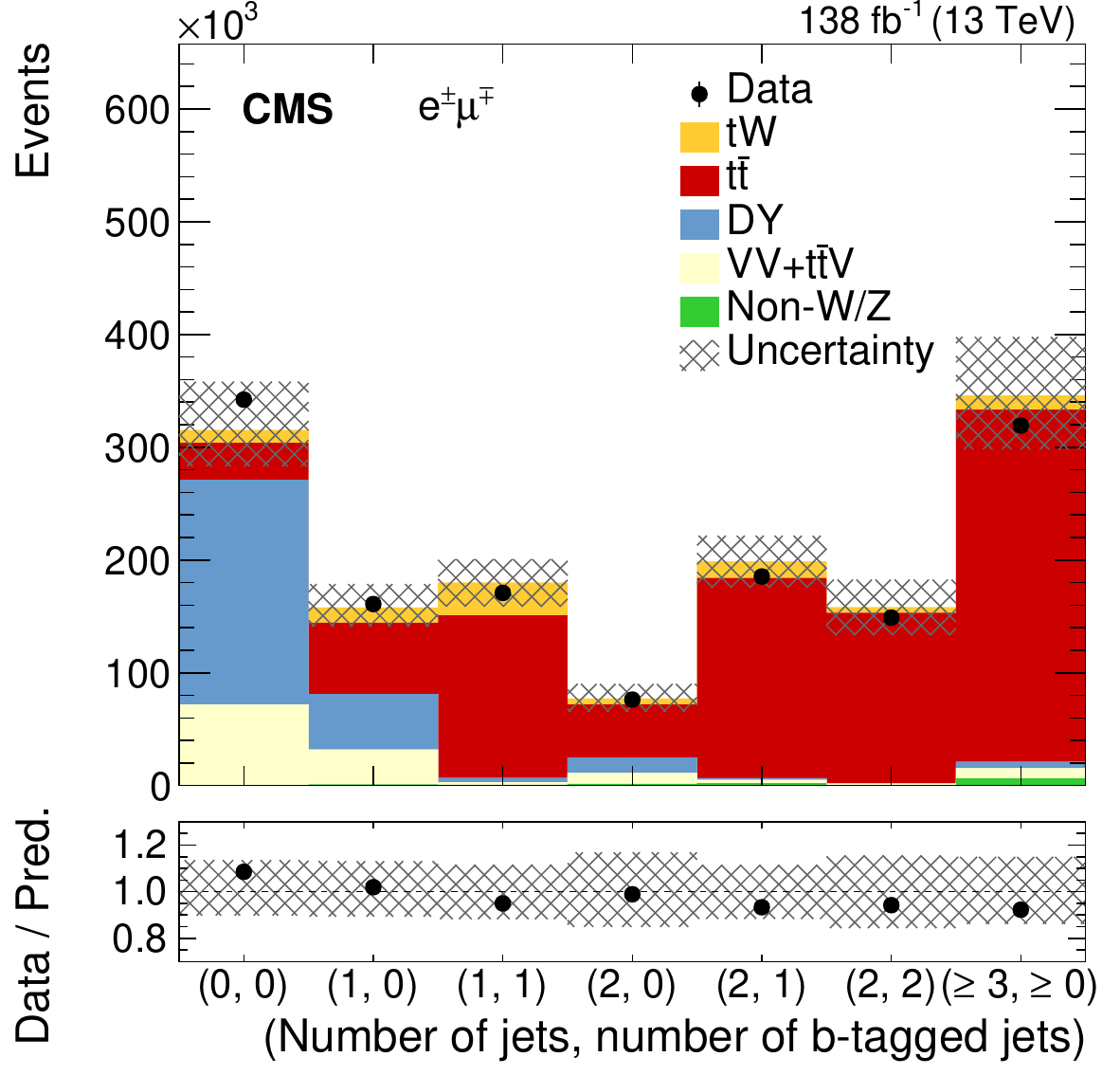}
\caption{\label{tWchannnel:categories} Categories in the number of jets and b-jets in the $tW$ dilepton analysis by ATLAS~\cite{ATLAS:2017quy} (\textbf{left}) and CMS~\cite{CMS:2022ytw} (\textbf{right}). }
\end{figure}

The dilepton decay channel for the $tW$ process refers to processes where one lepton arises from the top quark decay through $Wb$ and another lepton is produced by the associated $W$ boson decay. We describe features of the ATLAS~\cite{ATLAS:2017quy} and CMS~\cite{CMS:2022ytw} analyses measuring differential cross-sections at 13 TeV with the dilepton channel, where the leptons refer to electrons or muons. 
Nominal SM predictions for the $tW$ process use the DR scheme. 
For this analysis, the~main background contribution after event selection is the $t\bar{t}$ process in the dilepton decay channel, amounting to nearly 80\% of the event yield after selection. The~signal region is defined with exactly one reconstructed jet being tagged as a b-jet (so-called $1j1b$ region) to~remove contributions from doubly resonant diagrams. 
In general, a~selection on the transverse missing energy does not need to be applied (among recent measurements, the~ATLAS 13 TeV inclusive cross-section measurement~\cite{ATLAS:2016ofl} is an exception); this variable is used to reconstruct kinematic quantities and provide input to machine learning techniques. 
Figure~\ref{tWchannnel:categories} shows the number of events after selection, sorted in bins of the number of jets and b-jets. 
Two (three) regions that are defined depending on the number of jets and b-jets are used to measure the inclusive cross-section by ATLAS (CMS), with~dedicated~BDTs.

\vspace{-9pt}

The inclusive cross-section at 13 TeV is measured to be $\sigma_{tW} =$ 79.2 $\pm$ 0.9 (stat) $^{+7.7}_{-8.0}$~ (syst) $\pm$ 1.2 (lumi) pb at CMS using 138 fb$^{-1}$~\cite{CMS:2022ytw}, and~$\sigma_{tW} =$ 94 $\pm$ 10 (stat.) $^{+28}_{-22}$ (syst.) $\pm$ 2 (lumi.) pb by ATLAS using 3.2 fb$^{-1}$~\cite{ATLAS:2016ofl}, in~agreement with SM theoretical predictions. The~dominant systematic uncertainty is the jet energy scale, followed by the background normalization and the theory uncertainties on $tW$ process modeling. 

\begin{figure}[H]
\includegraphics[width=0.45\textwidth]{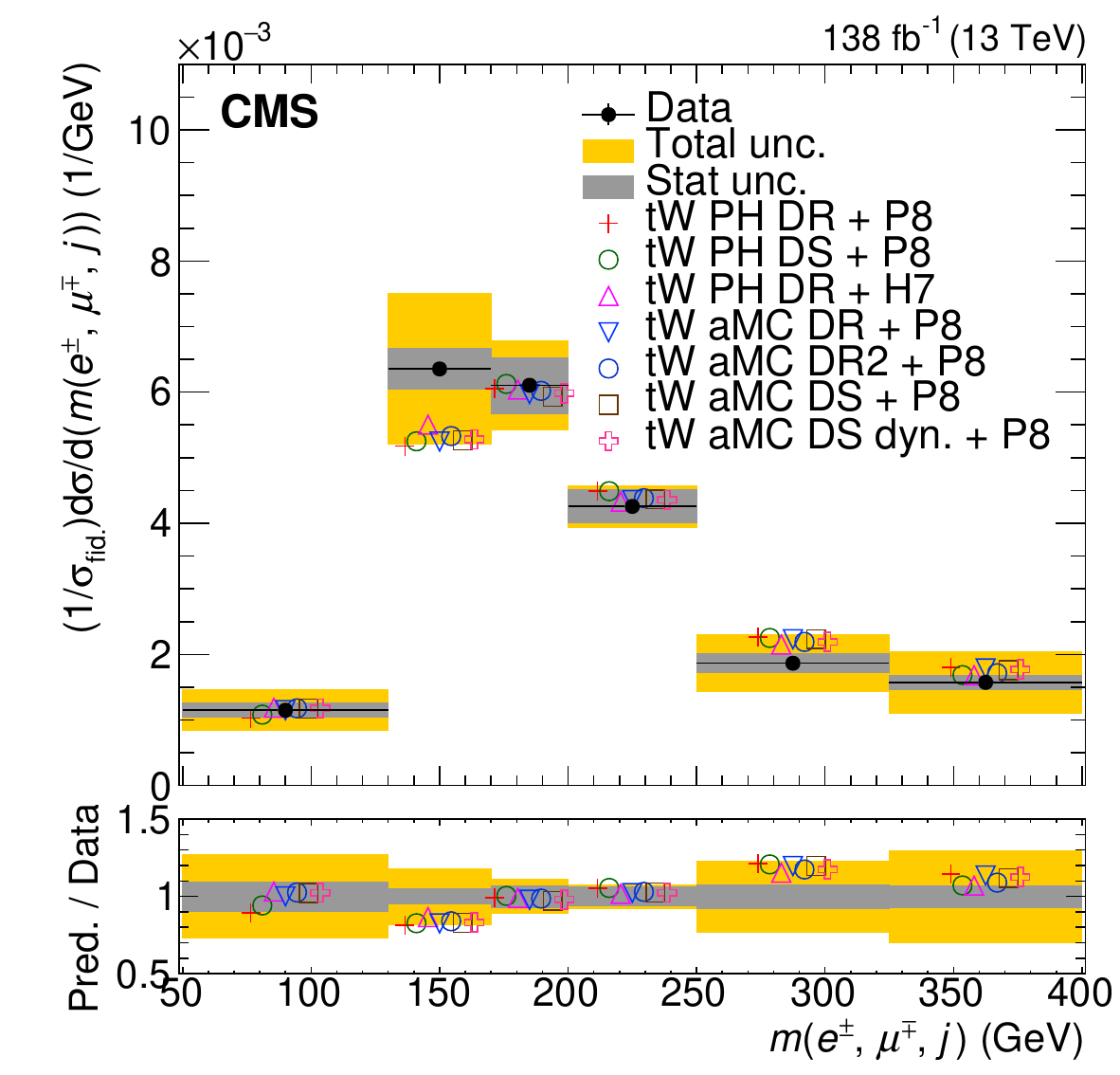}
\includegraphics[width=0.45\textwidth]{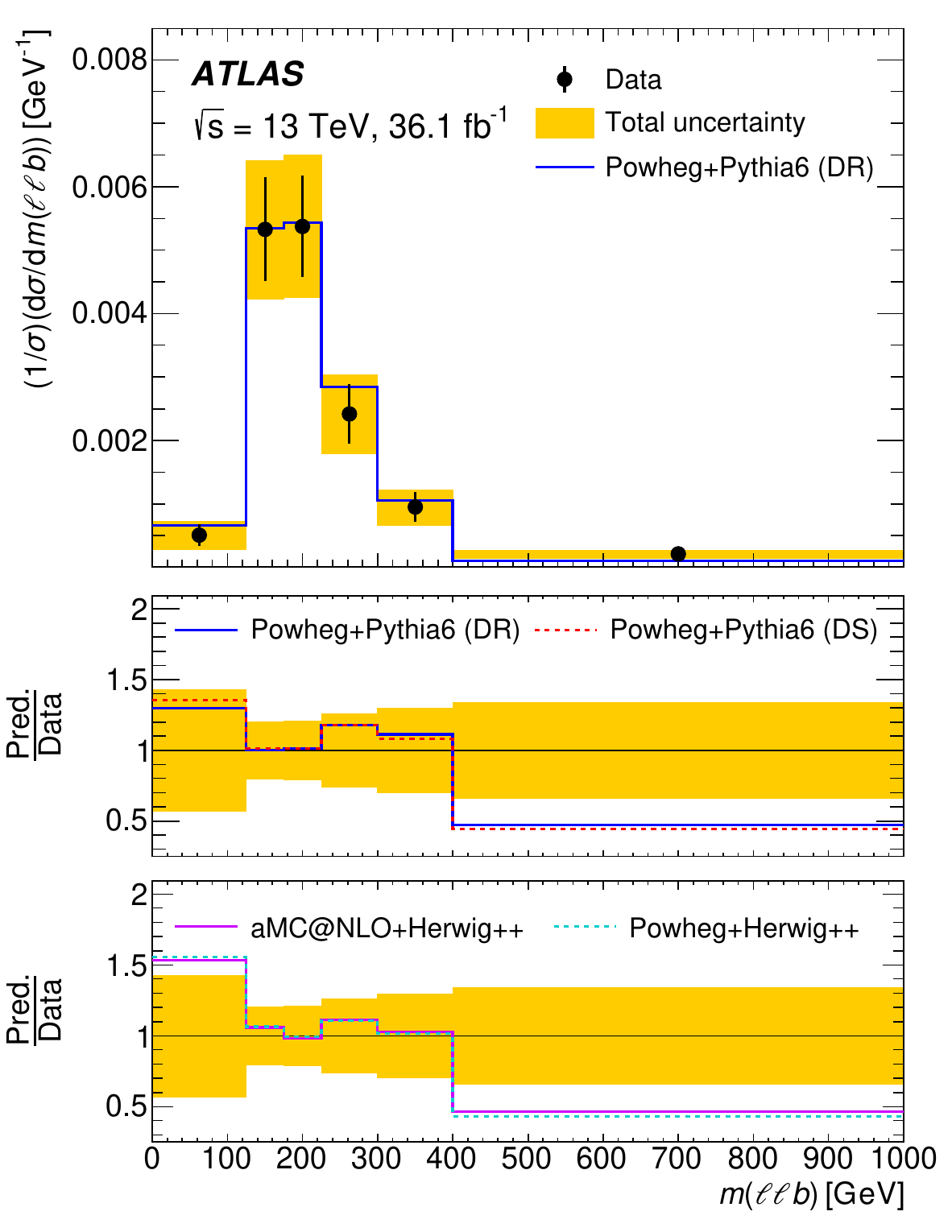}
\caption{\label{tWchannnel:mllb} Invariant mass of the dilepton and b-jet in the $tW$ dilepton analysis by ATLAS~\cite{ATLAS:2017quy} (\textbf{left}) and CMS~\cite{CMS:2022ytw} (\textbf{right}), comparing data and several predictions for $tW$ modeling.}
\end{figure}

The~$1j1b$ region---by both ATLAS and CMS---is used to extract the differential cross-sections. 
In the ATLAS analysis, an~additional selection is applied to the output of the BDT to increase the separation between the signal and backgrounds for the differential measurement. In~CMS, a veto on additional loose jets is also applied. The~data are corrected for detector effects and compared to theoretical predictions, such as the invariant mass of the dilepton and b-jet in Figure~\ref{tWchannnel:mllb}. 
In most of the measured bins, the data and simulations agree within less than 1$\sigma$; however, more data are needed to discriminate between the different ways of modeling the~signal.

\vspace{-6pt}

Measuring the $tW$ process in the lepton+jets channel, which targets a final state where one of the two $W$ bosons decays leptonically and the other hadronically, is a challenging task, owing to the prominent $t\bar{t}$ and $W$+jet backgrounds arising from the selection. 
Machine learning methods are used to enhance the signal-over-background ratio, the~NN by ATLAS~\cite{ATLAS:2020cwj}, and the BDT at CMS~\cite{CMS:2021vqm}. 
ATLAS extracts the signal using a two-dimensional distribution in the NN output and the invariant mass of the hadronically decaying $W$ in~events with at least three jets (including one b-jet). 
CMS employs the BDT outputs in three regions, whether there are two, three, or four jets in the event (including one b-jet). 
The analyses lead to evidence for the $tW$ process in the lepton+jets channel using 8 TeV data by ATLAS, and~an observation using 13 TeV data at CMS. The~measured inclusive cross-sections are in agreement with the SM predictions, and~the precision is already dominated by systematic uncertainties.
The main systematic uncertainties arise from the jet energy scale, background normalization, and $t\bar{t}$ or $tW$ modeling. 

The lepton+jets analysis shows that, nowadays, more difficult channels are used to measure the $tW$ process. One of the next steps would be to scrutinize the tails of kinematic distributions by using boosted jet tagging, allowing to access highly boosted regions~\cite{ATLAS:2015uhg,CMS:2021mku} that are sensitive to new resonances, like excited b-quarks appearing in theories beyond the SM, such as composite models~\cite{Baur:1989kv}.
Differential distributions are measured with the dilepton channel and will be investigated more differentially in the future. 
Despite having a smaller cross-section than the $t$-channel, the~$tW$ process could also be used to measure SM parameters.
Similar to measurements performed in the $t$-channel, measuring charge ratios would be interesting since they are sensitive to PDFs; this would require separating top from antitop contributions in $tW$ production with advanced techniques like the matrix element method~\cite{Brochet:2018pqf}. 

\subsubsection{Understanding the Interference between $tW$ and \ttbar\ Processes}

While the above-mentioned measurements of $tW$ process are designed to minimize the interference with the $t\bar{t}$ process by mostly selecting events with only one b-jet, a~recent analysis by ATLAS~\cite{ATLAS:2018ivx} targeted a phase space with exactly two b-jets, where the interference effect was maximized. 
This analysis utilized a variable defined as the invariant mass of a lepton and a b-jet as a proxy for the top quark mass. Since there is ambiguity in assigning leptons and b-jets to a given top quark, a~particular choice is made:
\begin{linenomath}
\begin{equation}m_{b\ell}^{minimax} = min \Big( max(m_{b_1\ell_1}, m_{b_2\ell_2}),max(m_{b_1\ell_2}, m_{b_2\ell_1}) \Big),
\end{equation}
\end{linenomath}
where particles 1 and 2 are interchangeable. This variable is defined in such a way that, at LO, $m_{b\ell}^{minimax} < \sqrt{m_t^2-m_W^2}$. The~cross-section above this value has increased sensitivity to the interference between single and double resonant~contributions.

Events are selected if there are two leptons and two jets satisfying a tight b-tagging criterion, with~a veto on further leptons using a loose requirement (which suppresses backgrounds arising from $t\bar{t}$ associated with heavy flavor jets). 
The analysis measures the normalized differential cross-section in a phase space at the  generator level as close as possible to the reconstructed level, as~a function of the $m_{b\ell}^{minimax}$ observable. 

\begin{figure}[H]
\includegraphics[width=0.6\textwidth]{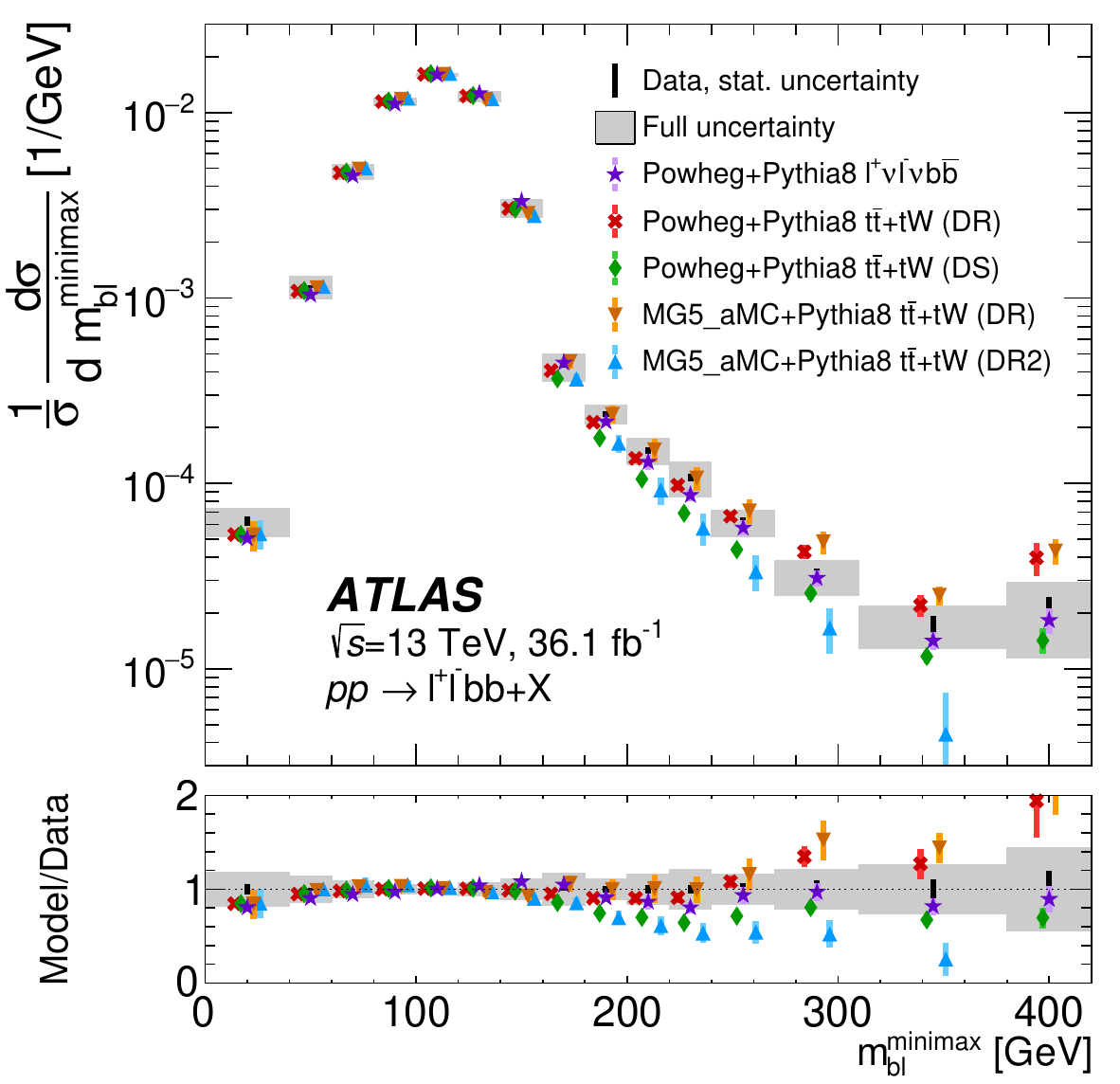}
\caption{\label{tWchannnel:bb4l_datamc} Data comparison of MC predictions for the normalized differential cross-section of the $tW$ process in a region maximizing the interference, as~a function of the $m_{b\ell}^{minimax}$ variable~\cite{ATLAS:2018ivx}. The~region sensitive to the interference lies above $\sqrt{m_t^2-m_W^2}$. }
\end{figure}

The data are compared to simulations at the particle level in Figure~\ref{tWchannnel:bb4l_datamc}, after the background subtraction and correction for detector effects. 
The simulation sample matching the best data across the entire range of $m_{b\ell}^{minimax}$  includes both $tW$ and $t\bar{t}$, as well as their interference with POWHEG~\cite{Jezo:2016ujg}. Samples featuring interference suppression with the DR or DS scheme do not reproduce the data at large values of $m_{b\ell}^{minimax}$. 

Due to the datasets expected at LHC Run 3 and the HL-LHC, one can expect new measurements to probe the nature of the interference in more depth. The~lepton+jets final state could be scrutinized as well for this purpose, since the theoretical predictions at NLO, including the interference, were recently made available~\cite{Jezo:2023rht}.

\vspace{-3pt}

\subsection{The Challenging $s$-Channel}

\begin{figure}[H]
\includegraphics[width=0.35\textwidth]{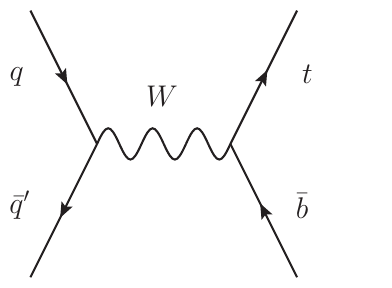}
\caption{\label{S-channel:Diagrams}Feynman diagram for single-top $s$-channel production at the LO in pQCD in the 5FS~\cite{ATLAS:2015jmq}.}
\end{figure}

The final state for top quark production in the $s$-channel is similar to that of the $t$-channel in Section~\ref{t-channel}, except~that the top quark is now produced with a $b$ or $\bar{b}$ quark in the final state instead of a light quark (in the 5FS). The~process occurs through the exchange of a time-like $W$ boson instead of a space-like $W$ boson, as~shown in Figure~\ref{S-channel:Diagrams}. 
The virtual $W$ boson has to be far away from its resonant mass to produce a top quark, and~this highly suppresses the corresponding cross-section, which makes the observation of the $s$-channel very challenging. 
The top quark is more likely to be produced with central b-jets than with a forward light~jet. 

\vspace{-3pt}

\begin{figure}[H]
  \resizebox{6cm}{!}{\includegraphics{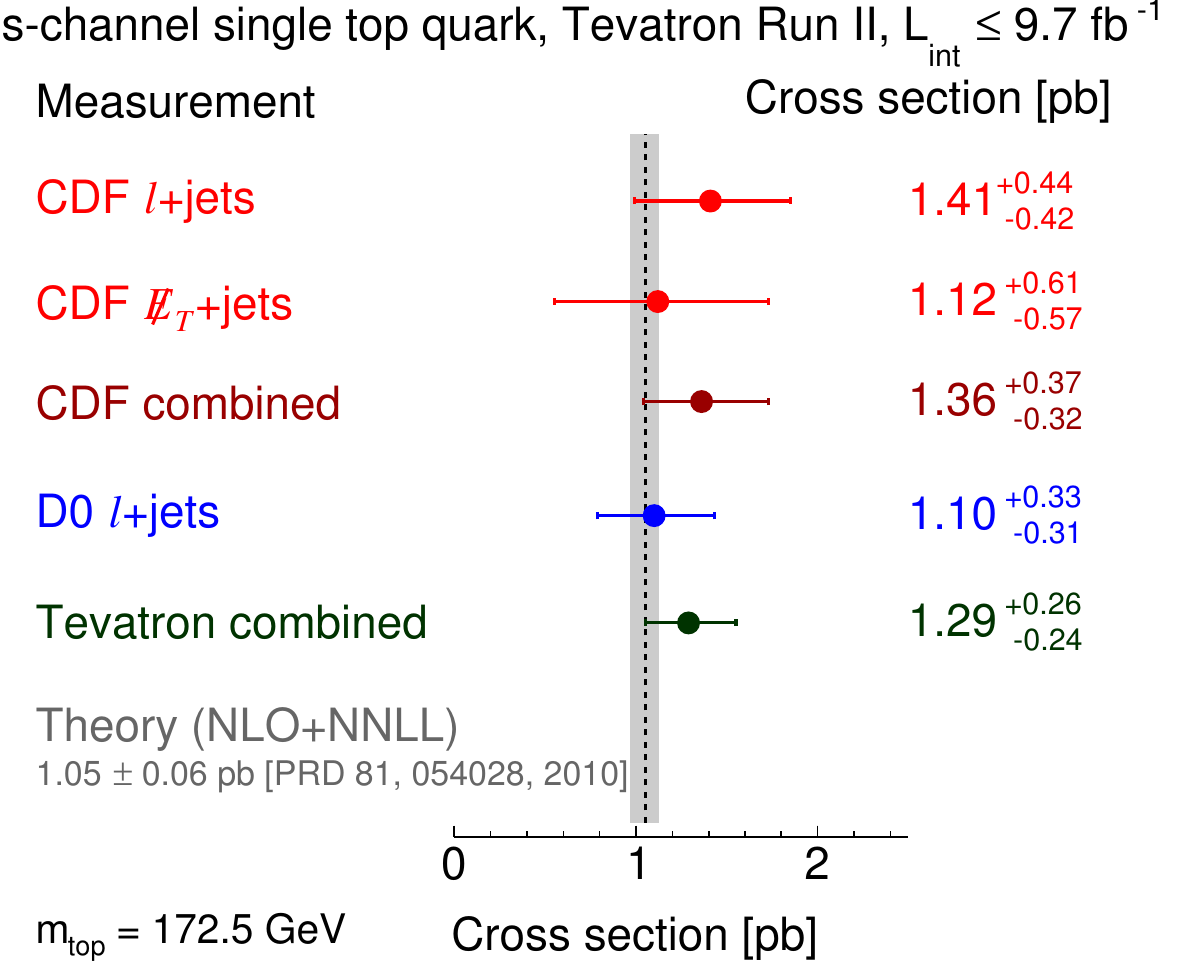}}
  \caption{Measured cross-sections for the $s$-channel at the Tevatron along with their combined results~\cite{CDF:2014uma}.} 
  \label{fig:s-channel_SoverB}
\end{figure}

The $s$-channel process is observed at the Tevatron~\cite{CDF:2014uma}, using 9.7 fb$^{-1}$ of proton--antiproton collisions collected at D0 and CDF at $\sqrt{s}=1.96$ TeV. This process remains to be observed at the LHC. 
At CDF, the~lepton+jets channel and \met+jets channel are used, while the lepton+jets channel is used at D0. Multivariate techniques are employed to identify the b-jets and reduce the contribution of background processes. 
Events are classified in categories depending on the number of jets, and~the number and quality of b-jets. 
Multivariate discriminants are built to extract the $s$-channel cross-section using a Bayesian statistical technique. 
The result is $\sigma_{s} = 1.29^{+0.26}_{-0.24}$ pb, which is in~agreement with the SM prediction of $\sigma = 1.05 \pm0.06$ pb at an approximate NNLO with NNLL accuracy~\cite{Kidonakis:2010tc} at the Tevatron. 
The $s$-channel process was observed at 6.3 $\sigma$ at the Tevatron. 
Figure~\ref{fig:s-channel_SoverB} shows the measured cross-section for each channel at the Tevatron along with their combined results.

Proton--antiproton collisions at the Tevatron mainly provide a quark and an antiquark in the initial state while the LHC does not. Furthermore, the~b-quark content in the proton is larger at the LHC. As~a result, the~ratio of the $s$-channel to the $t$-channel cross-section decreases from the Tevatron to the LHC. 
Furthermore, the~ratio of signal-to-background is quite favorable at the Tevatron relative to the LHC. 
When the energy in the p--p center of mass increases, this search becomes more difficult: the quark luminosity increases at a 
slower pace than the gluon luminosity when increasing the center-of-mass energy. As~an example, the~ratio of the $s$-channel to the $t\bar{t}$ cross-section changes from 2.1\% at 8 TeV to 1.2\% at 13 TeV~\cite{ATLAS:2022wfk}. 

For all of these reasons, searches for the $s$-channel are very challenging at the LHC. 
The first search by ATLAS using 8 TeV data resulted in a significance of 1.3$\sigma$~\cite{ATLAS:2014hvq} (with 1.4$\sigma$ expected); however, a subsequent search on the same dataset employed the matrix element method (MEM~\cite{Kondo:1988yd}, which has been used since the early measurements of the top quark mass at the Tevatron~\cite{D0:2004rvt,CDF:2006lnv}), leading to an observed significance of 3.2$\sigma$~\cite{ATLAS:2015jmq} (with 3.9$\sigma$ expected). 
CMS analyzed Run 1 data using the 7 and 8 TeV datasets, resulting in an observed significance of 2.5$\sigma$ with an expected significance of 1.1$\sigma$~\cite{CMS:2016xoq}. 
Recently, ATLAS performed a search using the same analysis techniques with the MEM as in their 8 TeV paper, analyzing Run 2 data at 13 TeV~\cite{ATLAS:2022wfk}. Despite the unfavorable signal-to-background ratio at 13 TeV compared to 8 TeV, a~similar observed (3.3$\sigma$) and expected (3.9$\sigma$) significance was~achieved. 

\begin{figure}[H]
  \resizebox{6cm}{!}{\includegraphics{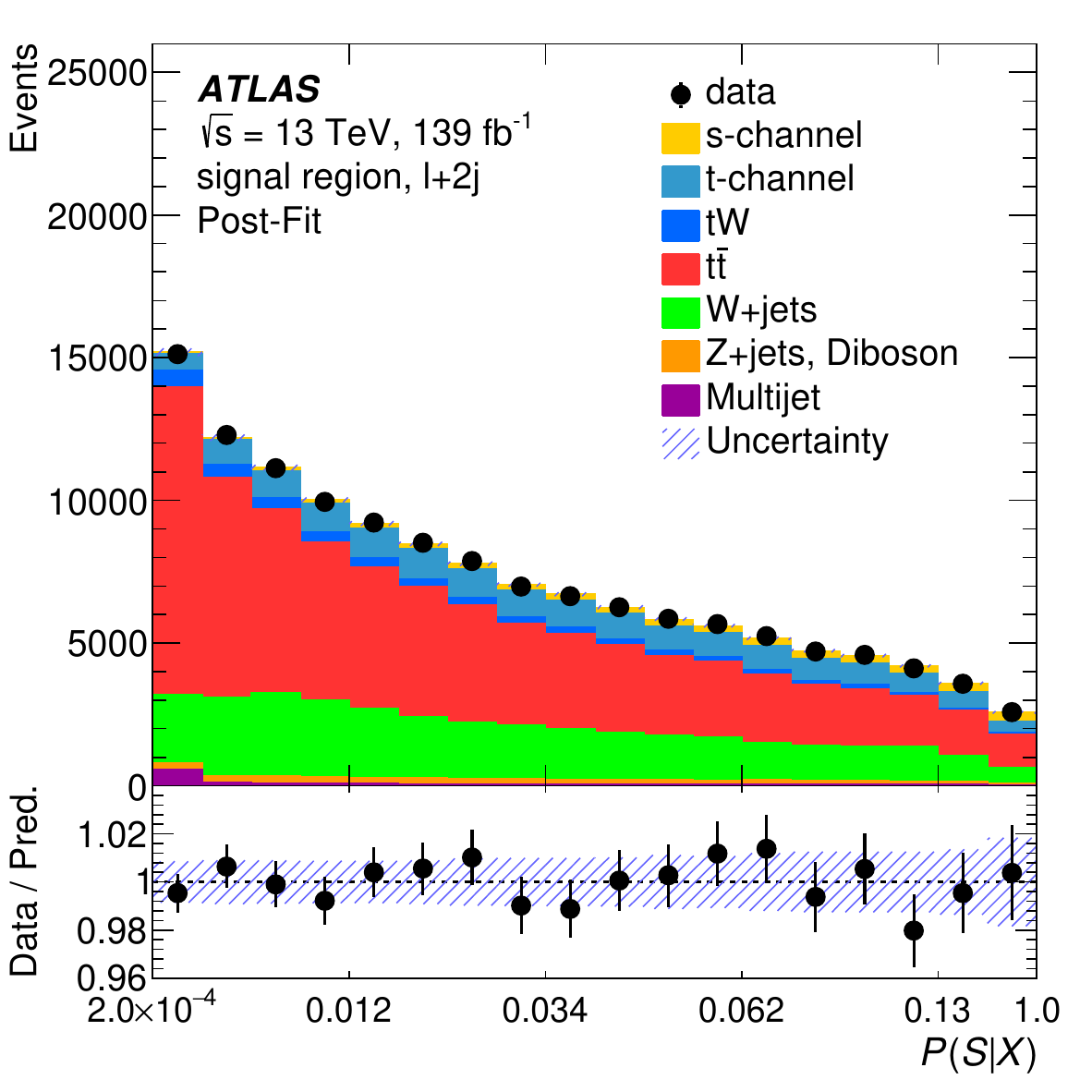}}
  \resizebox{6cm}{!}{\includegraphics{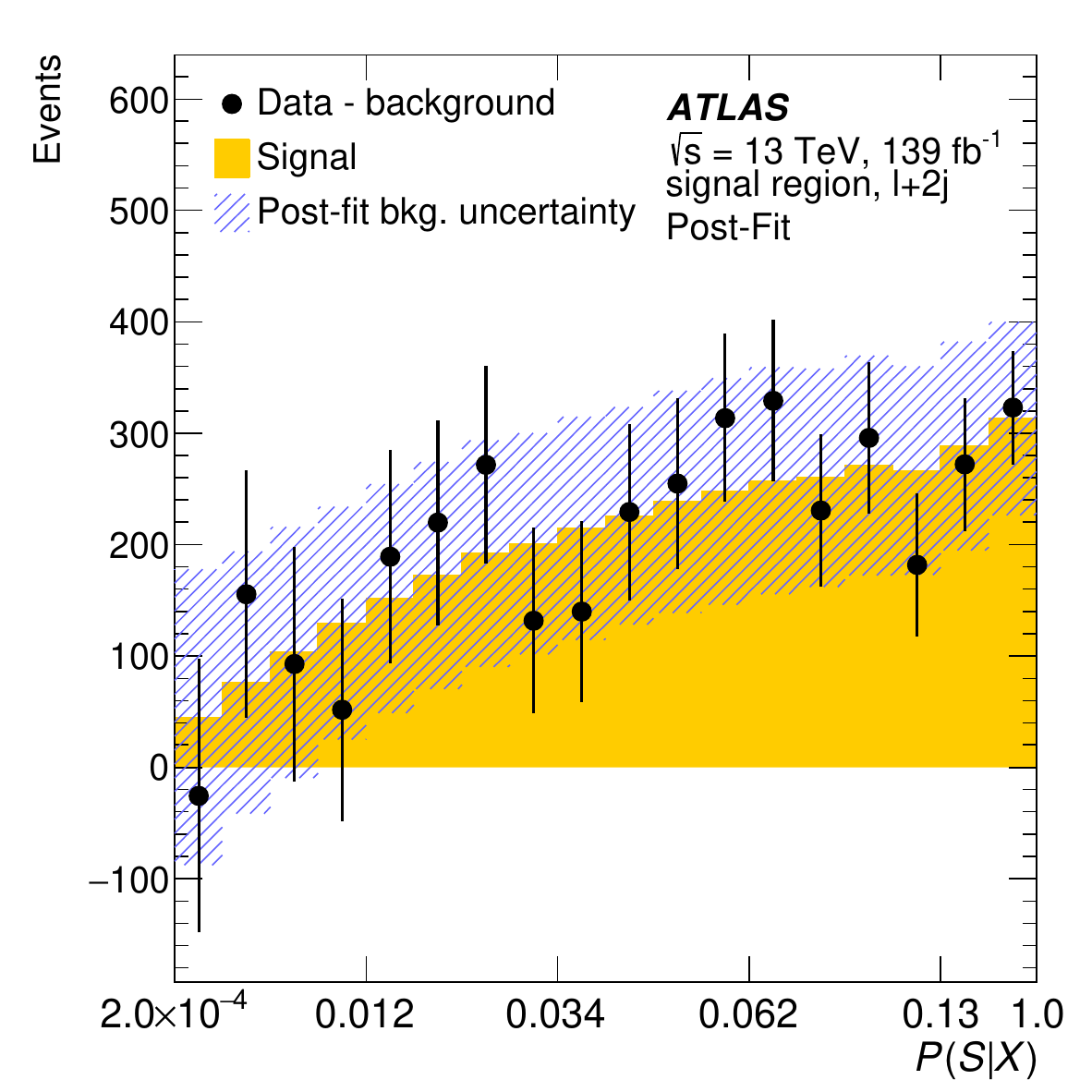}}  
  \caption{Results of the $s$-channel searches by ATLAS: (\textbf{left}) post-fit distribution comparing data and simulation for the MEM likelihood, and~(\textbf{right}) signal distribution after the background subtraction~\cite{ATLAS:2022wfk}.} 
  \label{fig:s-channel_ATLASresult}
\end{figure}

Since the ATLAS result is the latest, with~the largest observed significance, and~the only one published using 13 TeV data, we will provide details on this analysis. 
The lepton+jet channel is analyzed, with~one electron or muon having $p_T > 30$ GeV and at least two jets with $p_T > 25$ GeV. Events from multijet production are reduced by requiring \mbox{\met~> $35$ GeV} and $m_{T,W}>30$ GeV. In~the signal region, exactly two jets are required, and~both of them must be b-tagged. A~validation region targets the $W$+jets process, where one of the jets must fail the b-tag requirement. Events are also validated using two regions enriched in the \ttbar\ process, with~three or four jets, among~which, two must be b-tagged. 
The normalization for multijet production is estimated from the data, while the other background processes are taken from the simulation. 
A dedicated method, the~MEM, is employed to further reduce the backgrounds. The~MEM consists of calculating a probability density, representing the compatibility of each event with signal and background hypotheses, using exact calculations at the LO in pQCD. Hypotheses for the $s$-channel, $t$-channel, \ttbar\ production, and~$W$ boson production are considered. A~likelihood is built by combining these hypotheses, and~the less likely events are discarded. The~shape of the likelihood distribution in the signal region is then used to extract the $s$-channel cross-section. The~post-fit distribution is shown in Figure~\ref{fig:s-channel_ATLASresult}, left, and the~signal after background subtraction is shown in Figure~\ref{fig:s-channel_ATLASresult}, right. 
The measured cross-section is $\sigma = 8.2\pm 0.6$(stat)$^{+3.4}_{-2.8}$(syst) pb, in~agreement with the SM prediction of $\sigma_{SM}=10.32^{+0.40}_{-0.36}$ pb at the NLO accuracy in pQCD.
As a side note, predictions at the NNLO in pQCD are available~\cite{Liu:2018gxa} and could be used by the LHC~experiments.

As a conclusion, the~observation of the $s$-channel remains to be achieved at the LHC. A~result for a CMS analysis using Run 2 data is desired. 
The analysis is already systematic-dominated; therefore, new techniques should be employed to reduce the uncertainties. 
A simultaneous fit using signal and control regions could be used to further constrain the background contributions. 
An involved analysis technique beyond the MEM, like a deep NN (DNN), could also improve the significance. 
Despite maintaining an unfavorable signal-to-background ratio compared to the Tevatron, the~searches should be pursued at the LHC with Run 3 data and at the HL-LHC to make an~observation.

\section{\label{SingleTopBoson}Associated Production of a Single-Top Quark with a Neutral~Boson}
\unskip

\subsection{A Newcomer: Associated Production of a Single-Top Quark with a Photon ($t\gamma$) }

\begin{figure}[H]
\includegraphics[width=.85\textwidth]{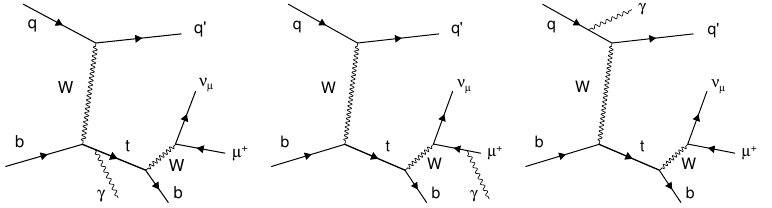}
\caption{\label{Tgammaq:Diagrams}Examples of Feynman diagrams for $t\gamma q$ processes at the LO in pQCD~\cite{CMS:2018hfk}.}
\end{figure}

\textls[-25]{The production of a photon in association with a top quark ($t\gamma$) is a rare process, accessible at the LHC. 
The cross-section predicted at NLO in pQCD with Madgraph5\_aMC@NLO~\cite{Alwall:2014hca}} is $ 2.95 \pm 0.13 \mathrm{(scale)} \pm 0.03 \mathrm{(pdf)}$ pb (as quoted by CMS~\cite{CMS:2018hfk}), requiring the photon $p_T$ to be greater than 10 GeV before~the top quark decay in~the 5FS. 
The cross-section is dominated by $t$-channel diagrams with the radiation of a photon ($t\gamma q$), featuring a forward jet due to the electroweak nature of the $t$-channel. 
The cross-section for the $t\gamma q$ production is known at the approximate NNLO~\cite{Kidonakis:2022ocq}. 
Measuring the $t\gamma$ process extends the landscape of the measured top quark process and is an experimental challenge, owing to its low cross-section. 
The $t\gamma$ final states are also powerful tools used to constrain the FCNC~\cite{universe8110609}. Together with $t\bar{t}\gamma$ processes, they can be used to constrain the top-$\gamma$ coupling. 
Examples of Feynman diagrams are shown in Figure~\ref{Tgammaq:Diagrams}. The~photon can be emitted in the initial state, final state, or in the top quark~decay.

\begin{figure}[H]
\includegraphics[width=0.55\textwidth]{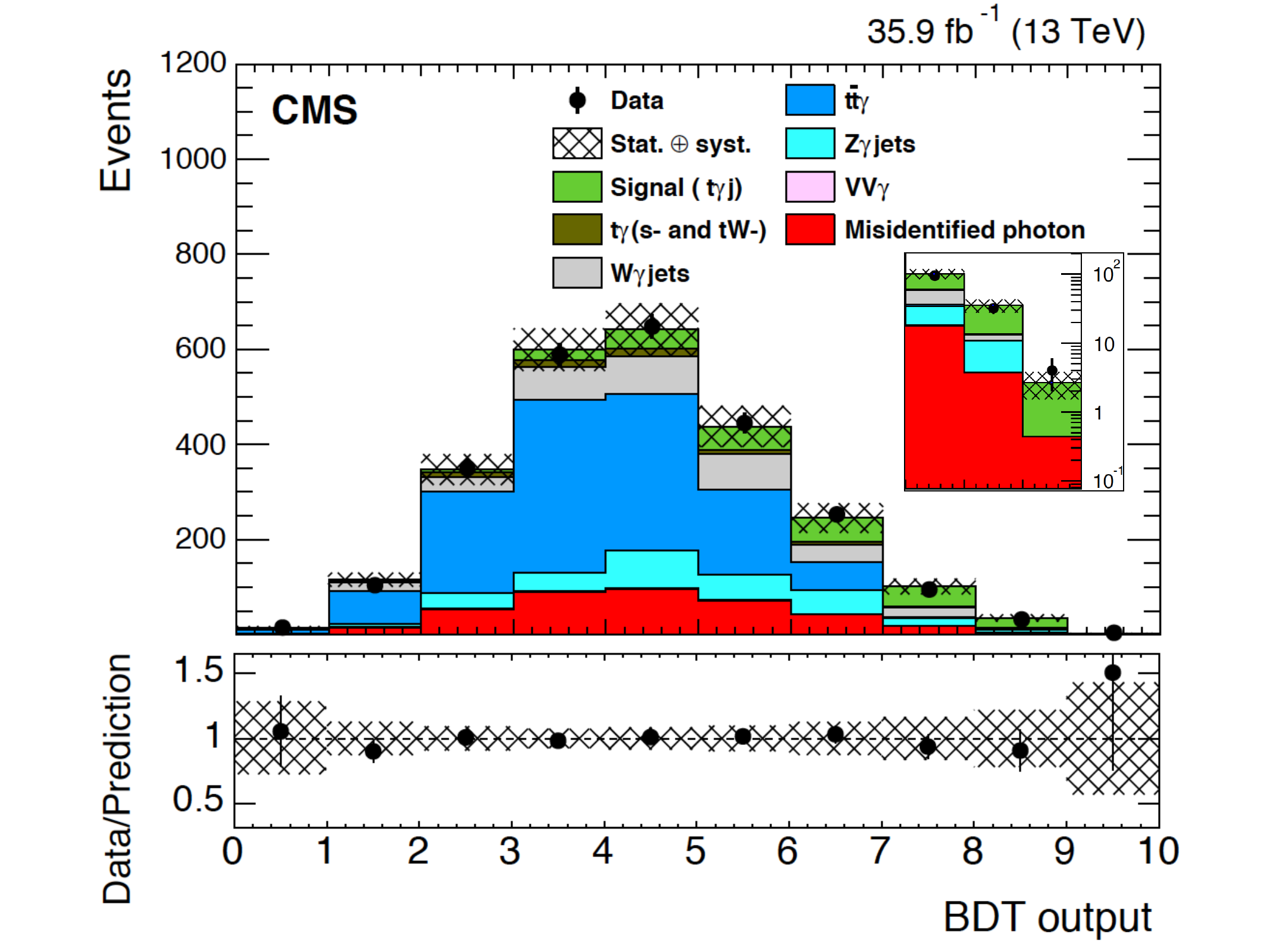}
\includegraphics[width=0.4\textwidth]{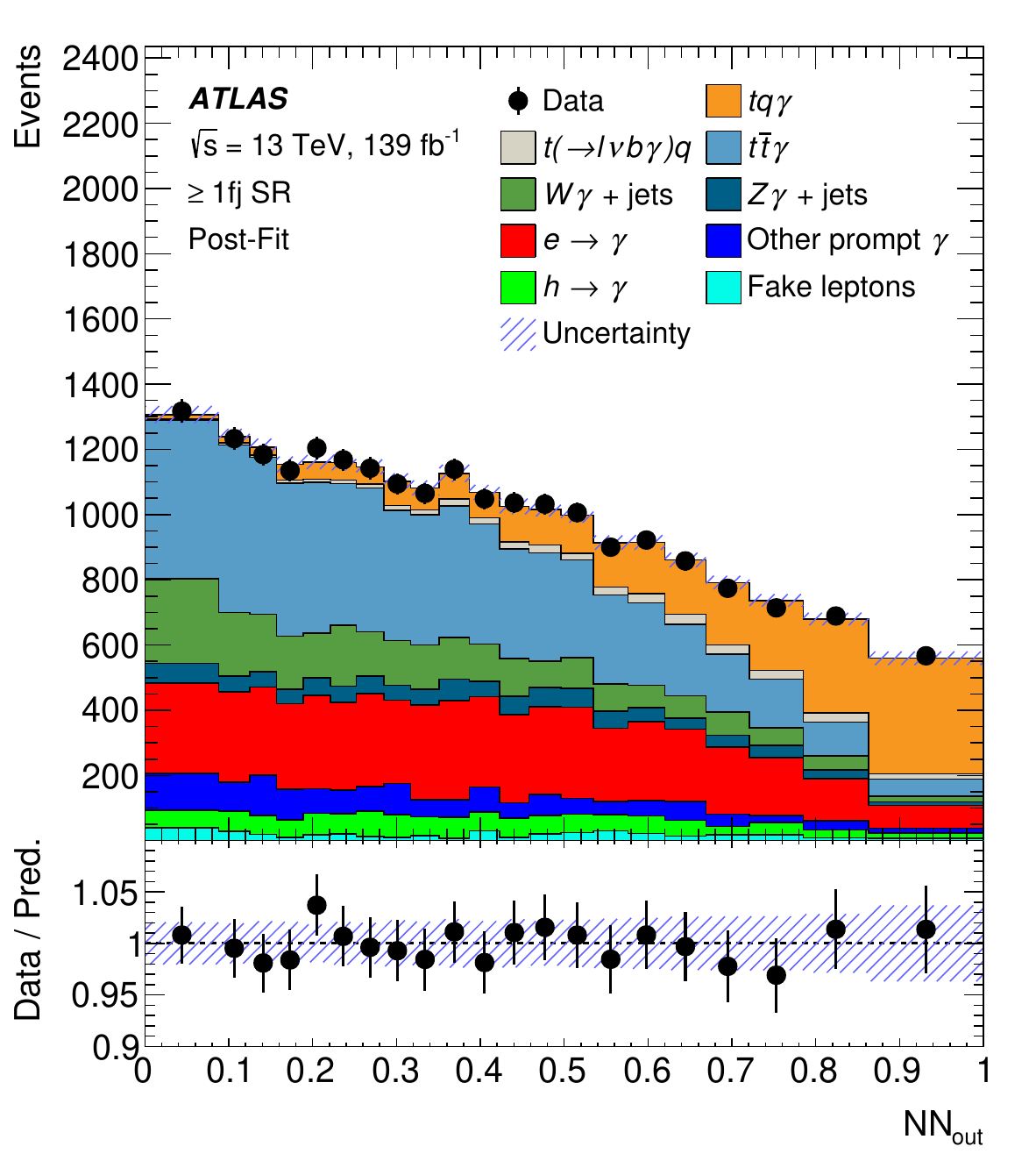}
\caption{\label{Tgammaq:discriminants} Discriminants used in the $t\gamma q$ signal extraction: BDT output at CMS~\cite{CMS:2018hfk} (\textbf{left}) and DNN output by ATLAS~\cite{ATLAS:2023qdu} (\textbf{right}) in the signal region with one forward~jet.}
\end{figure}

The searches for the $t\gamma q$ process led to evidence at CMS~\cite{CMS:2018hfk} using 36 fb$^{-1}$ of Run 2 LHC data, and~an observation by ATLAS with the full Run 2 dataset~\cite{ATLAS:2023qdu}. 
Special care is needed in single-top $t$-channel MC samples to remove photons produced in the parton shower since they could be double-counted with photons produced at the matrix element level in the $t\gamma q$ signal samples. 
There is also some freedom in the signal definition: photons arising from the top quark decay are treated as backgrounds in the ATLAS analysis~\cite{ATLAS:2023qdu}. 
The dominant backgrounds contain prompt leptons and photons, like $t\bar{t}\gamma$ and $W+\gamma$ processes, and~processes involving jets or electrons misidentified as photons (hereafter denoted as ``fake photons''). 
A control region is defined to measure the $t\bar{t}\gamma$ background. The~$W+\gamma$ process also benefits from a control region in the ATLAS analysis. The~fake photon backgrounds are two-fold, either arising from the misidentification of an electron as a photon, or~of a jet as a photon. In~the ATLAS analysis, both are estimated with dedicated methods from the data, while in the CMS analysis, only the backgrounds made of jets misidentified as photons are estimated from the data. 
To maximize the sensitivity to the signal, the~signal extraction is performed by constructing a discriminant with a BDT (CMS) and a DNN (ATLAS). 
Both analyses make use of the forward jet to discriminate the signal against the backgrounds, including the pseudorapidity as an input variable to the machine learning algorithm. The~discriminants are shown in Figure~\ref{Tgammaq:discriminants}. 

\vspace{-6pt}

The observed (expected) significance obtained is 4.4$\sigma$ (3.0$\sigma$) at CMS~\cite{CMS:2018hfk} and 9.1$\sigma$ (6.7$\sigma$) by ATLAS~\cite{ATLAS:2023qdu}. 
With such a large significance, the~$t\gamma$ processes can provide sufficient statistics for a first differential cross-section measurement at LHC Run 3. Complementing the $t\bar{t}\gamma$ channel with the $t\gamma q$ channel to probe the top-photon coupling will become especially relevant at the HL-LHC~\cite{Fael:2013ira}. 
As a side note, the~single-top $tW\gamma$ process has also been measured simultaneously with the $t\bar{t}\gamma$ process~\cite{ATLAS:2020yrp} because of the interference at NLO, similar to the $tW$ and $t\bar{t}$ processes, but  in a $t\bar{t}\gamma$ phase space chosen without particular enhancement of the $tW\gamma$ process or the~interference.

\subsection{A Path toward Top-$Z$ Coupling: Single-Top Quark Production with a Z Boson ($tZ$)}

The first process observed for single-top quark production in association with a neutral boson is actually the single-top quark production with a Z boson ($tZ$), due to the datasets made available at the LHC. 
In general, the~$tZ$ processes refer to the production of a single-top quark in association with a $Z$ boson, including the interferences between on-shell and off-shell $\gamma^{*}$ or $Z$ bosons. 
Similar to the $t\gamma$ process, the~process with the largest cross-section is provided in the $t$-channel ($tZq$).

\begin{figure}[H]
  \resizebox{13.8cm}{!}{\includegraphics{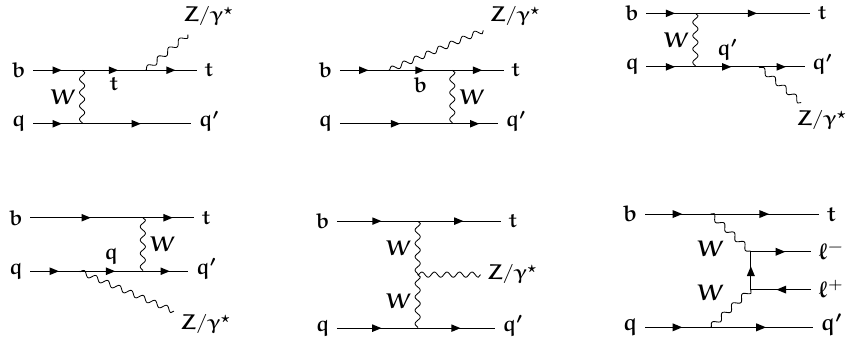}}
  \caption{Representative Feynman diagrams for the $tZq$ production at the LO in pQCD~\cite{JHEP_TOP_20_10}.} 
  \label{fig:tZqDiagram}
\end{figure}

\textls[-15]{The Feynman diagrams for $tZq$ production at the LO in pQCD can be seen in \mbox{Figure~\ref{fig:tZqDiagram}}. 
The inclusive $tZq$ cross-section predicted at NLO in the SM, as~calculated with MG5\_aM@NLO}, is 800 fb $^{+6.1\%}_{-7.4\%}$~\cite{Aaboud_2018_tZq1}. 
Because of its clear signature and interesting signal-to-background ratio, the~$tZq$ process is measured in the three-lepton channel. The~cross-section for $tZq$ production in the three-lepton decay channel, calculated at NLO with MG5\_aMC@NLO, and including a dilepton invariant mass cut of $m_{\ell\ell}$ > 30 GeV, is 94.2$^{+1.9}_{-1.8}$ (QCD \mbox{scale) $\pm$ 2.5 (PDF)} fb~\cite{Sirunyan_2018_tZq2}. 
The cross-section for $tZq$ production could be updated with the latest predictions at the approximate NNLO~\cite{Kidonakis:2022ljg} in future~measurements.

\vspace{-3pt}

The $tZq$ production has several interesting features. 
In a similar way with the $t$-channel process without an associated $Z$ boson, the~top and antitop quarks from the $tZq$ production are strongly polarized, making this process an excellent probe for studying $t-Z$ couplings,~particularly in the context of EFT measurements.
It is also sensitive to triple-gauge couplings $WWZ$, in~a complementary manner with the diboson production. Both are potentially sensitive to physics beyond the~SM.

Data are selected with a combination of single-lepton or double-lepton triggers. 
Events are selected events if they contain three well-identified and isolated leptons (electrons or muons possibly arising from $\tau$ lepton decays).
A pair of same-flavored opposite-charged leptons, compatible with a $Z$ boson decay, is then required. 
Because the $tZq$ process is a $t$-channel process, it contains a light jet preferentially produced at large $|\eta|$, a~b-tagged jet arising from the top quark decay, and missing transverse energy arising from the neutrino from the $W$ boson~decay.

Similar to other analyses presented in this review, the~signal is extracted from signal and control regions defined by the number of jets and b-tagged jets. 
The first signal region requires $N_j=2$ or $N_j=3$ with $N_b=1$ (so-called $2j1b$ and $3j1b$ regions). These regions contain most of the signal with the $WZ$+jets process as the dominant background, and~with contributions from other diboson processes. 
For larger jet multiplicities and b-tagged jet multiplicities ($N_j\geq3$, $N_b\geq 2$), the~dominant background source arises from $t\bar{t}Z$ events, with~contamination from $tt+W,H$ processes. 
A control region with $N_b=0$ allows constraining the diboson~contribution.

Other background sources are from non-prompt lepton events in $t\bar{t}$ or $Z$+jets processes. 
While backgrounds presenting three prompt leptons are estimated from simulations  and~constrained from the data in the likelihood fit, events containing at least one non-prompt lepton are not well-described by simulations and are, therefore, more difficult to estimate. 
In the CMS observation paper~\cite{Sirunyan_2019tZqOBs}, the~analysis uses a fully data-driven technique, where the probabilities for measuring a non-prompt lepton are measured from a region where one lepton fails the lepton isolation. 
The ATLAS observation paper~\cite{Aad_2020_tZqObs} uses a semi data-driven technique, where the normalization of the non-prompt background is estimated from the data in control regions, and~the kinematic distributions are determined from simulations of \ttbar+$tW$ and $Z$+jets events, by~replacing b-jets with non-prompt leptons and accounting for the needed~corrections.

\begin{figure}[H]
  \resizebox{6cm}{!}{\includegraphics{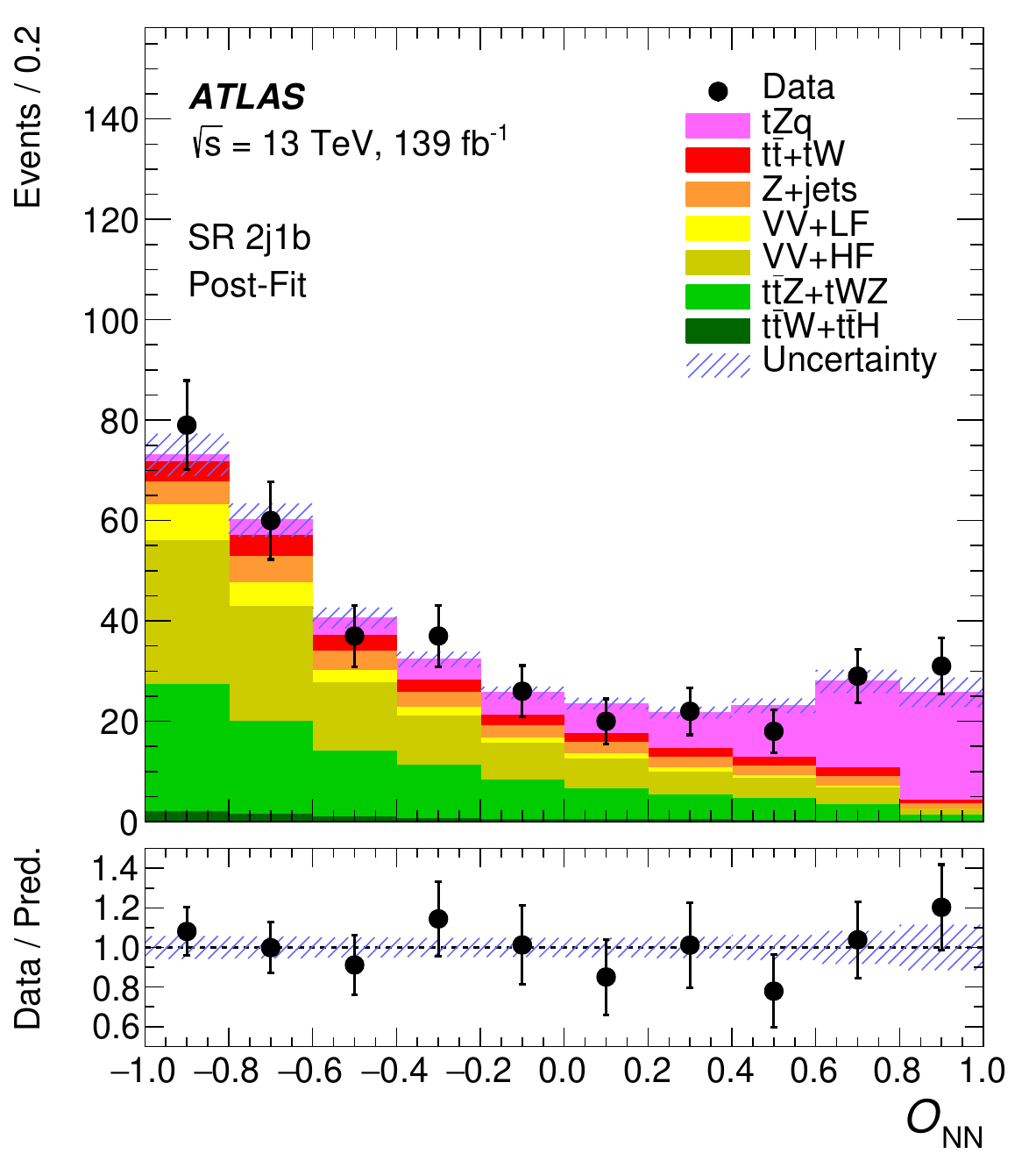}}
  \resizebox{6cm}{!}{\includegraphics{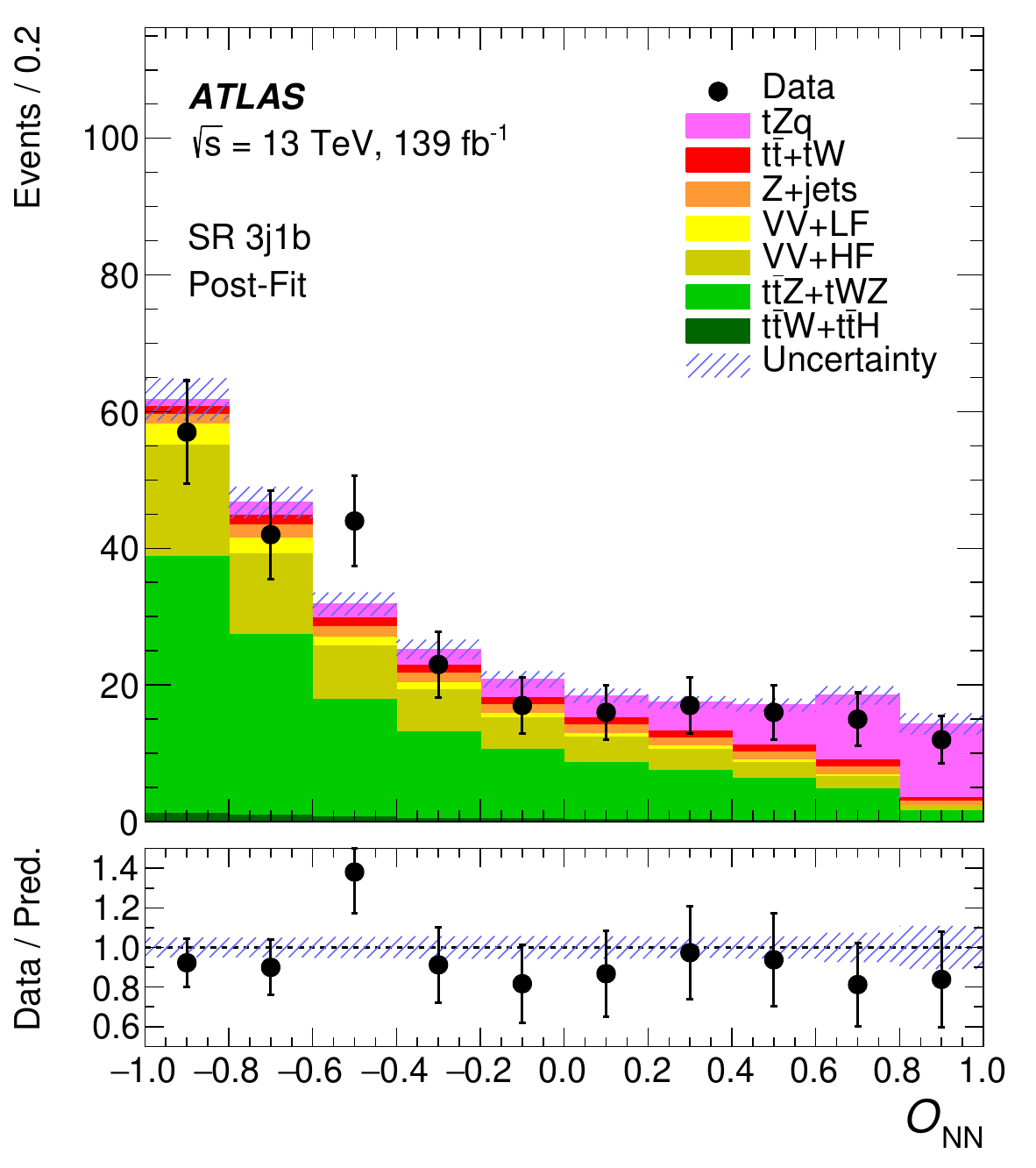}}
  \caption{Distributions of the NN output in the signal regions $2j1b$ (\textbf{left}) and $3j1b$ (\textbf{right}) in the ATLAS analysis~\cite{Aad_2020_tZqObs}.} 
  \label{fig:tZqATLASNN}
\end{figure}

The discriminating variables used in the fit are based on multivariate discriminants (BDT or NN), which include kinematic variables related to the reconstructed $Z$ bosons or top quarks, the~pseudorapidity of the spectator jet $|\eta_{j'}|$, dijet invariant mass, or~kinematic variables related to the lepton from the $W$ decay. Examples of NN output distributions from ATLAS~\cite{Aad_2020_tZqObs} can be found in Figure~\ref{fig:tZqATLASNN}.

The most recent inclusive $tZq$ cross-sections measured by ATLAS~\cite{Aad_2020_tZqObs} and CMS~\cite{JHEP_TOP_20_10} are compatible with the SM. The~precision is still dominated by statistical uncertainties, although~with the CMS results, systematic and statistical uncertainties are almost of the same level.
The dominant sources of systematic uncertainties are mainly experimental, and~are not identical between ATLAS and CMS (while the analysis techniques are relatively similar). For ATLAS, the dominant uncertainties arise from background modeling and normalization, the~jet energy scale, and~the lepton selection; for CMS, the~systematic uncertainties are dominated by the QCD scale uncertainties in the signal modeling, non-prompt lepton background estimation, and $WZ$ process normalization. All these uncertainties have similar orders of magnitude. 
Differences in the relative sizes of the systematic uncertainties by ATLAS and CMS are probably explained by differences in the analysis strategy, noticeably in the treatment of the background~estimates.

\begin{figure}[H]
  \resizebox{6cm}{!}{\includegraphics{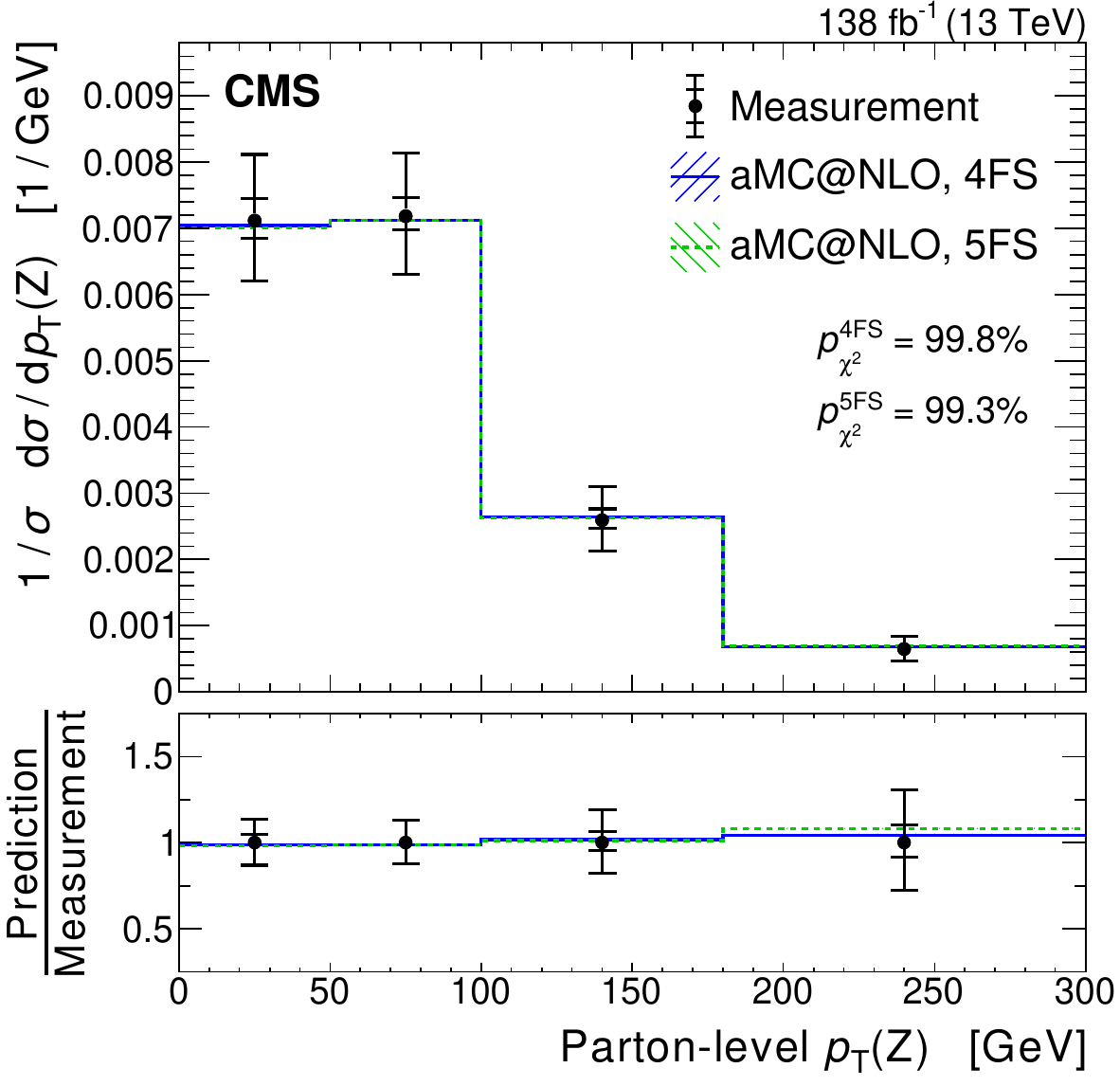}}
  \resizebox{6cm}{!}{\includegraphics{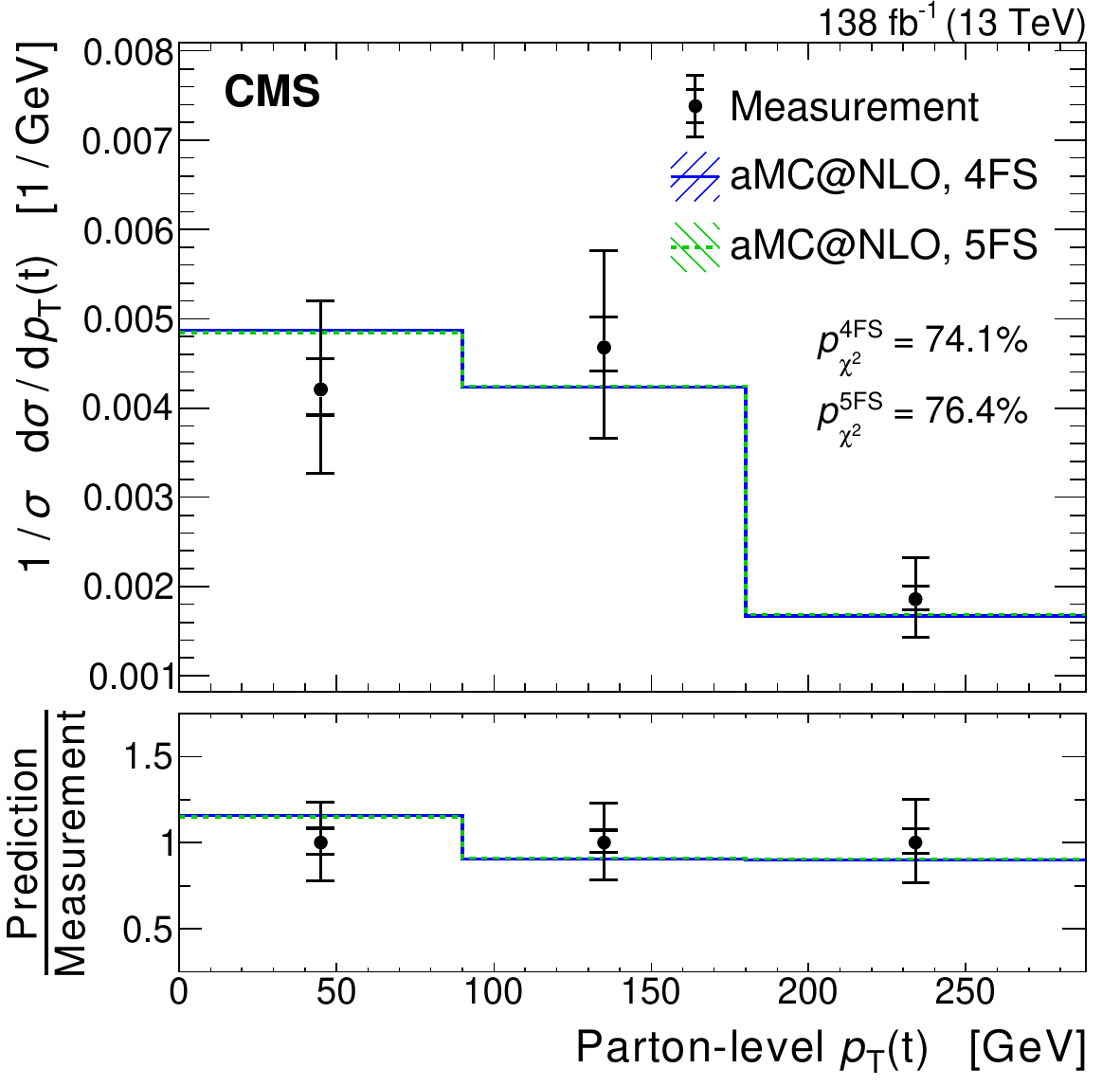}}
  \resizebox{6cm}{!}{\includegraphics{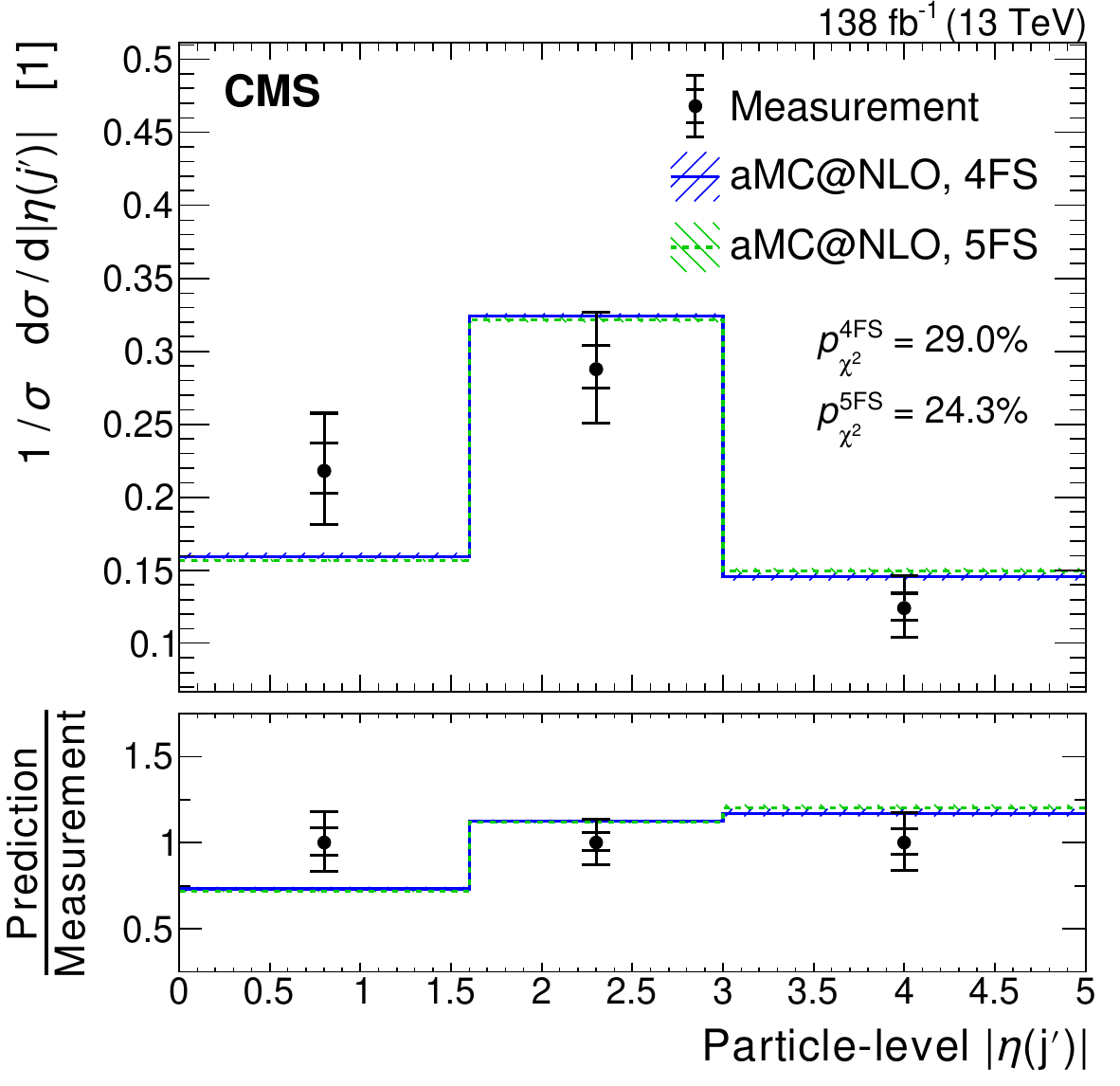}}
 \hspace{48pt} \resizebox{6cm}{!}{\includegraphics{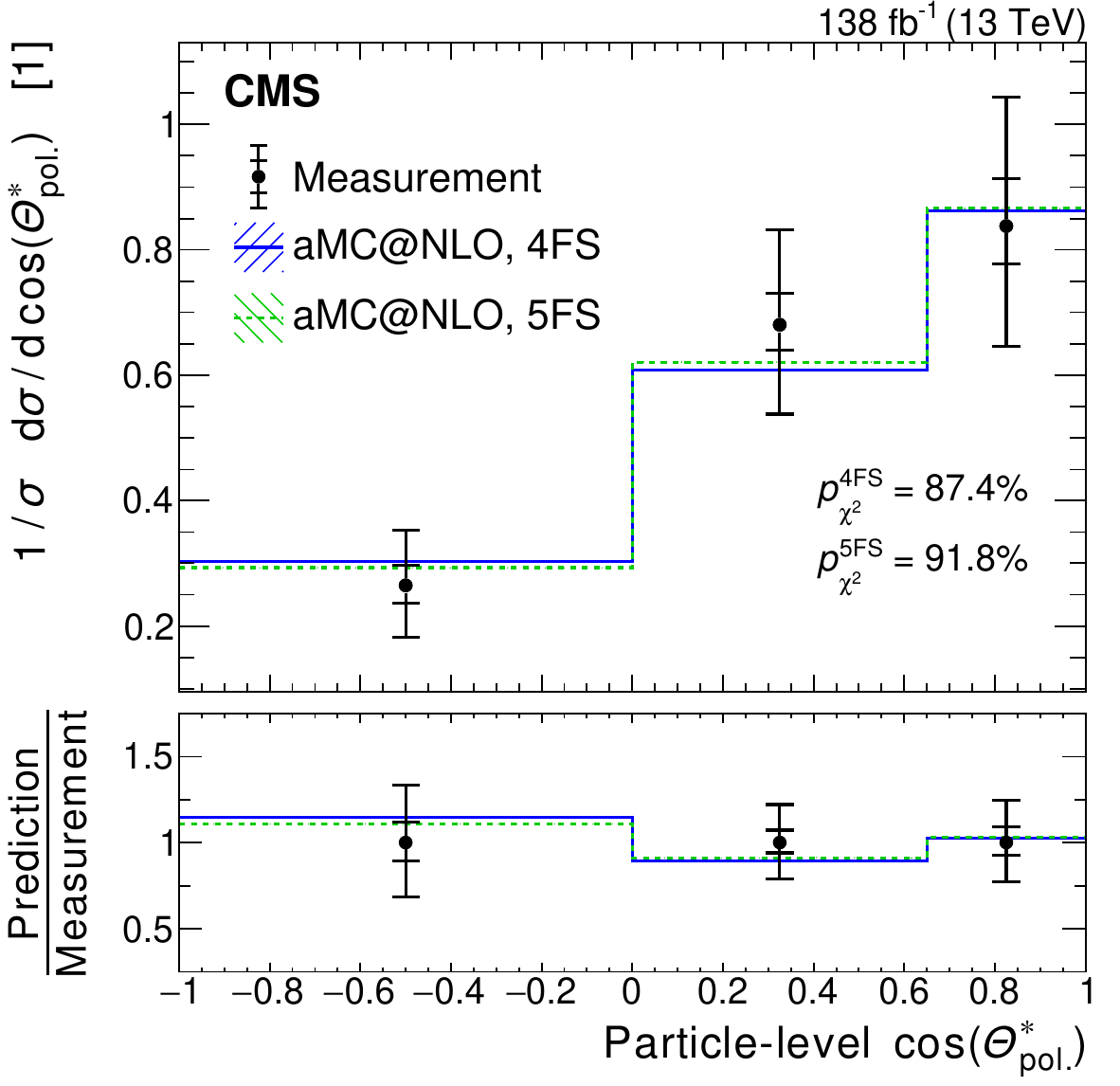}}
  \caption{Normalized differential cross-sections measured at the parton level as a function of $p_T(Z)$ (\textbf{top left}) and $p_T(t)$ (\textbf{top right}) at the parton level, and~as a function of $|\eta(j')|$ and $\cos{\theta^*_{pol.}}$ at the particle \textls[-15]{level~\cite{JHEP_TOP_20_10}. The~inner and outer vertical bars represent the systematic and total uncertainties,~respectively. }} 
  \label{fig:tZq_diffDistrib}
\end{figure}

Due to the large integrated luminosity provided by the LHC, it is now possible to measure differential cross-sections for $tZq$ production~\cite{JHEP_TOP_20_10}. 
This analysis follows a different approach to extract the signal. A~multi-class NN is used to separate the $tZq$ process from the $t\bar{t}Z$, $WZ$, and other $t+X$ processes. The~signal region is then sub-divided based on the bins of the observables of interest, at the~detector level. The~NN score of the $tZq$ node is used to extract the signal in each bin. Similarly, the~NN score of the $t\bar{t}Z$ node is used to constrain the $t\bar{t}Z$ background. An~unfolding procedure infers the particle- or parton-level distributions. Examples of differential cross-section measurements for $p_T(Z)$ and $p_T(t)$ at the parton level, and~$|\eta(j')|$ and $\cos{\theta^*_{pol.}}$ at the particle level are shown in Figure~\ref{fig:tZq_diffDistrib}. The~$\cos{\theta^*_{pol.}}$ variable is the cosine of the polarization angle of the top quark, defined as:
\begin{linenomath}
\begin{equation}
\cos{\theta^*_{pol.}} = \frac{\vec{p}(q'^*)\cdot \vec{p}(l_t^*)} { | \vec{p}(q'^*) || \vec{p}(l_t^*)| }  
\end{equation} 
\end{linenomath}
with $\vec{p}(q'^*)$ and $\vec{p}(l_t^*)$, the three momenta of the light jet and  the lepton from the top quark decay. A good agreement between data and predictions is observed. This very promising publication presents the first differential measurements of a rare single-top process, and~can serve as the basis for future studies. In~particular, it provides a clear procedure to perform a differential measurement, featuring~an interesting signal extraction based on a multi-class~discriminant.

Eventually, the~first measurement for the $tWZ$ process led to evidence~\cite{CMS-PAS-TOP-22-008} (presented as a conference note by CMS). This very rare process can be seen as a $tZ$ production in the $tW$ channel, where it shares similar modeling issues since it interferes at NLO with the $t\bar{t}Z$ process~\cite{ATLAS:2016xpx}. The~analysis techniques are similar to those of the $tZq$ analysis, using a multi-class NN, with~a multi-lepton signature targeted. This result opens up a new era for measuring top quark processes associated with~multi-bosons.

\subsection{\label{SingleTopTHQ}The $tH$ Processes: Companions for the top Quark Yukawa~Coupling}

\subsubsection{Introduction to the $tH$ Processes}

\begin{figure}[H]
\includegraphics[width=0.6\textwidth]{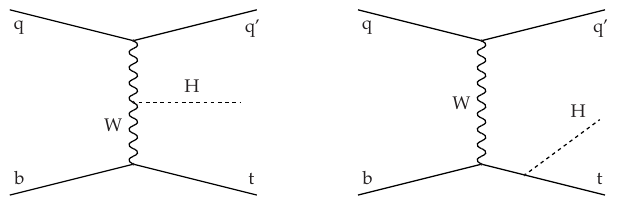}
\caption{\label{THq:Diagrams}Examples of Feynman diagrams for $tH q$ processes at the LO in pQCD~\cite{CMS:2015nrd}.}
\end{figure}

Among the processes involving a top quark and a boson in the final state, the~$tH$ processes are produced with the lowest cross-section predicted in the SM, of~approximately 71 fb and 16 fb at NLO for the $t$-channel ($tHq$ process) and the $tW$-associated production ($tHW$ process) with $\sqrt{s}=$13 TeV~\cite{LHCHiggsCrossSectionWorkingGroup:2016ypw}. 
The latest predictions for the $tHq$ cross-section are computed at NNLO in pQCD~\cite{Forslund:2021evo}. 
The $tHq$ processes share many properties with the $t\gamma q$ and the $tZq$ processes, noticeably their modeling in the 4FS or 5FS schemes, and~the production of an associated quark in the forward direction. The~Feynman diagrams for the production of $tHq$ are depicted in Figure~\ref{THq:Diagrams}.
The $tHW$ production is also considered in the~analyses.

\vspace{-3pt}

\begin{figure}[H]
\includegraphics[width=0.45\textwidth]{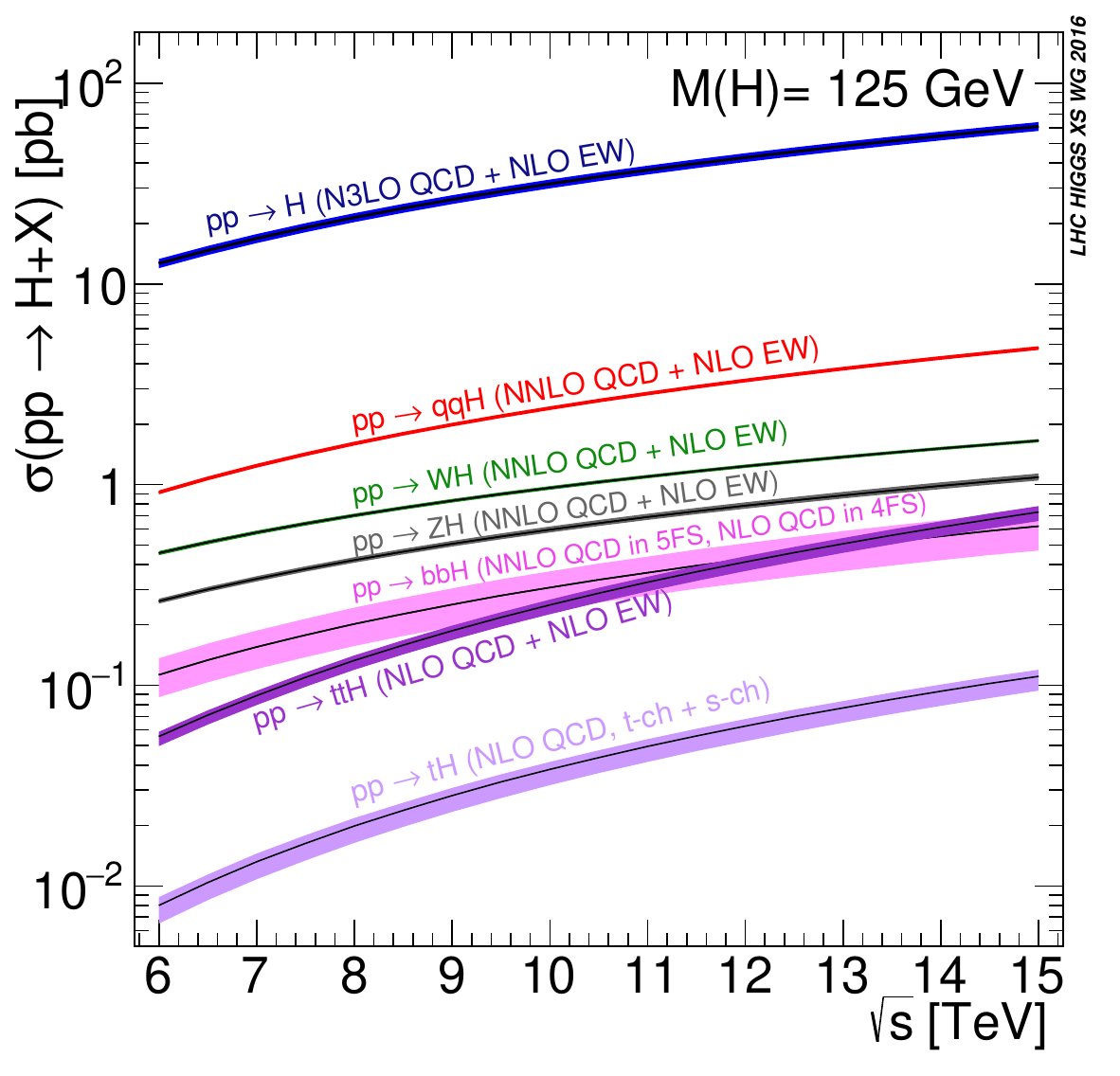}
\includegraphics[width=0.45\textwidth]{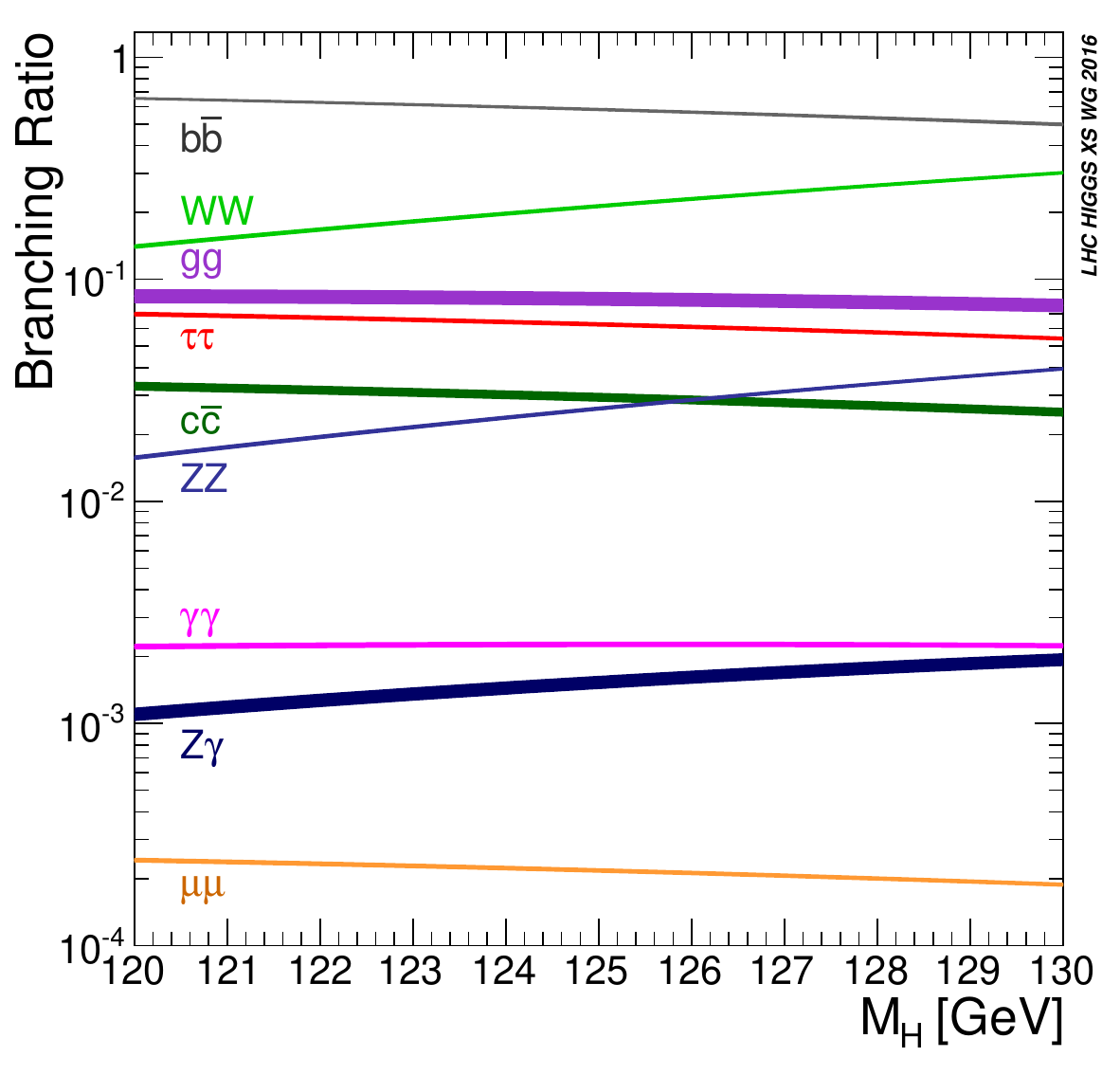}
\caption{\label{THq:HiggsXSvsEnergy}\textbf{Right}: Cross-section for the Higgs boson production as a function of $\sqrt{s}$. \textbf{Left}: branching ratio for the Higgs boson decay~\cite{LHCHiggsCrossSectionWorkingGroup:2016ypw}.}
\end{figure}

The search for $tH$ processes is traditionally performed in association with the search for the Higgs boson measurement in the $t\bar{t}H$ production mode, whose cross-section is larger than the cross-section of the $tHq$ process by a factor of 10, as~shown in Figure~\ref{THq:HiggsXSvsEnergy}. 
The amplitude for $tHq$ production features an interesting property, as it features interference between diagrams where the Higgs boson is emitted from a top quark line and those arising from $W$ boson exchange. This property makes the measurement of the $tHq$ process appealing since it provides access to the sign of the Yukawa coupling of the top quark. If~the sign of the Yukawa coupling $\kappa_t$ is negative, the~interference becomes constructive, for~instance, by  increasing the cross-section by a factor of approximately 12 if $\kappa_t = -1$~\cite{Farina:2012xp}. 
The $tH$ final states, on~equal footing with the $tZ$ and the $t\gamma$ final states, are also used in the FCNC searches~\cite{universe8110609}. 

\unskip

\subsubsection{Searches for the $tH$ Processes}

The early searches for the $tH$ processes at 8 TeV~\cite{CMS:2015nrd} attempted to directly measure the $tH$ production, while the $t\bar{t}H$ process was treated as a background. It was, however, realized that by varying the value of the top quark Yukawa coupling, the~cross-section for both the $tH$ and $t\bar{t}H$ processes would be modified in a correlated way. Nowadays, the~searches for the $tHq$ process are performed in a combined measurement with the $t\bar{t}H$ process, either targeting the measurement of the top quark Yukawa coupling or~simultaneously measuring the cross-section for the $tH$ and  $t\bar{t}H$ processes.

Branching ratios for the Higgs boson decay are shown in Figure~\ref{THq:HiggsXSvsEnergy}. The~ATLAS and CMS analyses target the main decay modes of the Higgs boson: $H\rightarrow\gamma\gamma$; $H\rightarrow WW$, $H\rightarrow ZZ$, and $H\rightarrow \tau\tau$ (grouped under the naming of the ``multilepton final state'' since $W$, $Z$, $\tau$, and associated top quarks can decay leptonically), and to a lesser extent, $H\rightarrow b \bar{b}$ (suffering from a lack of available luminosity to achieve similar sensitivity as the other channels). We present here the methodology and the latest~results.

The analysis of the $H\rightarrow\gamma\gamma$ decay channel with Run 2 data by ATLAS~\cite{ATLAS:2022tnm} and CMS~\cite{CMS:2021kom} follows a similar strategy to the measurements of the other production mechanisms of the Higgs boson. 
The small $H\rightarrow\gamma\gamma$ branching fraction (close to 0.2\% at $m_H=125$ GeV) is compensated by the excellent resolution of the electromagnetic calorimeters (the effective mass resolution on the Higgs boson is close to 1.5 GeV, depending on the analysis categories). The~background processes involving jets reconstructed as photons are reduced using photon isolation and information on the shape of the electromagnetic energy deposit, with~sequential criteria by ATLAS and a multivariate method at CMS. Several event classes are constructed, specifically targeting a given production mechanism. For~each event class targeting the $tH$ processes, the~background is reduced by the means of a BDT discriminant, which is subsequently fitted with a smoothly falling function. 
In the latest versions of the analysis~\cite{ATLAS:2022tnm,CMS:2021kom}, several subcategories are built to specifically target the $t\bar{t}H$ and $tH$ processes in kinematic bins, and~the fit is interpreted in the so-called ``simplified template cross-section'' framework (STXS)~\cite{Berger:2019wnu}. The~STXS framework is a convention used to provide results in kinematic bins at particle levels within a defined acceptance for each Higgs boson production mechanism. 
In the CMS analysis, a~category at the reconstructed level specifically targets $tH$ in the leptonic channel, and~a DNN discriminant is used to improve the separation between $t\bar{t}H$ and $tH$. Using this category, together with many reconstructed-level event classes in a simultaneous fit, the~cross-section for the $tH$ processes at the STXS level is quoted to be $6.3 ^{+3.4}_{-3.7}$ times the SM expectation (in the so-called ``maximal merging scenario'', where fewer  STXS categories are used at the particle level than in the ``minimal merging scenario''). 
In the ATLAS analysis, four reconstructed categories targeting $tH$ processes are defined, where two categories specifically target the $tHq$ processes with either a positive or a negative top quark Yukawa coupling (defined using the output of a NN),~one category targets the $tHW$ process, and~the remaining category gathers events with low-scores of the BDT for $tHq$ and $t\bar{t}H$. At~the STXS level, the~cross-section for the $tH$ processes is $2.1 ^{+4.2}_{-3.1}$ times the SM~expectation.

\begin{figure}[H]
\includegraphics[width=0.55\textwidth]{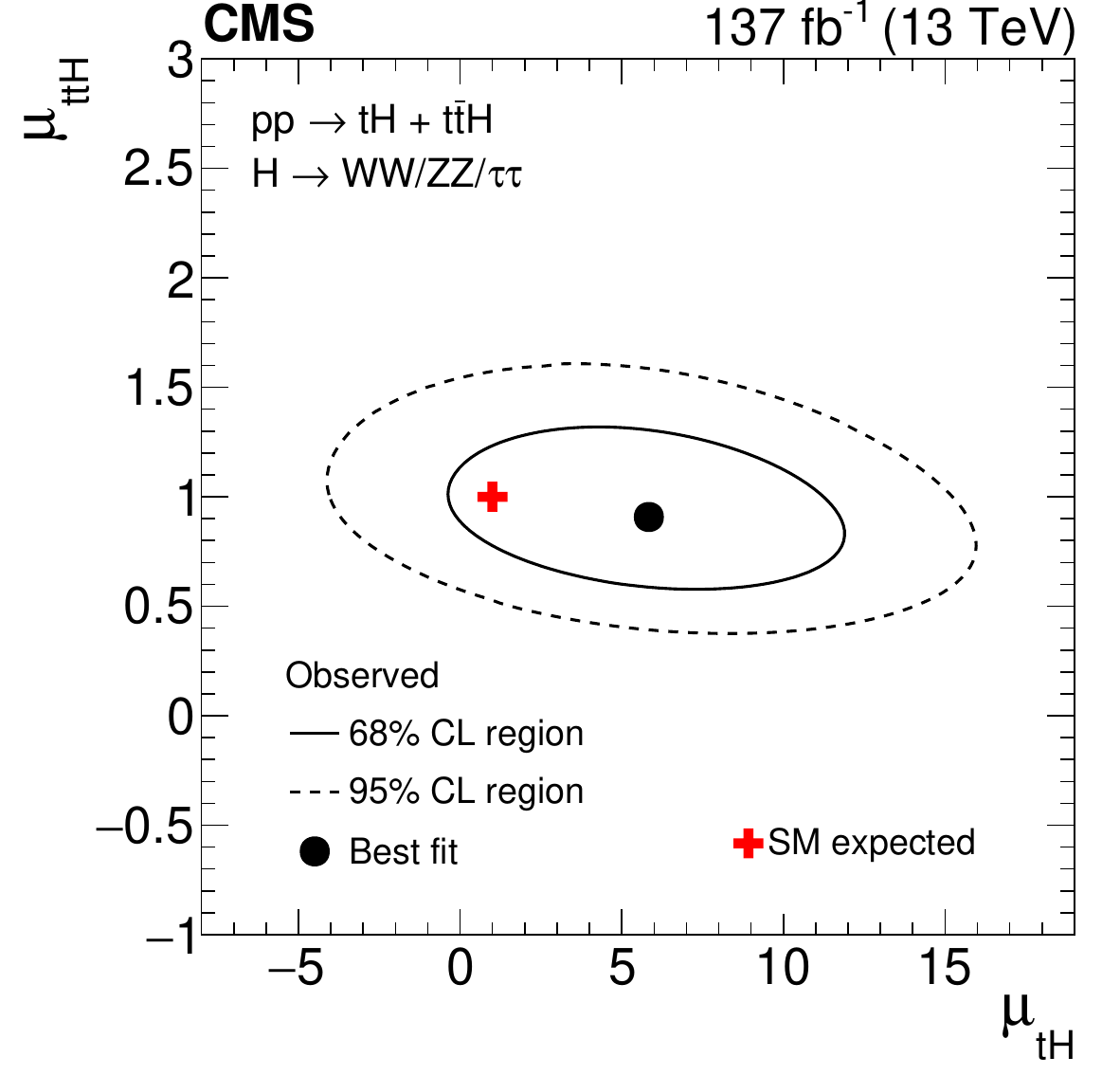}
\caption{\label{THq:CMSmuTTHvsTH}Likelihood contour as a function of the signal strengths for $t\bar{t}H$ and $tH$ processes in the multilepton analysis~\cite{CMS:2020mpn}.}
\end{figure}

Using the multilepton channel, CMS~\cite{CMS:2020mpn} reported measurements of the cross-section for $t\bar{t}H$ and $tH$ production simultaneously with Run 2 data. This analysis uses multiple final states. For leptonic top decay, the configurations are the same-sign $2\ell+0\tau_h$ (where $\ell=e,\mu$ and $\tau_h$ denotes hadronically decaying $\tau$), $3\ell+0\tau_h$, $2\ell+1\tau_h$ (both same-sign and opposite-sign), $1\ell+2\tau_h$, $4\ell+0\tau_h$, $3\ell+1\tau_h$ and $2\ell+2\tau_h$. For hadronic top decay, the configurations are $1\ell+1\tau_h$ and $0\ell+2\tau_h$. 
The sensitivity arises mainly from the same-sign channel $2\ell+0\tau_h$, $3\ell+0\tau_h$, and $1\ell+2\tau_h$. In~those main categories, the~analysis employs a multi-class DNN, separately providing discriminants for $t\bar{t}H$ and $tH$, while using simpler BDTs in the other categories. In~the same-sign channels, $2\ell+0\tau_h$, $2\ell+1\tau_h$, categories are further divided according to the lepton flavor and whether the b-jet number is larger or smaller than 2.
The jet-faking lepton background is estimated with a data-driven method by relaxing lepton identification criteria in a region enriched in multijet events. Backgrounds resulting from mismeasuring the lepton charge are determined using $Z\rightarrow ee$ events. The~dominant background arises from $t\bar{t}W$ and $t\bar{t}Z$ processes, estimated from simulation. The~background arising from the conversion of leptons in the detector is estimated from the simulation. 
The signal is extracted using bins in the multivariate discriminants. Several control regions with $3\ell$ and $4\ell$ final states are also used in the fit. Two parameters of interest are measured: the signal strength $\mu$ for $t\bar{t}H$ and for $tH$ processes. 
The signal strength for $tH$ production is \mbox{$5.7 \pm 2.7$(stat) $\pm 3.0$(syst)}. Additionally, a~2-dimensional distribution of the likelihood as a function of $\mu_{t\bar{t}H}$ and $\mu_{tH}$ is measured, as~shown in Figure~\ref{THq:CMSmuTTHvsTH}. 

A few results arising from combinations of final states are also reported. 
A dedicated CMS analysis specifically targeting $tH$ at 13 TeV using 36 fb$^{-1}$ of Run 2 data~\cite{CMS:2018jeh} employed the $H\rightarrow\gamma\gamma$ and multilepton final states, as~well as the final state $H\rightarrow b\bar{b}$ in the $VH$ production mode with the single-lepton decay of the associated boson. The~multilepton analysis uses simpler techniques than those previously described~\cite{CMS:2021kom,CMS:2020mpn}, and~trained multivariate methods with $tH$ processes as the signal. The~$H\rightarrow\gamma\gamma$ analysis reinterprets the content of the $t\bar{t}H$ categories of a previous analysis. The~$H\rightarrow b\bar{b}$ analysis brings little sensitivity and will not be described here. The~combined measurement results in an observed limit on the cross-section for the $tH$ production of 1.94 pb at 95\% CL in~the SM hypothesis. The results for the hypotheses with negative top quark Yukawa coupling are also reported.
In commemoration of the 10th anniversary of the Higgs boson discovery, grand combinations were performed by both CMS~\cite{CMS:2022dwd} and ATLAS~\cite{ATLAS:2022vkf}, including many final states. Categories specifically targeting  the $tH$ processes are taken from the $H\rightarrow\gamma\gamma$ channel by ATLAS, and~$H\rightarrow\gamma\gamma$ and the multilepton channel at CMS. The~CMS combination reports a measured signal strength of $\mu_{tH} = 6.05 ^{+2.66}_{-2.42}$.

Studies estimating the sensitivity to the $tH$ processes at the HL-LHC were expecting a relative uncertainty of 90\% on the $tH$ signal strength in the SM hypothesis~\cite{CMS:2018qgz} (with the $t\bar{t}H$ signal strength floating); however, these studies were based on early projections and would need to be updated with the latest ATLAS and CMS~results.

\subsubsection{Probing the Sign of the Top Quark Yukawa~Coupling}

With the analyses from ATLAS and CMS for $H\rightarrow\gamma\gamma$ final state~\cite{ATLAS:2022tnm,CMS:2021kom}, combined with~the CMS multilepton analysis~\cite{CMS:2020mpn} and the earlier CMS combination~\cite{CMS:2018jeh}---all of which include categories that specifically target the $tH$ processes---it is now feasible to determine the sign of the top quark Yukawa coupling, thanks to the interference observed between Feynman diagrams that showcase the Higgs boson coupling to both the top quark and the $W$ boson. The~modifier $\kappa_t$ of the top quark Yukawa coupling in the SM, $y_{t,SM}$, is defined as $\kappa_t = y_{t}/y_{t,SM}$. Furthermore, since a similar interference is also present in the $H\rightarrow\gamma\gamma$ decay between the top quark loop and the $W$ boson loop, further sensitivity is gained in this channel. Sensitivity to the positive values remains dominated by the $t\bar{t}H$ process in direct measurements, and~by the $gg\rightarrow H$ process (involving a top quark loop) in indirect measurements because of the larger cross-section. 

\begin{figure}[H]
\includegraphics[width=0.45\textwidth]{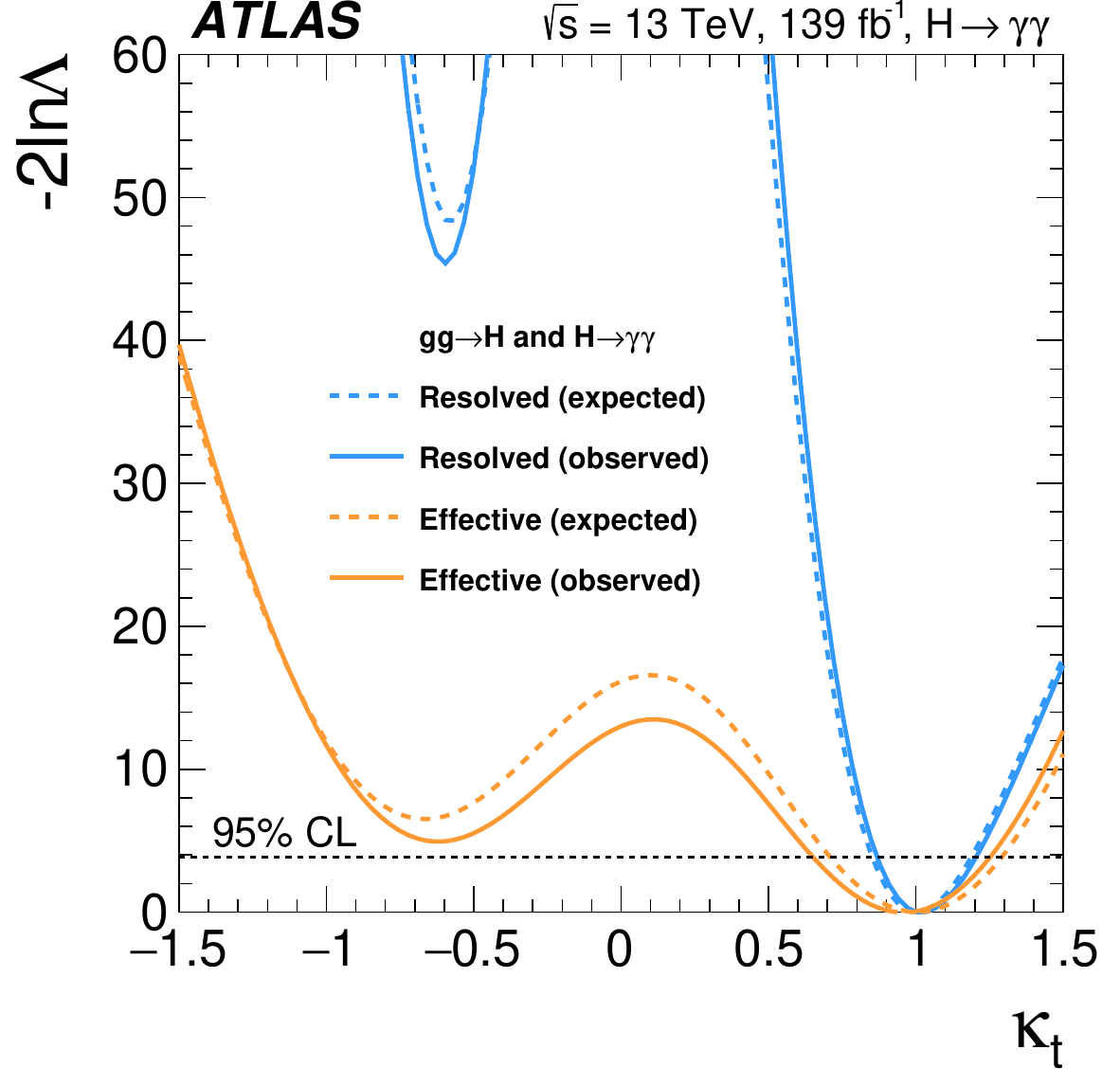}
\includegraphics[width=0.45\textwidth]{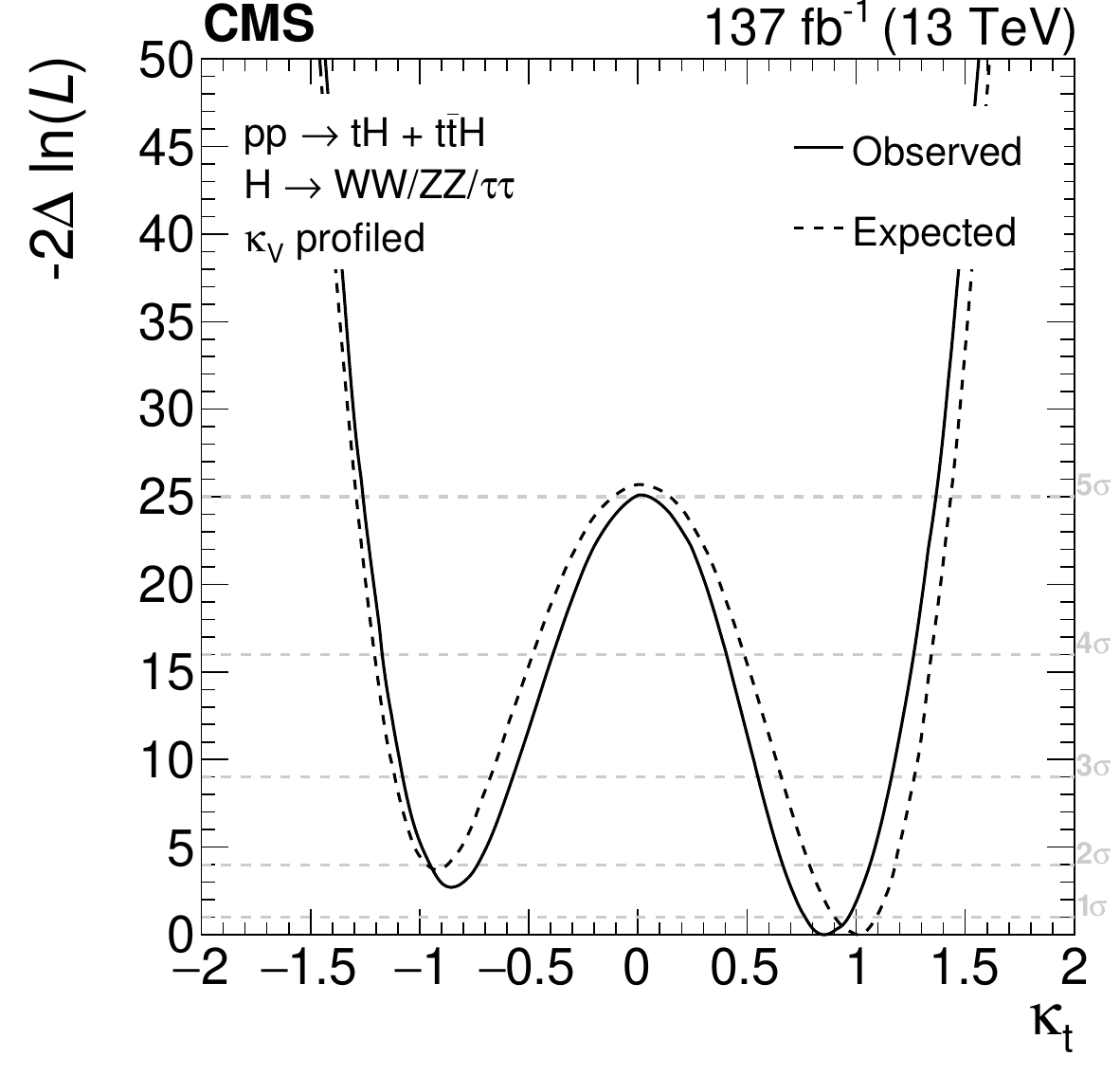}
\caption{\label{THq:KappaTopScans}Log-likelihood as a function of the $\kappa_t$ parameter, in~the ATLAS $H\rightarrow\gamma\gamma$ analysis~\cite{ATLAS:2022tnm} (\textbf{left}), and~in the CMS multilepton analysis~\cite{CMS:2020mpn} (\textbf{right}).}
\end{figure}

Figure~\ref{THq:KappaTopScans} shows the likelihood fit value as a function of the $\kappa_t$ parameter. The~best-fit value is positive and close to 1, while a second minimum of the likelihood is found at a value close to -1. As~shown on the left side of Figure~\ref{THq:KappaTopScans}, including the parameterization of the gluon fusion mechanism as a function of $\kappa_t$ in the likelihood provides more weight to the positive value of $\kappa_t$. On~the contrary, when only the $tH$ and $t\bar{t}H$ processes are included, more sensitivity is gained on the sign of $\kappa_t$. Values outside of $0.65 < \kappa_t < 1.25$ in the first case and~$0.87 < \kappa_t < 1.20$ in the second case are excluded at 95\% CL by the $H\rightarrow\gamma\gamma$ analysis by ATLAS. The~CMS multilepton analysis results in $-0.9 < \kappa_t < -0.7$ or $0.7 < \kappa_t < 1.1$ at 95\% CL.

Projections for the measurement of the top quark Yukawa coupling at the HL-LHC are reported by CMS~\cite{CMS:2022dwd} without emphasis on a possible negative coupling. A~precision on the order of 3-4\% on $\kappa_t$ would be achievable, while a precision on the order of 10\% is achieved today~\cite{CMS:2022dwd,ATLAS:2022vkf}. 

The $tH$ processes, together with the $t\bar{t}H$ process, can also be used to set constraints on a CP-odd top quark Yukawa coupling. Such measurements were performed by ATLAS with the $H\rightarrow\gamma\gamma$~\cite{ATLAS:2020ior} and $H\rightarrow b \bar{b}$~\cite{ATLAS:2023cbt} channels, and~at CMS with the $H\rightarrow\gamma\gamma$~\cite{CMS:2020cga} and multilepton~\cite{CMS:2022dbt} channels. Since the $t\bar{t}H$ process has a larger cross-section than the $tH$ processes, most of the sensitivity will come from the former, and~these measurements will not be described~here.

\section{\label{SingleTopProperties}Discovery Potential of Property Measurements and~Interpretations}

The large number of single-top events produced at the LHC and the high precision obtained in single-top measurements allow for the measurements of top quark properties, which can be seen as tests of the SM or a search for physics beyond the SM. 
Although single-top quark production has a lower cross-section compared to  \ttbar\ production, the production and subsequent decay of single-top quarks to $Wb$ engage the $Wtb$ vertex twice, during both the top quark production and its decay. This interesting feature can be used to measure several interconnected properties: the couplings of the $Wtb$ vertex, including the CKM matrix element $|V_{tb}|$, the~$W$ polarization, and the top quark polarization. Precision measurements of the $Wtb$ couplings can be expressed in terms of CP-even and CP-odd effective couplings or within the EFT. Apart from the $|V_{tb}|$ measurement, which can be inferred from the single-top cross-section, the~general experimental strategy for measuring all other properties consists of performing various angular analyses of the top quark decay, and choosing suitable angular distributions to measure the parameters of interest. 
Additional couplings can be probed within the EFT, including four-fermion couplings, and~couplings between the top quark and neutral~bosons. 

This review will not discuss the top quark mass measurement using the single-top $t$-channel or the CPT symmetry tests comparing top and antitop quark masses in single-top events (for a recent result, see~\cite{CMS:2021jnp}), since the precision is not yet at the required level for competing with \ttbar\ measurements.
This section will cover the other above-mentioned top quark properties using the single-top quark as a probe, reaching a precision similar to or better than that achieved in \ttbar\ measurements.

\subsection{\label{VtbSection}Measurement of the CKM Matrix Element $|V_{tb}|$}

\begin{figure}[H]
  \resizebox{11cm}{!}{\includegraphics{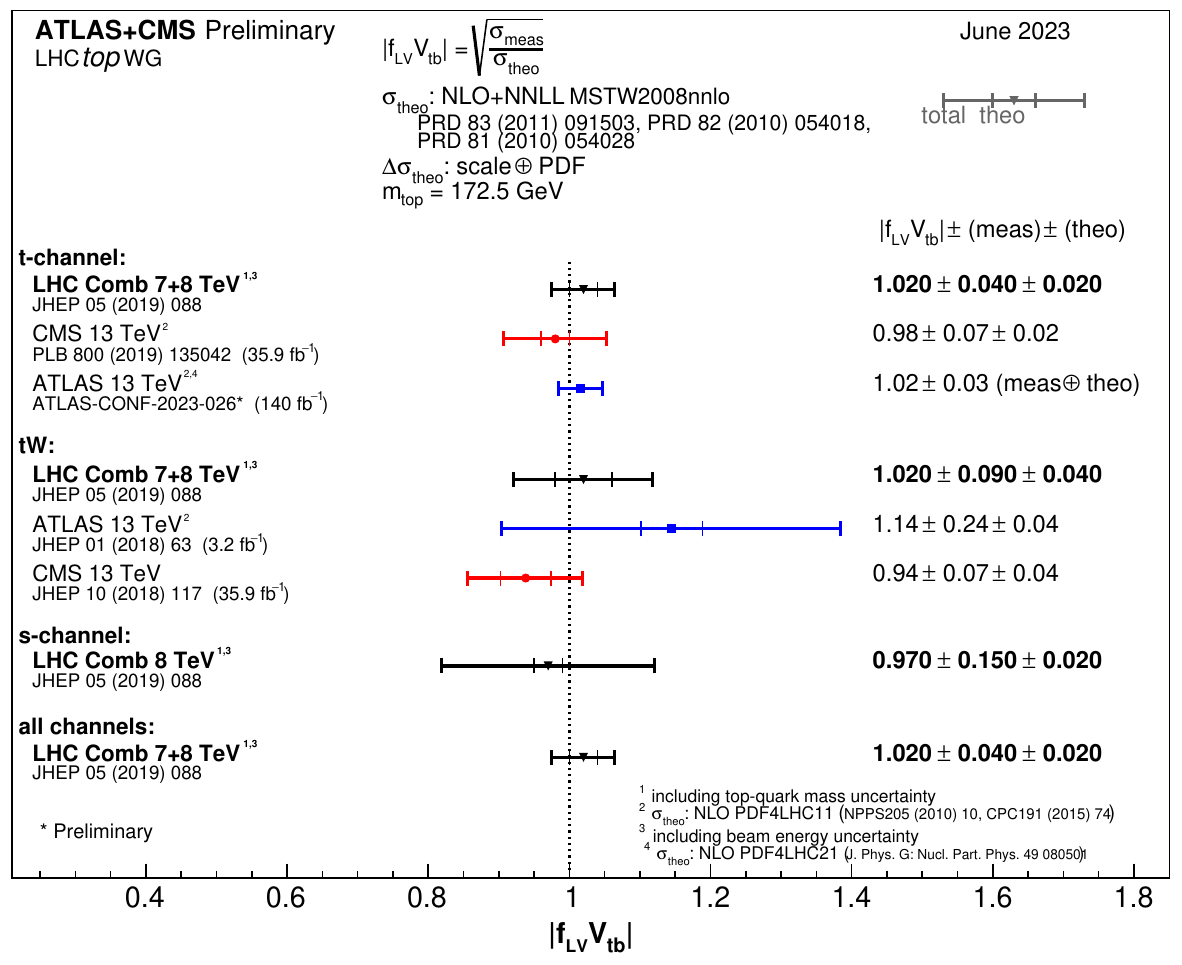}}
  \caption{Summary of the ATLAS and CMS extractions of the CKM matrix element $V_{tb}$ from single-top quark measurements~\cite{JHEP05.2019.088}, compared with theoretical predictions at NLO+NNLL accuracy~\mbox{\cite{Kidonakis:2011wy,Kidonakis:2010ux,Kidonakis:2010tc}}.} 
  \label{fig:tchan_vTB}
\end{figure}

Because the $V_{tb}$ CKM matrix element is close to unity in the SM, the~measurement of $V_{tb}$ is particularly intriguing, and~its study is an excellent way to better understand the SM and search for signs of new physics. 
The measurement of the $V_{tb}$ CKM matrix element is strongly related to the electroweak nature of the single-top production. 
The cross-section for the single-top production can be used to test the unitarity of the CKM matrix. Assuming the values of $|V_{td}|$ and $|V_{ts}|$ are much smaller than those of $|V_{tb}|$, the~measured single-top cross-section can be used to determine $|V_{tb}|$ according to the following formula~\cite{JHEP05.2019.088}:
\begin{linenomath}
\begin{equation}
|f_{L_V}V_{tb}| = \sqrt{ \frac{\sigma^{meas}}{\sigma^{theo}} },
\end{equation}
\end{linenomath}
with $\sigma^{meas}$ denoting the measured cross-section, $\sigma^{th}$ denoting the SM theoretical cross-section, assuming \mbox{$|V_{tb}|=1$}, and~$f_{LV}$ denoting an anomalous form factor (of the kind vectorial left-handed, as~in the SM), which can be different from 1 in new physics models. 
Such a method was used to reinterpret several single-top cross-sections at 7 and 8 TeV. Their combinations, including ATLAS and CMS results for $t$-channel, $tW$ production, and $s$-channel, were performed in the context of the LHC$top$WG,~leading to the most precise $|f_{LV}V_{tb}|$ measurement to date, as~shown in Figure~\ref{fig:tchan_vTB}. One can see that the $t$-channel measurements dominate the combination. 
The latest $t$-channel measurement at 13 TeV~\citep{ATLAS-CONF-2023-026} improves over this combination by approximately 30\% in precision, with~$|V_{tb}|=1.014 \pm 0.031$ reported.

It is possible to release the assumption that $|V_{td}|$ and $|V_{ts}|$ are negligible compared to $|V_{tb}|$. 
Such a method has also been pursued, consisting of measuring $|V_{tb}|$, $|V_{td}|$ and $|V_{ts}|$ in a model-independent way, using single-top $t$-channel-enriched events~\cite{CMS:2020vac}. The~main principle of the analysis relies on considering several single-top $t$-channel signals, according to the presence of a $tWb$ vertex in single-top production ($ST_{b, q}$), in~top quark decay ($ST_{q, b}$), or~in both ($ST_{b, b}$). Several signal regions, based on the jet and b-tagged jet multiplicities, can be defined and fitted simultaneously. Further discrimination between $ST_{b, q}$, $ST_{q, b}$, and $ST_{b, b}$ is obtained using kinematic and angular properties of the involved processes, using the fact that (1) PDFs are different for each of them, and~(2) the presence of an additional b-jet from gluon-splitting can affect top quark reconstruction. Using the constraint of CKM unitarity ($|V_{tb}|^2 + |V_{ts}|^2 +|V_{td}|^2 = 1$), a~precision similar to that of the combination~\cite{JHEP05.2019.088} is achieved~\cite{CMS:2020vac} with an integrated luminosity of 35.9 fb$^{-1}$ of 13 TeV proton--proton collisions. The~method allows performing the measurements under the constraints of BSM scenarios. The results are compatible with previous measurements and the SM~predictions.

\subsection{$W$ Boson Polarization~Fractions}

\begin{figure}[H]
\hspace{-3pt}\includegraphics[width=0.45\textwidth]{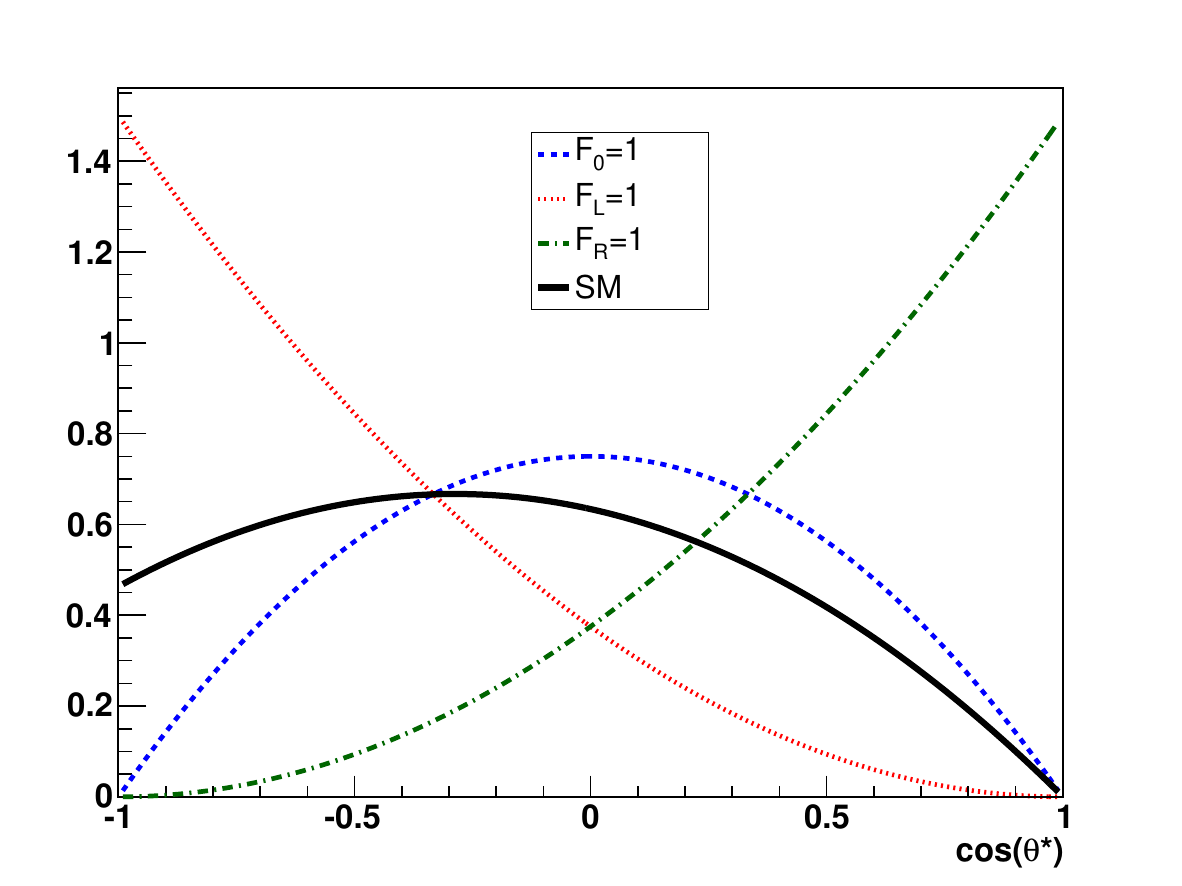}
\caption{\label{SingletopEFTprop:HelicityFractions} The differential decay rate as a function of $cos \theta^{*}$ in several scenarios for $W$ boson polarization~\cite{CMS:2016asd}.}
\end{figure}

The V--A structure of the electroweak theory, together with the mass of the particles involved, determine the fractions of longitudinal, left-handed, and right-handed $W$ boson polarization (sometimes called helicity fractions), denoted, respectively, as $F_0$, $F_L$, and $F_R$. Predictions for these fractions computed at NNLO in pQCD are~\cite{Czarnecki:2010gb} $F_0 = 0.687 \pm 0.005$, $F_L = 0.311 \pm 0.005$, and $F_R = 0.0017 \pm 0.0001$. Experimentally, the~fractions can be measured within the $W$ rest frame where the $W$ boson arises from leptonic top decay, using the angle $\theta^{*}$, defined as the angle between the direction of the charged lepton and the reversed direction of the b-quark. The~differential decay rate is:
\begin{linenomath}
\begin{equation}
\frac{1}{\Gamma}\frac{d\Gamma}{dcos \theta^{*}} = \frac{3}{4} (1-cos^2 \theta^{*}) F_0 + \frac{3}{8} (1-cos \theta^{*})^2 F_L + \frac{3}{8} (1+cos \theta^{*})^2 F_R,
\end{equation}
\end{linenomath}
with $F_0+F_L+F_R=1$. 
The differential decay rate as a function of $cos \theta^{*}$ is illustrated in Figure~\ref{SingletopEFTprop:HelicityFractions}.

\vspace{-9pt}

The fractions are obtained from a fit of the $cos \theta^{*}$ distribution to the data. 
The $W$ boson polarization fractions have been measured at CDF and D0~\cite{CDF:2012dup} with a precision on $F_0$ of the order of 10-15\%, using $t\bar{t}$ decay. 
At the LHC, the~single-top production in the $t$-channel and its large cross-section offer the possibility of measuring the polarization fractions in single-top decay in addition to $t\bar{t}$ decay. 
The fractions were measured at 8~TeV with CMS~\cite{CMS:2014uod}, as $F_L = 0.298 \pm 0.028 (stat) \pm 0.032 (syst)$, \mbox{$F_0 =0.720 \pm 0.039(stat) \pm 0.037(syst)$}, and~$F_R =-0.018 \pm 0.019(stat) \pm 0.011(syst)$. 
The precision achieved with single-top measurements justifies its inclusion in ATLAS and CMS combination of 8 TeV results~\cite{CMS:2020ezf}, leading to $F_0 = 0.693 \pm 0.014$, $F_L = 0.315 \pm 0.011$, and $F_R = -0.008 \pm 0.007$. 
The 7 TeV results were obtained by analyzing $t\bar{t}$ and were not considered since they were expected to bring about negligible~improvement. 

ATLAS also employed the ``generalized helicity fractions and phases'' formalism~\cite{Boudreau:2013yna} by~the means of amplitude decomposition in several angular distributions in the top quark rest frame. 
Among the parameters measured, the~transverse polarization fraction using single-top, and decaying at 7 and 8 TeV~\cite{ATLAS:2015ryj, ATLAS:2017rcx}, yields $F_T = F_L + F_R = 0.30 \pm 0.05$~\cite{ATLAS:2017rcx} as the best result. ATLAS also measures the phase between amplitudes for longitudinally and transverse $W$ bosons recoiling against left-handed b-quarks~\cite{ATLAS:2015ryj,ATLAS:2017rcx}, providing no sign of CP violation. From~this formalism, left- and right-handed fractions could in principle be~calculated. 

\subsection{$Wtb$ Effective Couplings and Interpretation in the SM~EFT}

The $Wtb$ effective couplings were also measured, either at CMS (as extracted from the $W$ boson polarization fractions~\cite{CMS:2014uod} or measured directly~\cite{CMS:2016uzc}) or by ATLAS, by analyzing the single-top amplitudes~\cite{ATLAS:2015ryj, ATLAS:2017rcx} or measuring various angular asymmetries~\cite{ATLAS:2017ygi}. The~Lagrangian describing the $Wtb$-effective couplings reads~\cite{Aguilar-Saavedra:2006qvv}:
\begin{linenomath}
\begin{equation}
\label{WtbCouplingsLagrangian}
L_{Wtb} = -\frac{g}{\sqrt{2}} \bar{b} \gamma^{\mu} (V_L P_L + V_R P_R) t W_{\mu}^{-} -\frac{g}{\sqrt{2}} \bar{b} \frac{i \sigma^{\mu\nu} q_{\nu}}{m_W} (g_L P_L + g_R P_R) t W_{\mu}^{-} + h.c.
\end{equation}
\end{linenomath}
where $V_L$, $V_R$ denote the vectorial left-handed and right-handed $Wtb$ couplings, and~$g_L$, $g_R$ denote the tensorial left-handed and right-handed $Wtb$ couplings (sometimes called, respectively, $f_V^L$, $f_V^R$, $f_T^L$, $f_T^R$, depending on the convention~\cite{CMS:2016uzc}). In~the SM at the LO in pQCD, $V_L = V_{tb}$ while $V_R=g_L=g_R=0$. The~couplings $V_R, g_L, g_R$ are complex and can be CP-odd if their imaginary part is non-zero. The~$V_{tb}$ CKM matrix element is inferred from the single-top cross-section measurement, as~discussed in Section~\ref{VtbSection}. 

The analyses have moderate sensitivity to the right-handed vectorial coupling and left-handed tensorial coupling. With~a simultaneous fit of both parameters, ATLAS reports $| V_R / V_L | < 0.37$ and $ | g_L / V_L | < 0.29 $ at 95\% CL~\cite{ATLAS:2017rcx}, and~CMS reports $f_V^R < 0.16$ and \mbox{$f_T^L < 0.057 $} at 95\% CL~\cite{CMS:2016uzc}, including $f_V^L$ in the fit using inclusive cross-section information. 
The best sensitivity on the $Wtb$ couplings is obtained on the $g_R$ coefficient. ATLAS obtained with a simultaneous fit $-0.12 < Re(g_R/V_L)< 0.17$ and $-0.07 < Im(gR /V_L) < 0.06$ at 95\% CL~\cite{ATLAS:2017rcx}. If~using single-top cross-section information and assuming a null imaginary part, CMS obtains $|Re(f_T^R)| < 0.046 $. 
These results can be compared with the combination of an 8 TeV $W$ boson polarization fraction (including $t\bar{t}$ channels)~\cite{CMS:2020ezf}: $-0.11 < Re(V_R) < 0.15$, $-0.08 < Re(g_L) < 0.05$, and $-0.04 < Re(g_R) < 0.02$. 
Since the imaginary part of $g_R$ cannot be accessed easily from the \ttbar\ process and would need a dedicated analysis~\cite{Aguilar-Saavedra:2006qvv}, the~single-top measurements, such as~\cite{ATLAS:2015ryj,ATLAS:2017rcx}, are~irreplaceable. 

The results obtained in the effective coupling formalism can be translated into the modern framework of the SM EFT~\cite{Buckley:2015lku}, adding all operators to the SM Lagrangian and respecting gauge invariance. The~$Wtb$ couplings considered in Equation~(\ref{WtbCouplingsLagrangian}) ($V_L$, $V_R$, $g_L$, $g_R$) are, respectively, related to the following four dimension-6 operators:
\begin{linenomath}
\begin{equation}
O_{\phi q}^{(3)} = \frac{c_{\phi q}^{(3)}}{\Lambda^2} i (\phi^{\dagger} \overleftrightarrow{D_{\mu}}^I \phi) (\bar{q}\gamma^{\mu} \tau^I q) ,
\end{equation}
\begin{equation}
O_{\phi t b} = \frac{c_{\phi t b}}{\Lambda^2} (\phi^{\dagger} \overleftrightarrow{D_{\mu}}^I \phi) (\bar{t}\gamma^{\mu} \tau^I b),
\end{equation}
\begin{equation}
O_{tW} = \frac{c_{tW}}{\Lambda^2} (\bar{q}\sigma^{\mu\nu} \tau^I t) \widetilde{\phi} W_{\mu\nu}^I,
\end{equation}
\begin{equation}
O_{bW} = \frac{c_{bW}}{\Lambda^2} (\bar{q}\sigma^{\mu\nu} \tau^I b) \phi W_{\mu\nu}^I,
\end{equation}
\end{linenomath}
using notations from~\cite{Grzadkowski:2010es}. 
Results from the combination of the $W$ boson polarization at 8 TeV are~\cite{CMS:2020ezf}: $-3.48 < Re(c_{\phi t b}) < 5.16$, $-0.48 < Re(c_{tW}) < 0.29$, and $-0.96 < Re(c_{bW})< 0.67$. A~translation from the best measurement of $Im(g_R)$~\cite{ATLAS:2017rcx} to the EFT formalism using~\cite{Buckley:2015lku} gives $-0.82 < Im(c_{bW}) < 0.70$. 

\subsection{Top Quark~Polarization}

\begin{figure}[H]
\includegraphics[width=0.5\textwidth]{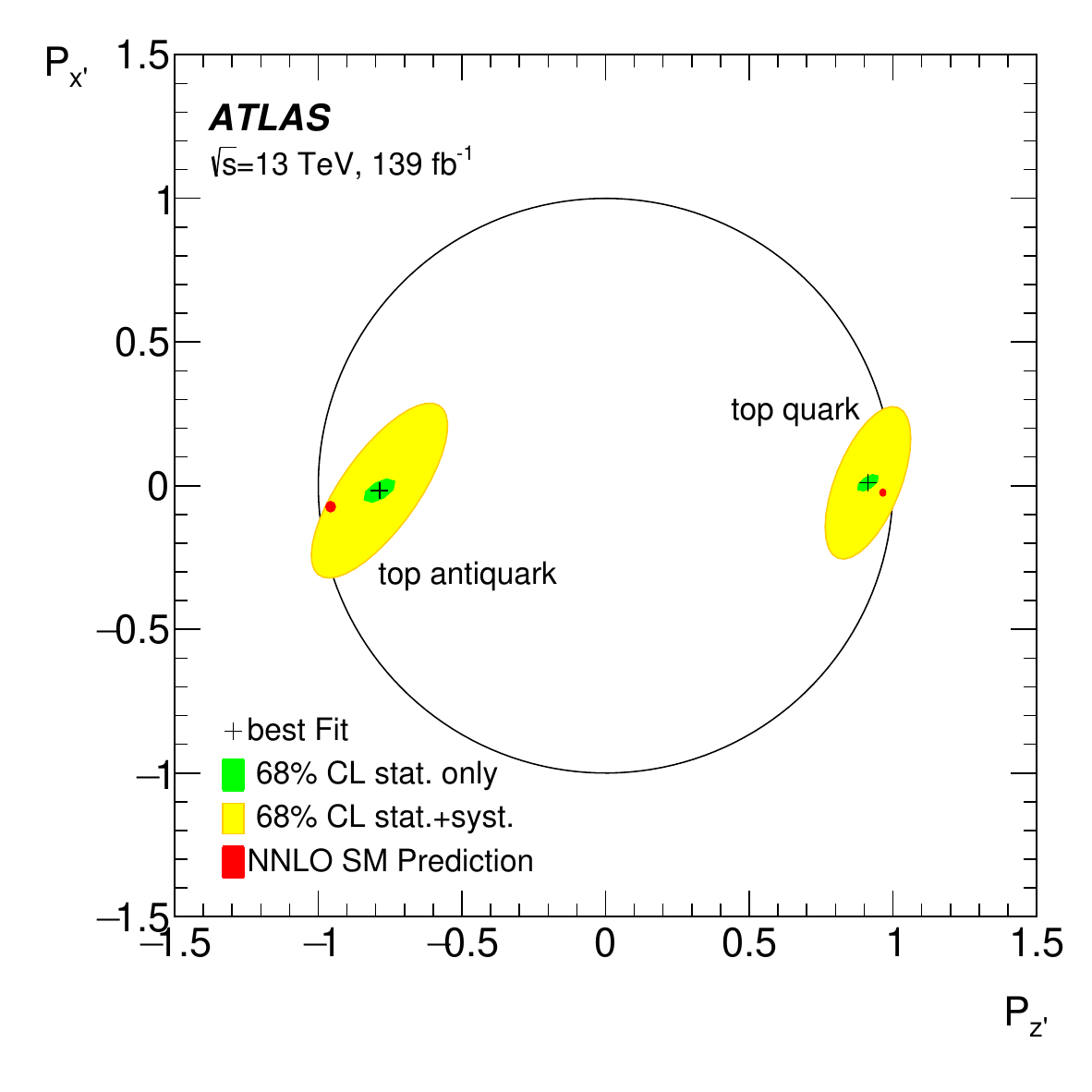}
\caption{\label{SingletopEFTprop:Polarization}Top quark polarization in the single-top $t$-channel production along $x'$ and $z'$ directions for the top and antitop quarks~\cite{ATLAS:2022vym}.}
\end{figure}

Recently, via~an analysis of the top quark polarization, ATLAS directly measured  the coefficient $Im(c_{tW})$ for the first time~\cite{ATLAS:2022vym}, using the full Run 2 dataset at 13 TeV. 
Because of parity conservation in QCD, top quarks in $t\bar{t}$ production are unpolarized, while top quarks are mostly polarized in single-top production. The~polarization vector $\vec{P}$ is defined with components $P_i = 2 < S_i >$, where $S_i$ is the top quark spin along the $i$ direction~\cite{Aguilar-Saavedra:2014eqa}, in~the top quark rest frame, where the $z'$ direction is defined as the $W$ boson direction, the~$x'$ direction is defined as the spectator quark direction projected on the transverse plane, and the $y'$ axis completes the direct basis. 
On this basis, the~values of the polarization vectors are close to $(-0024,0,0.965)$ for the top quark and $(-0.073,0,-0.957)$ for top antiquark produced in the $t$-channel at NNLO in pQCD~\cite{ATLAS:2022vym}. 
The top quark polarization can be extracted from angular distributions of top decay products defined in the top quark rest frame, given by the following general formula: 
\begin{linenomath}
\begin{equation}
\frac{1}{\Gamma}\frac{d\Gamma}{dcos \theta_{X}} = \frac{1}{2}(1+\alpha_X P_X cos \theta_X),
\end{equation}
\end{linenomath}
where $\theta_X$ is the angle between the top quark spin axis and the direction of motion of the chosen decay $X$, $\alpha_X$ is the spin analyzing power associated with the $X $, and $P_X$ is the top quark degree of polarization along the direction of $X$. 
The measurement of top quark polarization in~\cite{ATLAS:2022vym} is performed using the angular distributions related to the charged lepton (shown to have the largest spin analyzing power, close to 1) arising from the top decay and projected in the previously defined directions. 
If the top quark polarization had previously been measured at the LHC along the $z$ direction (for instance in~\cite{ATLAS:2017ygi}), the~measurement~\cite{ATLAS:2022vym} is the most precise and includes $x'$ and $y'$ directions. For top and antitop quarks, it leads, respectively, to $P_x'=0.01 \pm 0.18$, $P_y'=-0.029 \pm 0.027$, $P_z'=0.91 \pm 0.10$, and $P_x'=-0.02 \pm 0.20$, $P_y'=-0.007 \pm 0.051$, $P_z'=-0.79 \pm 0.16$. The~polarizations along directions $x'$ and $z'$ are also reported in Figure~\ref{SingletopEFTprop:Polarization}. 
Using the same angular distributions, ATLAS reports $-0.9 < Re(c_{tW}) < 1.4$ and $-0.8 < Im(c_{tW}) < 0.2$ at 95\% CL. 
\vspace{-9pt}
\unskip

\subsection{Discussion on other Couplings with Single-Top Quark Measurements in the SM~EFT}

If anomalous coupling measurements in single-top quark processes are primarily interesting for $Wtb$ couplings, other couplings are also actively measured, e.g., the coupling between heavy quarks and light quarks, the~coupling between heavy quarks and neutral bosons, and the coupling between heavy quarks and leptons. 
The discussion in this section excludes the FCNC (for a review, see~\cite{universe8110609}). 

In general, single-top production with a boson can help constrain the coupling between top quarks and neutral bosons. 
The top-Z (resp. top-Higgs) coupling impacts the single-top quark produced in association with a $Z$ boson (resp., a Higgs boson). 
The top-gluon coupling impacts the $tW$ channel (since $tW$ channel LO diagrams feature one gluon in the initial state) and any production channel considered at NLO, where gluons can be emitted from top quarks. 
The process of single-top production accompanied by a photon has just been observed and could be used in the near future for measuring the top-$\gamma$ coupling~\cite{Fael:2013ira}. 
It has also been emphasized that the $tZq$ and $tHq$ processes can be greatly impacted by some of these couplings~\cite{Degrande:2018fog}. 
However, the~cross-sections for processes of single-top production in association with bosons ($t+V$) are lower than those of top pairs produced in association with bosons ($t\bar{t}+V$); therefore, analyses of $t+V$ final states are generally swamped by $t\bar{t}+V$ backgrounds. As~a consequence, measuring the $tZ$, $tH$, and $tg$ couplings requires, for~consistency,~the modeling of the anomalous couplings in $t+V$ and $t\bar{t}+V$ simulation samples, which will help in constraining the couplings. 
It is difficult to disentangle what is the exact contribution of single-top production to the sensitivity in these couplings. We will, therefore, limit ourselves to providing some examples, where the contributions of single-top processes are explicitly included. 
Generic searches for measuring top quark couplings in the multi-lepton lepton final state define many event classes, targeting a great number of EFT operators that impact $tZq$ and $tHq$ processes~\cite{CMS:2020lrr}, e.g., nine operators involving two quarks and one or more bosons (with some impacting the $Wtb$ vertex considered at the production level only), as~well as seven operators involving two heavy quarks and two leptons. An~updated analysis~\cite{CMS-PAS-TOP-22-006} involving more operators needs to be published. 
The top gluon coupling was considered in~\cite{CMS:2020lrr} by including its impact on gluon radiation at LO. 
Measurements of EFT operators in $t\bar{t}Z+tZq$ final states~\cite{CMS:2021aly} include five operators involving two quarks and one or more bosons (including $Wtb$ vertex) and uses machine learning to maximize sensitivity. 
The Yukawa coupling is measured in $t\bar{t}H$ analyses by including its impact on $tHq$, as~discussed in Section~\ref{SingleTopTHQ}. 

A recent measurement of the $t$-channel process using full Run 2 data by ATLAS~\cite{ATLAS-CONF-2023-026} (to be published) sets constraints on the coupling between light and heavy quarks (the $C_{q,Q}^{(1,3)}$ coefficient within the SMEFT framework), in~a competitive manner with global fits reinterpreting LHC~data. 

The EFT is a consistent framework-preserving gauge invariance in a model-independent way~\cite{Degrande:2012wf} (as long as new physics appears at a high energy scale);~therefore, there is a tendency to employ the EFT framework more  widely, replacing previous anomalous coupling frameworks. 
The LHC$top$WG, together with the LHC EFT WG, are working on prescriptions toward the combination of direct top quark EFT~measurements.

\section{\label{Conclusions}Conclusions}

\begin{figure}[H]
  \includegraphics[width=.9\textwidth]{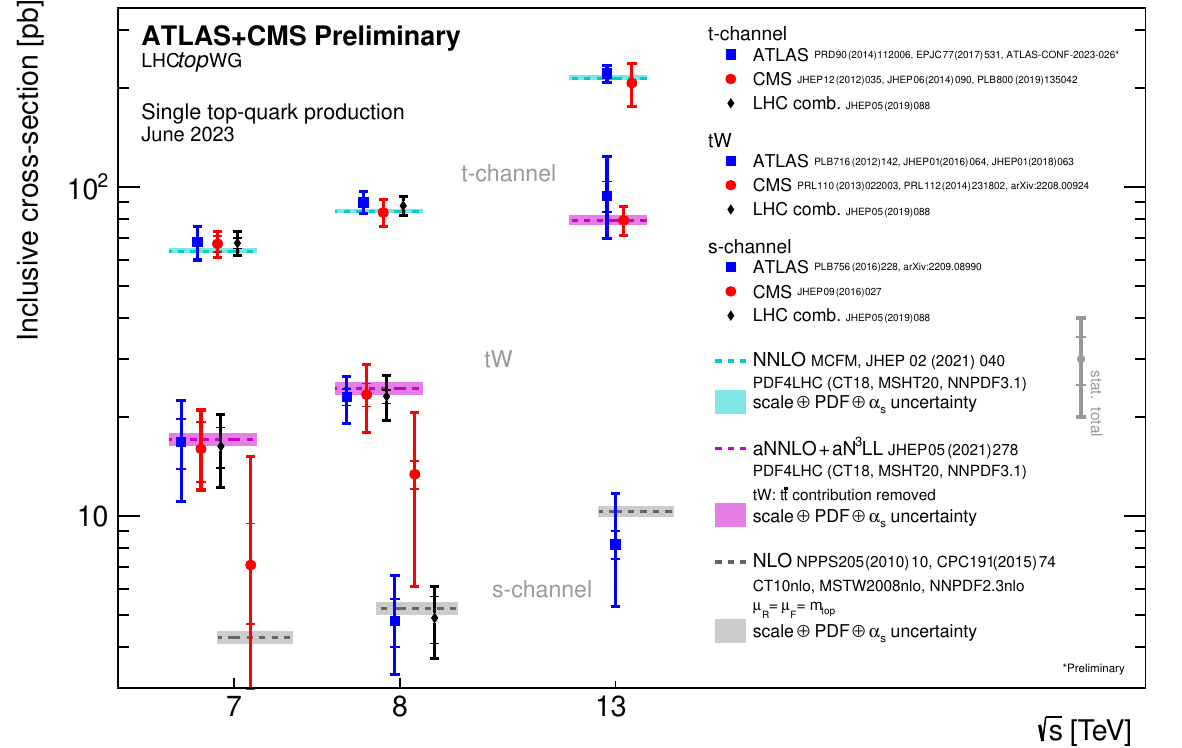}
  \caption{Summary of single-top quark cross-section measurements by ATLAS and CMS in the $t$-channel, $tW$ production, and $s$-channel, as~functions of the center of mass energy, compared with theoretical predictions at NNLO~\cite{Campbell:2020fhf}, approximate NNLO+N$^{3}$LL~\cite{Kidonakis:2021vob}, and NLO~\cite{Campbell:2010ff,Kant:2014oha} accuracy, provided by the LHC$top$WG~\cite{LHCtopWGsummaryFigures}.} 
  \label{fig:tXsummaryXsec}
\end{figure}

After more than 10 years of data-taking with the LHC, the~understanding of the physics involving single-top quark processes has undergone a spectacular change. 
Prior to the LHC, a~single-top production was discovered, singling out the $t$- and $s$-channels. 
Nowadays, 14 years on, the~differential cross-sections for the $t$-channel and $tW$ production modes are measured in great detail. The~$t$-channel is routinely used for top quark property measurements. This ranges from the structure of the $Wtb$ vertex to the~$W$ boson and top quark polarization, not to mention the top quark mass measurement. The~$tW$ production is employed to probe delicate interference effects with the $t\bar{t}$ process. The~$s$-channel process remains to be observed at the LHC, but~initial evidence suggests that such an observation is on the horizon.
The cross-sections measured for single-top quark production in the $t$-channel, $tW$ channel, and $s$-channel by ATLAS and CMS are compared with theoretical predictions in Figure~\ref{fig:tXsummaryXsec}.

\begin{figure}[H]
  \includegraphics[width=1.0\textwidth]{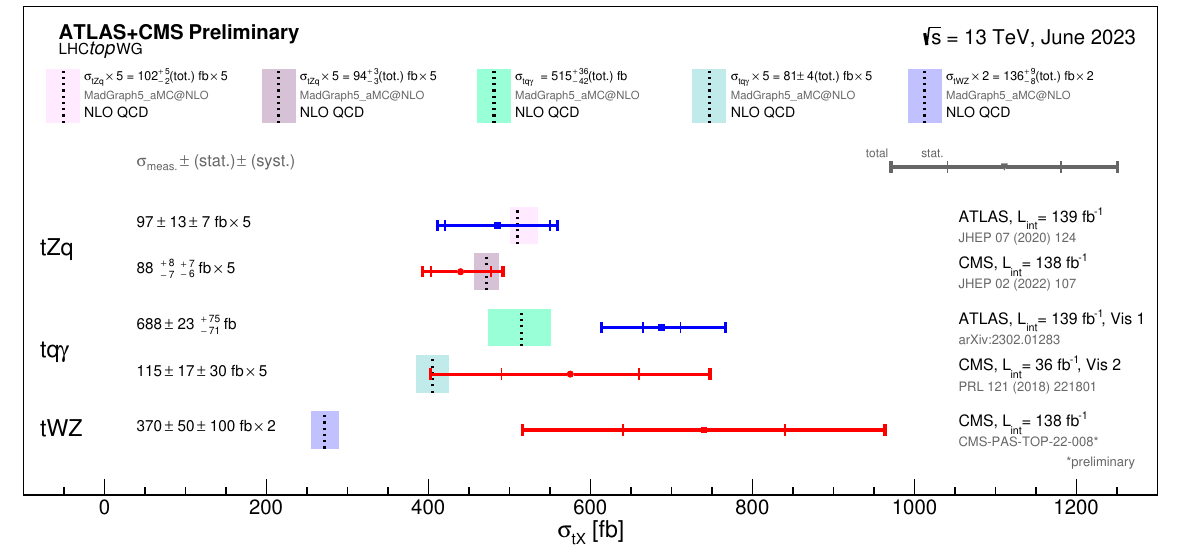}
  \caption{Summary of single-top quark cross-section measurements by ATLAS and CMS in the associated production with a $Z$ boson or a photon compared with theoretical predictions at NLO~\cite{Alwall:2014hca} accuracy; provided by the LHC$top$WG~\cite{LHCtopWGsummaryFigures}.} 
  \label{fig:tBosonsummaryXsec}
\end{figure}

Run 2 of the LHC offered a new opportunity for observing and exploring the associated production of top quarks and neutral bosons. 
After its observation, the~production of a single-top quark with an associated $Z$ boson was measured differentially for the first time. It is now employed as a probe of various couplings within~the SM EFT framework. The~associated production with a photon has also been observed with the full Run 2 dataset. The~$tH$ processes are used to probe the sign of the top quark Yukawa coupling; however,  they have not been observed as yet. 
The cross-sections measured by ATLAS and CMS for single-top quark production associated with a $\gamma$ or $Z$ boson are compared with theoretical predictions in Figure~\ref{fig:tBosonsummaryXsec}.

The Run 3 of the LHC is ongoing, with~a center-of-mass energy of 13.6 TeV, resulting in a predicted increase of about 11\% in the inclusive cross-section relative to 13 TeV for the three main production modes~\cite{LHCtopWGnnloCrossSections,Campbell:2020fhf,Kidonakis:2021vob,Liu:2018gxa} (and a similar increase for the $t\bar{t}$ process~\cite{LHCtopWGttbarCrossSections,Czakon:2011xx}). 
One can expect measurements for all of the processes discussed in this review to be performed at this unprecedented energy, verifying if the data still agree with the SM predictions. 
The luminosity is expected to accumulate during Run 3 in a way that is at least comparable to Run 2, producing a new dataset that is larger by a factor of 1.4 (around 140 fb$^{-1}$, collected separately by ATLAS and CMS during Run 2, and~200 fb$^{-1}$ during Run 3). 
Statistically dominated measurements will profit from this step in center-of-mass energy and expected luminosity, such as the measurement of the $t\gamma q$ process, where a first differential cross-section can be targeted. 
Hopefully, some of the limitations of the previous measurements will be lifted to improve the precision, provided that additional work on the systematic uncertainties is carried out. 
For instance, more work is needed on the topic of the parton shower modeling since it is now the largest source of uncertainties in the $t$-channel measurements and an important source in the measurement of the $tW$ production. 
Global efforts will also be needed to reduce the systematic uncertainties in the $tZq$ measurement, which is dominated by several sources of large experimental uncertainties. 
Most of the measurements of the top quark properties in the single-top quark area employ either the $t$-channel or the $tZ$ production and are impacted by the above uncertainty sources. 
For the top quark coupling measurements ($Wtb$ vertex and top-boson couplings), it is expected that the movement toward the generalized usage of the EFT will be pursued, allowing the search for new physics in precision measurements in a unified way and allowing for the combination of complementary measurements. 
Finally, there is hope that the $s$-channel process could be observed at LHC Run 3 by~reducing the uncertainties and refining the analysis techniques. 
On the other hand, the~search for the $tH$ processes will continue, although~their observation will have to be postponed to the HL-LHC, where it will remain a challenge~\cite{CMS:2018qgz}. 

Beyond these extensions of the already engaged single-top quark program, new possibilities can be explored at the HL-LHC. 
Using boosted top quarks with a jet substructure is one of them (already used for \ttbar\ measurements~\cite{CMS:2021vhb} or in Ref.~\cite{CMS-PAS-TOP-22-008}), since more events will be available in the tails of the distributions to search for new physics~\cite{Aguilar-Saavedra:2019ptp}. 
The process of producing three top quarks is occasionally categorized within single-top quark physics. It stands as a minor background in the four top quark process measurements~\cite{ATLAS:2023ajo,CMS:2023ftu}, and~deserves a direct search~\cite{Ahmed:2022hce}. 
The measurement of the $tWZ$ production at Run 2 was the first of its kind, featuring a single-top quark accompanied by two bosons; its observation could be within reach, likely at the HL-LHC, where it could be used to probe the top-boson couplings~\cite{Faham:2021zet}. 
And even rarer processes can be reached, i.e., single-top quark production with a combination of two $W$,$Z$ bosons or photons could be measured beyond $tWZ$; some studies suggest that the production of single-top quarks through vector boson fusion is another rare process to explore, offering high sensitivity to new top quark couplings~\cite{Maltoni:2019aot}. 
In general, the~program of measuring the top quark couplings within the EFT is still in its infancy. 
One can foresee that the couplings in which the single-top quark area is relevant will be measured systematically at the HL-LHC~\cite{Durieux:2022cvf}. For~instance, searching for a possible CP violation in the top-Higgs boson coupling will be conducted, where separating $tH$ from $t\bar{t}H$ will be crucial~\cite{Bahl:2020wee}, or~measuring the top quark couplings to the gauge bosons, such as the top-$\gamma$ coupling~\cite{Fael:2013ira}. 
Combining measurements from other top quark production modes, along with insights from~B physics, electroweak, and Higgs boson measurements, will certainly be essential, and could lead to the observation of statistical deviations indicative of physics beyond the~SM.

\vspace{6pt} 



\authorcontributions{Writing -- review and editing, J.A. and N.C. All authors have read and agreed to the published version of the manuscript. 

\funding{This research received no external~funding.}

\dataavailability{No new data were created or analyzed in this study. Data sharing is not applicable to this review. 
} 




\conflictsofinterest{The authors declare no conflict of~interest.}


\begin{adjustwidth}{-\extralength}{0cm}

\reftitle{References}
\PublishersNote{}
\end{adjustwidth}
\end{document}